\documentclass[floatfix,prd,twocolumn,letterpaper,lengthcheck,superscriptaddress,showpacs,amssymb,amsmath,amsfonts,aps,altaffilletter,nofootinbib,nopreprintnumbers,longbibliography]{revtex4-1}

\usepackage[dvipsnames]{xcolor}
\usepackage{color}
\usepackage{hyperref}
\usepackage{amsmath}
\usepackage{amsthm}
\usepackage{multirow}
\usepackage{float}
\usepackage{graphicx}
\usepackage{tikz}
\usetikzlibrary{calc}
\usepackage{placeins}
\usepackage{xspace}
\usepackage{physics,enumerate,ulem}

\usepackage{mathrsfs}
\DeclareSymbolFontAlphabet{\mathrsfs}{rsfs}

\newcommand{\be}{\begin{equation}}
\newcommand{\ee}{\end{equation}}

\begin{document}

\title[]{The Interior MOTSs of Spherically Symmetric Black Holes}

\author{Robie A. Hennigar}
\affiliation{
    Department of Mathematics and Statistics, Memorial University,
    St. John's, Newfoundland and Labrador, A1C 5S7, Canada
}
\affiliation{
	Department of Physics and Astronomy, University of Waterloo, 
	Waterloo, Ontario, Canada, N2L 3G1
}
\affiliation{
	Department of Physics and Computer Science, Wilfrid Laurier University, 
	Waterloo, Ontario, Canada N2L 3C5
}

\author{Kam To Billy Chan}
\affiliation{
    Department of Physics and Physical Oceanography, Memorial University, 
    St. John's, Newfoundland and Labrador, A1B 3X7, Canada
}

\author{Liam Newhook}
\affiliation{
    Department of Physics and Physical Oceanography, Memorial University, 
    St. John's, Newfoundland and Labrador, A1B 3X7, Canada
}

\author{Ivan Booth}
\affiliation{
    Department of Mathematics and Statistics, Memorial University,
    St. John's, Newfoundland and Labrador, A1C 5S7, Canada
}

\begin{abstract}
    There are notable similarities between the marginally outer trapped surfaces (MOTSs) present in the interior of a binary black hole merger and those present in the interior of the Schwarzschild black hole. Here we study the existence and properties of MOTSs with self-intersections in the interior of more general static and spherically symmetric black holes and coordinate systems. Our analysis is carried out in a parametrized family of Painlev{\'e}-Gullstrand-like coordinates that we introduce. First, for the Schwarzschild spacetime, we study the existence of these surfaces for various slicings of the spacetime finding them to be generic within the family of coordinate systems we investigate. Then, we study how an inner horizon affects the existence and properties of these surfaces by exploring examples: the Reissner-Nordstr\"om black hole and the four-dimensional Gauss-Bonnet black hole. We find that an inner horizon results in a finite number of self-intersecting MOTSs, but their properties depend sensitively on the interior structure of the black hole. By analyzing the spectrum of the stability operator, we show that our results for two-horizon black holes provide exact-solution examples of recently observed properties of unstable MOTSs present in the interior of a binary black hole merger, such as the sequence of bifurcations/annihilations that lead to the disappearance of apparent horizons.
\end{abstract}

\maketitle

\section{Introduction}
\label{sec:intro}

With recent advances in gravitational wave astronomy, there remains little doubt that black holes are ubiquitous in Nature. But what is a black hole? The problem of defining a black hole often centers on understanding the boundary between the black hole region and the rest of the Universe: the \textit{horizon}. However, the practical problem of characterizing black hole boundaries is not trivial. 

The most familiar and often used characterization of a black hole is the event horizon. The event horizon is mathematically clean to define and obeys a number of useful theorems, for example, concerning its topology and area increase~\cite{hawking_ellis_1973}. The original formulation of the laws of black hole thermodynamics, which continue to provide foundational insights into quantum gravity, describe properties of the event horizon~\cite{Bardeen:1973gs}. But while theoretically very useful, the event horizon suffers from practical limitations. Foremost among these are that the event horizon is telelogical and depends on the global nature of spacetime (e.g.~the structure of infinity). As a consequence, event horizons themselves are generically unobservable~\cite{Visser:2014zqa}: they cannot be located at a moment of time. 

These drawbacks motivate alternative characterizations of black holes. While there exists a menagerie of alternatives to the event horizon~\cite{Booth:2005qc}, here our focus will be marginally outer trapped surfaces (MOTSs). MOTSs are natural candidates for black hole boundaries through their relationship with trapped surfaces. A two-dimensional closed, space-like surface in a four-dimensional spacetime is said to be trapped if light rays emitted normal to the surface converge in the future in both the inward and outward directions. Trapped surfaces are a property of strong gravitational fields; under suitable assumptions they exist only within the event horizon and imply the existence of a singularity in the future~\cite{Penrose:1969pc}. A MOTS is a limiting case of a trapped surface where the expansion in the outward direction vanishes (and no constraint is placed on the expansion in the inward direction). Therefore, a continuity argument suggests that the boundary of a trapped region will be a MOTS  (though not the converse: not all MOTS are boundaries of trapped regions). As such, it is not so surprising that the spatial slices of event horizons for stationary black holes are MOTSs. But unlike the event horizon, a MOTS can be located at a given moment of time, making them better suited for study in highly dynamical situations, such as those encountered in numerical relativity. In this context, the outer-most MOTS on a given slice is typically referred to as the apparent horizon.

There are additional good reasons to consider MOTSs as candidates for black hole boundaries. MOTSs can be assigned physical quantities such as mass and angular momentum~\cite{Dreyer:2002mx, Krishnan:2007va, Gupta:2018znn}, and the laws of black hole mechanics can be extended to hold also for MOTSs~\cite{Hayward:1993mw,Ashtekar:2002ag, Ashtekar:2003hk,Ashtekar:2004cn}. MOTSs also play a role in holography, where they are the surfaces that arise in defining a coarse-grained entropy in dynamical situations~\cite{Engelhardt:2017aux}.  Key diagnostic information about a MOTS is provided by the stability operator and its spectrum~\cite{Andersson:2005gq,Andersson:2007fh}. Much attention has been paid to the case of the principal (smallest) eigenvalue of the stability operator. When the principal eigenvalue is positive or zero, the MOTS is said to be stable, and otherwise said to be unstable. Stable MOTSs possess the properties expected of black hole boundaries: they serve as barriers between trapped and untrapped regions, and the area of these surfaces increases in time. Thus, as advocated for in~\cite{Booth:2021sow,Pook-Kolb:2021hvz}, we reserve the term ``apparent horizon'' for the case of stable MOTSs. If a given MOTSs is strictly stable (i.e. with positive principal eigenvalue), then the surface is guaranteed to evolve smoothly into the future, while moments of apparent horizon formation/annihilation will coincide with the vanishing of the principal eigenvalue.


In recent years, considerable effort has been dedicated to understanding the behaviour of MOTSs in highly dynamical situations, such as during a binary black hole merger. When there is a departure from stationarity, the event and apparent horizons can behave very differently. This led to questions about whether or not MOTSs are well-behaved in these circumstances, and whether they provide a physically reasonable description of the underlying physics. The culmination of this effort has been a detailed understanding of the analog of the ``pair of pants'' diagram for the apparent horizon~\cite{pook-kolb:2018igu,  PhysRevLett.123.171102,PhysRevD.100.084044, pook-kolb2020I, pook-kolb2020II, Pook-Kolb:2021jpd,Booth:2021sow,Pook-Kolb:2021hvz}. This is considerably more complicated than the case of the event horizon. The final picture involves a potentially infinite number of MOTSs whose world-tubes weave forwards and back through time, with the apparent horizons of the two original participants in the merger being annihilated by additional, exotic MOTSs that were generated  in the interior during the merger.

These studies have highlighted a number of features that merit further investigation, some of which will be addressed in this work.  In~\cite{PhysRevLett.123.171102,PhysRevD.100.084044, Pook-Kolb:2021jpd,Pook-Kolb:2021hvz} the physical importance of unstable MOTSs was highlighted. Despite being unstable, and therefore the smoothness of time evolution not being guaranteed by the established theorems~\cite{Andersson:2005gq,Andersson:2007fh}, in all cases explored numerically these surfaces evolved smoothly and with good behaviour, just like their stable counterparts. For example, in~\cite{Pook-Kolb:2021jpd,Pook-Kolb:2021hvz} it was observed that associated to every formation/annihilation of unstable MOTSs is the vanishing of one of the eigenvalues of the stability operator, though not necessarily the principal one. 
%
%
  This feature suggests an organizing principle by which the behaviour of unstable MOTSs may be understood analytically. It was shown in~\cite{PhysRevLett.123.171102} that inclusion of the world-tube traced out by the unstable inner common MOTS allows for a connected (albeit nonsmooth) sequence of MOTSs  between the initial and final states of the merger, thereby allowing physical quantities associated to the black holes to be tracked from the initial to final states of the merger~\cite{pook-kolb2020I, pook-kolb2020II}. In~\cite{Pook-Kolb:2021jpd,Pook-Kolb:2021hvz}, it was shown that the unstable MOTSs present in the interior of a merger are essential to understanding the final fate of the initial two black holes. It has long been known~\cite{hawking_ellis_1973} that the apparent horizons of the  two original black holes continue to exist even after the formation of the common apparent horizon. While there has long been speculation about the final fate of those horizons~\cite{BenDov:2004gh,Booth:2005ng}, it is now known that they are annihilated by additional, unstable MOTSs that form after the appearance of the common apparent horizon.

In~\cite{PhysRevLett.123.171102} a novel geometrical property of unstable MOTSs was pointed out when it was shown that the inner common MOTS present in a binary merger ultimately develops self-intersections. It was then shown that MOTSs possessing self-intersections are, in fact, quite common: an apparently infinite number of such surfaces were found in the interior of the Schwarzschild spacetime in a Painlev{\'e}-Gullstrand (PG) slicing~\cite{Booth:2020qhb}. It was also argued in that work that such surfaces may be relevant to understanding the interior dynamics of black hole mergers in the extreme mass ratio limit framework of~\cite{Emparan:2016ylg}. It is our purpose here to better understand the properties of self-intersecting MOTSs in static and spherically symmetric spacetimes. 

Our paper is organized in the following way. We begin by presenting what we believe is a novel generalization of PG coordinates for static and spherically symmetric line elements, allowing for the construction of horizon-penetrating coordinate systems in cases where the standard PG coordinates fail. We go on to present a derivation of the equations defining MOTSs in these spacetimes, along with a discussion of the stability operator. Our main results begin with a discussion of the Schwarzschild spacetime, reviewing the self-intersecting surfaces first found in~\cite{Booth:2020qhb}, and showing that these surfaces are generic within a large family of PG-like slicings. We then consider black holes with inner (Cauchy) horizons, focusing on two examples: the Reissner-Nordstr\"om  black hole, and the black hole of four-dimensional Gauss-Bonnet gravity~\cite{Glavan:2019inb, Lu:2020iav, Fernandes:2020nbq, Hennigar:2020lsl}. The presence of an inner horizon renders the number of interior MOTSs finite, but we find in general the properties of these surfaces depend sensitively on the interior structure of the black hole in question. In the case of Reissner-Nordstr\"om, when viewed as a function of charge, the self-intersecting MOTSs are ``absorbed'' by the inner horizon. In the case of the Gauss-Bonnet black hole, there are pairs of surfaces (with a given number of loops) that, as a function of the Gauss-Bonnet coupling, form through smooth bifurcations. In all cases, our results are supported by an analysis of the spectrum of the stability operator, providing exact-solution examples of the various features observed in~\cite{PhysRevLett.123.171102,PhysRevD.100.084044, Pook-Kolb:2021jpd,Pook-Kolb:2021hvz}.


\section{General Considerations}

\subsection{Generalizing Painlev{\'e}-Gullstrand Coordinates}

Painlev{\'e}-Gullstrand coordinates 
\be 
ds^2 = - \left(1- \frac{2M}{r} \right) d\tau^2 + 2 \sqrt{\frac{2M}{r}} d\tau dr + dr^2 + r^2 d \Omega^2 \, ,
\ee
provide a useful, horizon-penetrating foliation of the Schwarzschild spacetime. The hallmarks of this coordinate system are that the spatial sections are intrinsically flat and the observers who have proper time proportional to $\tau$, i.e. with four-velocity $u_{\rm PG} = - d\tau$, correspond to radially infalling geodesics who begin at infinity with zero velocity. 

In~\cite{Booth:2020qhb}, PG coordinates were used to study the axisymmetric interior MOTSs of the Schwarzschild black hole. The aim of that work was to draw a connection between recent advances in understanding the interior of a black hole merger~\cite{PhysRevLett.123.171102} and the exact description of the event horizon of a merger that is possible in the extreme mass ratio limit~\cite{Emparan:2016ylg}. It was found that the interior of the Schwarzschild spacetime contains an infinite number of MOTSs that possess self-intersections, like those found in~\cite{PhysRevLett.123.171102}, providing exact solution examples of this feature of a black hole merger. Our work here will build on the foundation of~\cite{Booth:2020qhb}, seeking to better understand the exotic MOTSs present in the interiors of black holes. To do so, we will continue to use a generalization of PG coordinates. There are at least a few reasons why PG coordinates are useful for this purpose. First, they are horizon-penetrating, allowing for both the interior and exterior regions to be studied consistently in a single chart. Second, they are non-static, which means that the expansion in the inward and outward direction are independent. As such, it seems reasonable to expect that PG slicings mimic ``generic'' slicings moreso than the often-used Schwarzschild coordinates.\footnote{In Schwarzschild coordinates, the expansion in the inward and outward directions are proportional. As such, MOTSs in Schwarzschild coordinates are  extremal surfaces.} Lastly, the surfaces of constant time are three-dimensional space-like surfaces. Therefore, any two-surface embedded in these surfaces will also be space-like.

The PG coordinate system can be quite easily generalized to any static, spherically symmetric spacetime. In the case where $g_{tt} g_{rr} = -1$, the resulting metric is 
\be
\label{ordinaryPG}
ds^2 = -f d\tau^2 + 2 \sqrt{1-f} d\tau dr +  dr^2 + r^2 d\Omega^2 \, .
\ee
While the above hallmark features still hold for this metric, an obvious problem with this form is that it is valid only in regions where $f \le 1$. What this means is that PG coordinates will ultimately fail in a number of spacetimes of interest, such as the Reissner-Nordstr\"om spacetime and (anti) de Sitter space. Of course, the reason for this failure can be easily understood. For example, any timelike, radially infalling geodesic in the Reissner-Nordstr\"om spacetime will reach a turning point in the radial coordinate before it reaches the singularity (effectively, the singularity is gravitationally repulsive). Similarly, in anti de Sitter space, no time-like geodesics reach the asymptotic boundaries. Therefore, it is not so surprising that PG coordinates, which are built from such geodesic observers, ultimately fail in these spacetimes. Further examples and analysis of this can be found in~\cite{Faraoni:2020ehi}, while other generalizations of PG-like coordinates can be found in~\cite{Volovik:2003ga, Nielsen:2005af, Lin:2008xg}.

It will be convenient for the work done here to have a PG-like horizon-penetrating coordinate system that works also in these cases. To do so, we begin with the line element of static, spherically symmetric spacetime written in Eddington-Finkelstein coordinates
\be 
ds^2 = -f  dv^2 + 2 dv dr + r^2 d\Omega^2 \, ,
\ee
and perform the coordinate transformation
\be 
v = \tau + g(r) \Rightarrow dv = d\tau + g' dr \, , 
\ee
with $g(r)$ an arbitrary function.  After this transformation, and the identification
\be 
p(r) \equiv g'(2- f g') \, ,
\ee
the metric becomes
\be\label{GenPG_metric} 
ds^2 = -f d\tau^2 + 2 \sqrt{1-p(r) f}d\tau dr + p(r) dr^2 + r^2 d\Omega^2 \, .
\ee
The function $p(r)$ appearing above is arbitrary, subject only to the condition $p(r) > 0$, to ensure that the constant time surfaces are space-like, and the condition $p f <1$, to ensure that the chart is everywhere valid. As it enters into the metric through a coordinate transformation, the Einstein equations are in no way dependent on this function, as can be easily confirmed through a direct computation. 

While the metric~\eqref{GenPG_metric} appears very similar to the standard PG coordinate system, there are important differences. First, for general function $p(r)$ the observers with four-velocity $u = - d\tau / \sqrt{p(r)}$ are no longer geodesic. The acceleration is given by 
\be 
|a| = \sqrt{(u^\alpha \nabla_\alpha u^\beta)(u^\gamma \nabla_\gamma u_\beta)} =  \frac{|p'|}{2 p^{3/2}} \, .
\ee
Therefore, we see that in the case where $p(r)$ is constant these observers are geodesic, but in general they are not. When $p(r)$ is a constant, the coordinate system reduces to the generalized PG slicing of Martel and Poisson~\cite{Martel:2000rn}. In that case, the observers  move along radially infalling geodesics that begin at infinity with initial velocity 
\be 
v_\infty = \sqrt{1-p} \, .
\ee
As such, the Martel-Poisson coordinate system interpolates between `ordinary' PG coordinates for $p(r) = 1$ and Eddington-Finkelstein coordinates for $p(r) = 0$.

The second main difference is that the constant time surfaces are no longer flat. This is the case even when $p(r)$ is constant but not equal to one. However, for any choice of $p(r)$, it is straight-forward to show that the Cotton tensor vanishes. Thus, in all cases the three-metric is conformally flat. This can be seen explicitly via the following change of coordinates:
\be\label{conform} 
z = \frac{r}{\Omega} \cos \theta \, , \quad \rho = \frac{r}{\Omega} \sin \theta
\ee
with 
\be 
\Omega = r \exp \left[-\int \frac{\sqrt{p(r)}}{{r}} dr \right] \, ,
\ee
after which the three-metric becomes
\be 
ds_3^2 = \Omega^2 \left(dz^2 + d\rho^2 + \rho^2 d \phi^2 \right) \, .
\ee

The main advantage of the coordinate system we have introduced is that,
by a suitable choice of $p(r)$, it is possible to construct PG-like coordinate systems for spacetimes in which the usual radial geodesics turn around at some radius. As mentioned, this will generally mean that the observers associated to the coordinate system will be accelerating. However, as explained earlier, this is not unnatural since in spacetimes like Reissner-Nordstr\"om, the radially infalling time-like geodesics are `repelled' by the singularity. A natural idea for this case would be to construct a coordinate system based on an infalling \textit{charged} test particle moving in the Reissner-Nordstr\"om spacetime. This possibility is actually included within the family of coordinate systems we have introduced above. For example, the Lorentz force equation
\be 
u^\alpha \nabla_\alpha u^\beta  = \mu F^\beta{}_\alpha u^\alpha
\ee
(with $\mu \equiv q/m$ the charge to mass ratio of the test partial) is satisfied for the choice
\be\label{LorentzP} 
p(r) = \frac{r^2}{(C r - \mu Q)^2} \, ,
\ee
where $C$ is an arbitrary constant and $Q$ is the black hole charge. The integration constant $C$ is related to the velocity at infinity:
\be 
\lim_{r \to \infty} p(r) = \frac{1}{C^2} =  \sqrt{1-v_\infty^2} \, ,
\ee
and we see that in the cases $\mu = 0$ or $Q=0$ (i.e. either the test particle or the black hole is uncharged), then this choice of $p(r)$ reduces to the Martel-Poisson family of generalized PG metrics.
Taking $C > 0$, and provided that $\mu Q < 0$, so that the test particle carries the opposite charge of the black hole, then the combination $p f$ has a maximum of $p f  = 1/\mu^2$, which occurs at $r = 0$. Thus, for example, taking the charge to mass ratio of the test particle to be unity, the PG-like chart extends all the way to $r = 0$.\footnote{The magnitude of the acceleration in this case is $a = |Q||\mu|/r^2$ --- exactly what would be expected for a charged particle moving in a Coulomb field.} While~\eqref{LorentzP} provides a physically motivated choice for $p(r)$ for the charged black hole, we stress that this is not necessary, and in practice the most convenient choice of $p(r)$ for the given problem can be chosen.

\subsection{Expansion \& Marginally Outer Trapped Surfaces}

Now we will determine the conditions under which surfaces in this spacetime are MOTSs. To do so, we will use the method introduced in~\cite{Booth:2021sow} based on the concept of MOTSodesics. Here we are concerned with two-dimensional space-like surfaces $S$ in a four-dimensional static and spherically symmetric spacetime $(g_{\alpha \beta}, \nabla_\alpha)$. At any point on such a surface, the normal space can be spanned by two null vectors, which we will label as $\ell^+$ and $\ell^-$. We take both of these vectors to be future-pointing with $\ell^+$ being outward-directed and $\ell^-$ inward-directed. We cross-normalize these vectors such that $\ell^+ \cdot \ell^- = -1$. The spacetime metric will induce a space-like two-metric $q_{AB}$ on the surface $S$: 
\be 
q^{AB} e_A^\alpha e_B^\beta = q^{\alpha\beta} = g^{\alpha \beta} + \ell^{+\alpha}\ell^{-\beta} + \ell^{-\alpha}\ell^{+\beta} \,  
\ee
and the associated expansions are then written as
\be 
\theta^\pm = q^{\alpha\beta} \nabla_\alpha \ell^\pm_\beta \, . 
\ee

To exploit the insights of~\cite{Booth:2021sow}, we first need to introduce some further concepts. We foliate the four-dimensional spacetime in terms of surfaces of constant time. The induced metric on these surfaces will be denoted $h_{ij}$ and the future-pointing time-like normal is
\be 
u = - \frac{d\tau}{\sqrt{p(r)}} \, .
\ee
The hypersurface metric $h_{ij}$ can further be decomposed as
\be 
h_{ij} = N_{i} N_{j} + T_{i} T_{j} + \hat{\phi}_i \hat{\phi}_j
\ee
where $N$ is the unit normal to $S$, $T$ is the unit tangent to $S$, and $\hat{\phi}$ is the unit vector in the azimuthal direction. The surface $S$ is given by ascribing $(r,\theta) = (r(s), \theta(s))$, and the metric on this surface can be written in terms of the same quantities defined just above:
\be 
q_{ij} = T_i T_j + \hat{\phi}_i \hat{\phi}_j \, , 
\ee 
while the null normals take the form
\be 
\ell^+ = \frac{1}{2} \left[ u + N \right] \, , \quad \ell^- = u - N \, .
\ee
For the metric~\eqref{GenPG_metric}, these quantities have the explicit forms:
\begin{align}
T &= \dot{r} \partial_r + \dot{\theta} \partial_\theta \, , 
\\
N &= r \sqrt{p(r)} \left[\frac{\dot{\theta}}{p(r)} \partial_r - \frac{\dot{r}}{r^2} \partial_\theta \right] \, ,
\\
\hat{\phi} &= \frac{\partial_\phi}{r \sin \theta} \, ,
\end{align}
and we are working in an arclength parameterization defined by the condition
\be 
p(r) \dot{r}^2 + r^2 \dot{\theta}^2 = 1 \, .
\ee

The approach introduced in~\cite{Booth:2021sow} exploits axisymmetry to reduce the problem of finding a two-surface that satisfies $\theta^+ = 0$ to the problem of finding a curve that satisfies the MOTSodesic equations:
\be 
T^i D_i T^j = \kappa N^j
\ee
where 
\be 
\kappa = k_u - N_j \hat{\phi}^i D_i \hat{\phi}^j \, ,
\ee
with $k_u$ the trace of the extrinsic curvature of $u$ in $S$. The full MOTS $S$ is then obtained as the surface of revolution of the MOTSodesic.

The components of the extrinsic curvature $K_{ij} = e_i^\alpha e_j^\beta \nabla_\alpha u_\beta$ (with $e_i^\alpha \equiv \partial x^\alpha/\partial y^i$) can be worked out to be
\begin{align}
K_{rr} &= \frac{p^2 f' + p'}{2 \sqrt{p} \sqrt{1-pf}} \, , 
\\
K_{\theta\theta} &= - \frac{r \sqrt{1-pf}}{\sqrt{p}} \, , 
\\
K_{\phi \phi} &= - \frac{r \sin^2 \theta \sqrt{1-pf}}{\sqrt{p}} \, ,
\end{align} 
which then gives
\be 
k_u = q^{ss} K_{ss} + q^{\phi \phi} K_{\phi \phi} = \dot{r}^2 K_{rr} + \dot{\theta}^2 K_{\theta\theta} + \frac{K_{\phi \phi}}{r^2 \sin^2 \theta} \, .
\ee
A simple calculation shows that
\be 
N_j \hat{\phi}^i D_i \hat{\phi}^j = \frac{1}{r \sqrt{p}} \left[p \dot{r} \cot \theta  - r \dot{\theta} \right]
\ee
which, when combined with the above, gives
\begin{widetext}
\be 
\kappa = - \frac{1}{r \sqrt{p}} \left[p \dot{r} \cot \theta  - r \dot{\theta} \right] + \frac{1}{2r \sqrt{p(1-pf)} } \left[r p^2 \dot{r}^2 f' + r \dot{r}^2 p' - 2(r^2 \dot{\theta}^2 + 1)(1 - p f)  \right] \, .
\ee
\end{widetext} 

We further find that
\be 
T^i T^k \Gamma_{ik}{}^j  \partial_j = \left(\frac{p' \dot{r}^2 - 2 r \dot{\theta}^2}{2 p} \right) \partial_r + \frac{2 \dot{r} \dot{\theta}}{r} \partial_\theta ,
\ee
from which it is then easy to work out the MOTSodesic equations:
\begin{align}
\label{motsodesics}
\ddot{r} &= - \frac{p' \dot{r}^2 - 2 r \dot{\theta}^2}{2 p}  + \frac{r \dot{\theta} \kappa}{\sqrt{p}}  \, ,
\\
\ddot{\theta} &= - \frac{2 \dot{r} \dot{\theta}}{r} - \frac{\sqrt{p} \dot{r} \kappa}{r} \, . 
\end{align}

A simple check shows that spherical surfaces for which $f(r_0) = 0$,
\be 
r = r_0 \, , \quad \theta = \frac{s}{r_0} \, ,
\ee
always solve the above system, independent of the arbitrary function $p(r)$. This confirms that necessary fact that event and Cauchy horizons are MOTSs. A second check confirms that, using the above equations, 
\be 
p' \dot{r}^3 + 2 p r \dot{r} \ddot{r} + 2 r \dot{r} \dot{\theta} + 2 r^2 \dot{\theta} \ddot{\theta} = 0 \, , 
\ee
as required for the system to be consistent with the arclength parameterization.

\subsection{The Stability Operator}

Important information about a MOTS is provided by the stability operator and its spectrum~\cite{Andersson:2005gq, Andersson:2007fh}. A more complete discussion of the stability operator in the notation and conventions we employ here can be found in~\cite{Booth:2021sow,Pook-Kolb:2021jpd}. The stability operator governs deformations of a given MOTS in directions normal to that surface. In the case of non-spinning, axisymmetric geometry, the stability operator takes the form:
\be 
L_\Sigma \psi = - \Delta_\mathcal{S} \psi + \left(\frac{1}{2} \mathcal{R} - 2 |\sigma_+|^2  - 2 G_{++} - G_{+-} \right)  \psi \, .
 \ee
The function $\psi$ above fully characterizes the deformation, parameterizing it in the directions normal to the surface. Here, $\Delta_\mathcal{S}$ is the Laplacian on the surface $\mathcal{S}$, $\mathcal{R}$ is the scalar curvature of that surface, $|\sigma_+|^2$ is the square of the shear $\sigma^+_{AB} = \nabla_A \ell^+_B - \frac{1}{2} \theta_+ q_{AB}$, and the remaining terms are components of the Einstein tensor projected along the null directions:
\be 
G_{++} = G_{\alpha\beta}(\ell^+)^\alpha (\ell^+)^\beta \, , \quad G_{+-} = G_{\alpha\beta}(\ell^+)^\alpha (\ell^-)^\beta \, .
\ee
These latter terms will vanish in the case that the spacetime is vacuum. We will further write
\be 
\psi(s, \phi) = \sum_{m=-\infty}^{\infty} \psi_m(s) e^{im\phi} \, ,
\ee
after which the eigenvalue problem $L_\Sigma \psi = \lambda \psi$ reduces to
\be 
(L_\Sigma + m^2 q^{\phi \phi}) \psi_m = \lambda \psi_m \, .
\ee
The eigenvalues will be labelled by two integers: $l$ and $m$. The integer $l$ is chosen to run from $l = - m$ to $l = m$, with the eigenvalues in ascending order.

The spectrum contains important information about the properties of the MOTS in question, with special attention often given to the principal (smallest) eigenvalue~\cite{Andersson:2005gq, Andersson:2007fh}. For example, in cases where the principal eigenvalue is positive, it is guaranteed that the MOTS will evolve smoothly into the future. This is important for understanding whether, in dynamical settings, a given MOTS will evolve smoothly into a marginally outer trapped tube, as it should if the MOTSs are to serve as a black hole boundary. MOTSs that possess positive principal eigenvalue are said to be \textit{strictly stable}, MOTSs with non-negative principal eigenvalue are \textit{stable}, and those with negative principal eigenvalue are said to be \textit{unstable}. Stable MOTSs possess basic properties one would expect of surfaces that describe black hole boundaries. These surfaces serve as barriers for nearby MOTSs contained completely within a small neighbourhood of them. In other words, they serve as boundaries between trapped and untrapped regions. Strictly stable MOTSs are also those which evolve into everywhere spacelike world tubes.

An interesting connection between the stability operator and MOTSodesics was discovered in~\cite{Booth:2021sow}. If one considers the deviation of nearby MOTSodesics, in analogy with geodesic deviatation, then the MOTSodesic deviation equation contains precisely the same information as the stability operator. This allows for a `geometrization' of the stability operator, and sheds some conceptual clarity on its meaning. In this picture, a MOTS is stable if no nearby MOTSodesics intersect it. More generally, the number of intersections of nearby MOTSodesics with a given MOTS provides an exact count of the number of $m=0$ negative eigenvalues of the stability operator. This is analogous to the Jacobi operator for geodesics, where the number of conjugate points is related to the number of negative eigenvalues. 

In general, we will need to evaluate the spectrum of the stability operator numerically. An exception to this is the case where the MOTS is a geometric sphere. In this case, for any static and spherically symmetric metric of the form~\eqref{GenPG_metric}, the spectrum of the stability operator can be determined exactly as
\be 
\lambda_{l, m} = \frac{l(l+1) + r_0 f'(r_0)}{r_0^2} \, ,
\ee 
where $r_0$ is the radius of the sphere (e.g. $r_0 = r_+$ for the event horizon). Note that for geometric spheres there is no contribution from the parameter $p$ --- the final result is the same independent of the coordinate system used. How $p$ affects the more complicated multi-looped MOTSs will be explored numerically in later sections.

%

As mentioned, in general we must resort to numerical methods to determine the spectrum. For this purpose, we employ the pseudo-spectral method~\cite{Boyd, canuto2007spectral}. We consider here the case where a MOTS is described by a MOTSodesic with arclength parameterization $s \in [0, s_{\rm max}]$. We expand the eigenfunction $\psi_m$ in a Chebyshev series of functions $\phi_n = \cos (n \pi s/s_{\rm max})$,
\be 
\psi(s) = \sum_{n = 0}^N a_n \cos \left( \frac{n \pi s}{s_{\rm max}} \right) \, ,
\ee
and discretize the interval $[0, s_{\rm max}]$ into $N+1$ discrete, equally spaced points $s_i$. The advantage of working in this representation is that the regularity of the function $\psi(s)$ at the poles $s = 0, s_{\rm max}$ is automatically incorporated, and it is not necessary to implement additional boundary conditions. We then construct a matrix approximation to the operator equation:
\be 
L_\Sigma \psi = \lambda \psi \to L_{ij} a_j = \lambda \Phi_{ij} a_j \, ,
\ee
where $L_{ij} = (L_\Sigma \phi_j)(s_i)$ and $\Phi_{ij} = \phi_j(s_i)$. The problem of determining the spectrum of $L$ reduces to finding the eigenvalues of the matrix $\mathbf{M} = \mathbf{\Phi}^{-1} \mathbf{L}$, which we do using built-in methods in Mathematica. In practice, we find that good convergence requires a discretization of $N = 100$ or larger. We have validated our numerics by comparing the output with known analytical results, along with a comparison with the numerical methods of~\cite{pook_kolb_daniel_2021_4687700} in vacuum cases.

\subsection{Finding MOTSs}

In this work, we will primarily study axisymmetric MOTSs in a variety of different spacetimes and scenarios. Therefore, it will be convenient  to have a more automated method for determining the location of these surfaces. While we have employed a range of options in this work, here we present our basic method.

The key observation is that the closed, self-intersecting MOTSs are symmetric: the points at which they intersect the $z$-axis are symmetric about the $\rho$-axis. Therefore, a simple method to determine whether a given MOTSodesic is a candidate for a closed MOTS is to shoot the MOTSodesic perpendicularly from the $z$-axis at some initial point $z_0$ and then evaluate the corresponding value of $\rho$ for the MOTSodesic at the parameter value at which it intersects a line emanating from $-z_0$. In the case of a closed MOTS, this corresponding value of $\rho$ will be zero.  

\begin{figure}[htp]
\centering
\includegraphics[width=0.45\textwidth]{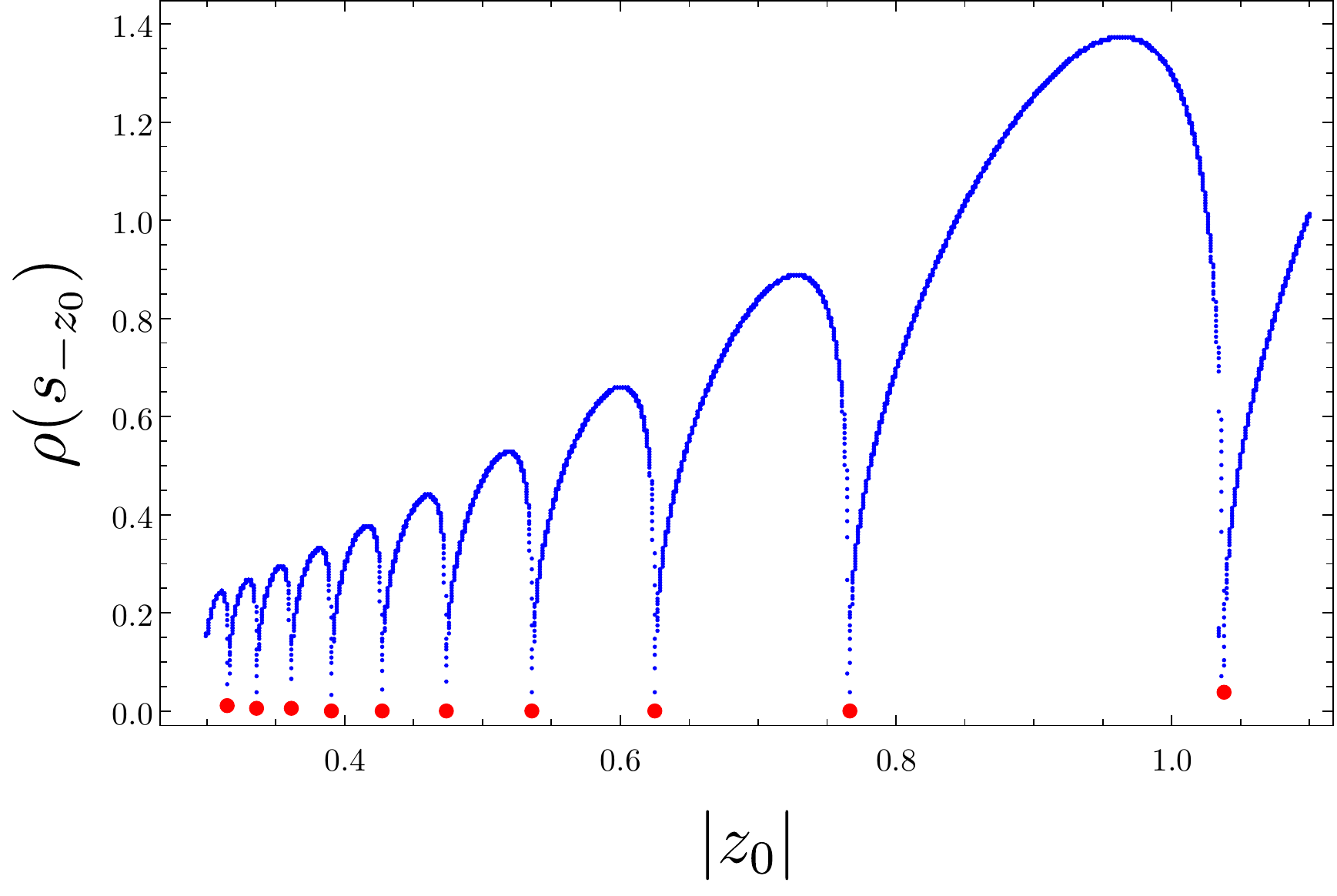}
\caption{Plot of the distance from $z$-axis of a MOTSodesic launched horizontally from the $z$-axis at a value $z_0$. Here $s_{-z_0}$ represents the parameter value at which $z(s_{-z_0}) = -z_0$. In the case where the MOTSodesic is closed, the corresponding value $\rho(s_{-z_0}) = 0$. These points are indicated with red circles on the plot. In this figure, we have considered the Schwarzschild solution with $M=1$ and $p=1$, corresponding to standard PG coordinates.}
\label{algorithm1}
\end{figure}

We show the result of this computation for the Schwarzschild black hole in standard ($p=1$) PG coordinates in Figure~\ref{algorithm1}. To produce this plot we have first solved the MOTSodesic equation in Mathematica, launching the solution horizontally from the $z$-axis from an initial location $z_0$. This gives a parametric solution for the MOTSodesic $(\rho(s), z(s))$. We then solve, using Mathematica's ``FindRoot'' routine the condition $z(s_{z_0}) = -z_0$ to determine the parameter value $s_{z_0}$. We then evaluate $\rho(s_{z_0})$ to determine how close to the axis the MOTSodesic is when it is located symmetrically from the starting point. It is this that is plotted in Figure~\ref{algorithm1} for various values of $z_0$.

After this distribution of distances has been determined, the MOTS can be obtained by determining the minima, shown as red dots on the lower plot. 
One method of finding minima is to generate plots similar to Figure~\ref{algorithm1} with a high $z_0$ resolution.

\begin{figure}[htp]
\centering
\includegraphics[width=0.45\textwidth]{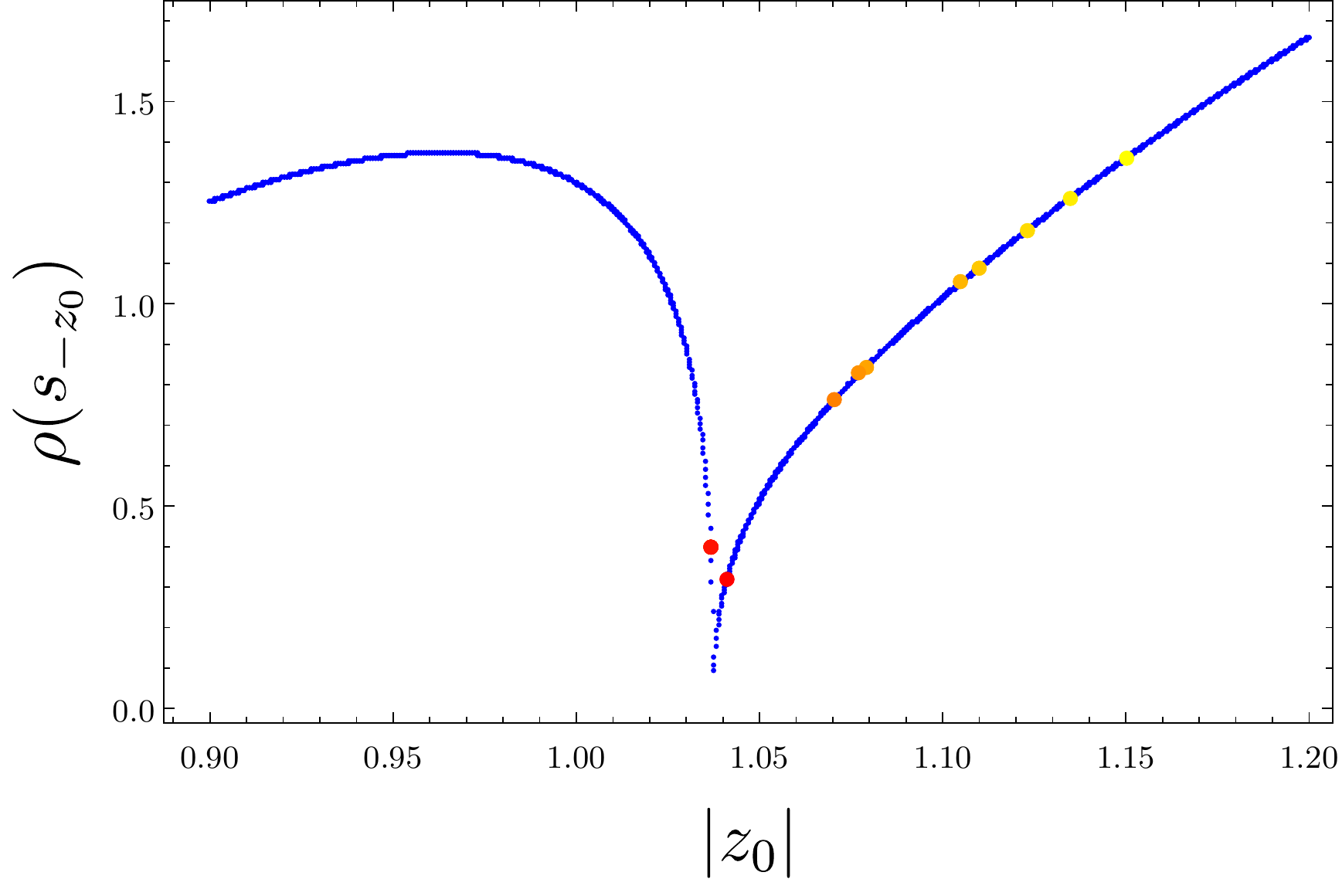}
\caption{Plot of the $\rho(s_{z_0})$ distribution for the Schwarzschild black hole in Painlev\'e-Gullstrand coordinates. The yellow-orange-red dots depict the the first 15 steps of our Monte-Carlo walker initiated at $z_0=1.15$ sampling with a standard deviation of $0.06$ (with yellow being the first position and red being the last).}
\label{GenPG_walkdemo}
\end{figure}

The supplementary method we have incorporated is a Monte-Carlo-type minimum descender. We initiate a `walker' where it samples a $\tilde{z}_0$ value from a Gaussian distribution centered on its current $z_0$. If the proposed $\tilde{z}_0$ yields a $\rho(s_{\tilde{z}_0})$ that is less than $\rho(s_{z_0})$, the walker will move to the proposed position ($z_0=\tilde{z}_0$). This is repeated as the walker descends the $\rho(s_{z_0})$ distribution until it reaches a tolerable minimum. Figure~\ref{GenPG_walkdemo} demonstrates the pathing of such a walker.

The brute-force tactic for finding minima is straightforward but its demand for a high resolution strains computational resources. A Monte-Carlo walker allows for a drastically more efficient convergence to a minimum. We have further optimized our walker adding a directional bias and a progressively decreasing standard deviation to hone in on the minimum. Our MOTS-finding algorithm uses a combination of brute-forcing minima on the $\rho(s_{z_0})$ distribution and using Monte-Carlo walkers to pinpoint the minimum to upwards of ten decimal places. In practice, we find this method to be faster than using built-in Mathmatica routines to minimize a function.

While what have have just described assumes that the MOTSs are launched perpendicular to the $z$-axis, it is straightforward to generalize the method to other situations as well. For example, in our analysis of Reissner-Nordstr\"om below, it was necessary to consider surfaces launched from the $\rho$-axis as well.

\section{The Schwarzschild Spacetime}
We first look at the Schwarzschild solution in generalized Painlev\'e-Gullstrand coordinates. It was shown in~\cite{Booth:2020qhb} that
standard PG time slices of the Schwarzschild black hole each contain an infinite number of closed, self-intersecting MOTSs 
inside the event horizon. Here we will show that this feature persists --- and is qualitatively identical --- in generalized PG coordinate systems for several choices of constant $p$, thereby showing that the result is robust, at least within this family of slicings. 

\begin{figure*}[htp]
\centering
\includegraphics[width=\textwidth]{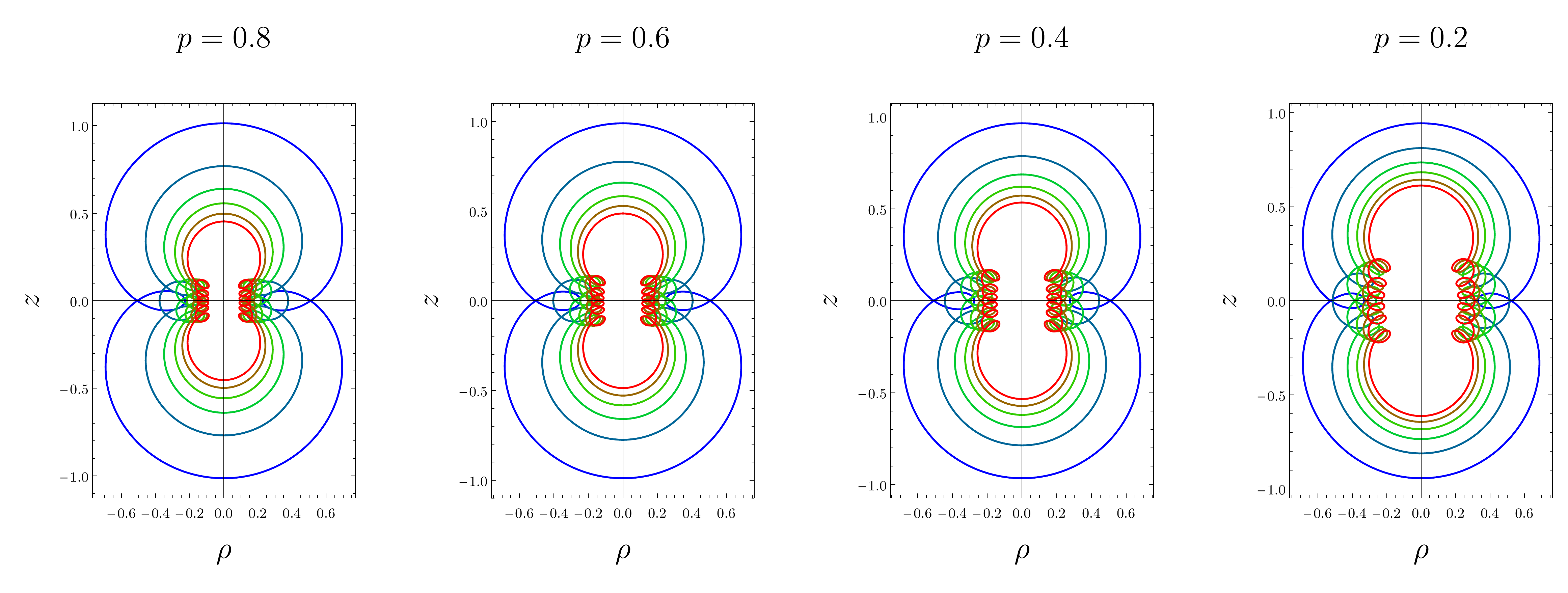}
\caption{Plots of the outer-most eight MOTS found in the Schwarzschild black hole for $M=1$. The colours blue-green-red indicate different one, two, three, up to eight-loop MOTS respectively. The left-most plot, $p=0.8$ cases are closest representative of what was found in~\cite{Booth:2020qhb}. 
}
\label{GenPG-loops}
\end{figure*}

The usual Schwarzschild metric function reads
\be
	f=1-\frac{2M}{r} \, .
\ee
We substitute this into the MOTSodesic equations~\eqref{motsodesics}, and using the methods just described, search for closed MOTSs in the black hole interior. We show the outer-most eight MOTSs, including the $r=2M$ event horizons, in Fig.\ref{GenPG-loops}. The coordinates in these plots have been mapped into the conformally flat coordinates given in Eq.~\eqref{conform}. From the plots, the self-intersecting feature is persistent for $p>0$, with the looping structures qualitatively identical to those found in the 	`ordinary' PG case of $p=1$.  In the limiting case of Eddington-Finkelstein coordinates, we do not find any looping MOTSs in the null hypersurface $\Sigma_\tau$. 

\begin{figure}[htp]
\centering
\includegraphics[width=0.45\textwidth]{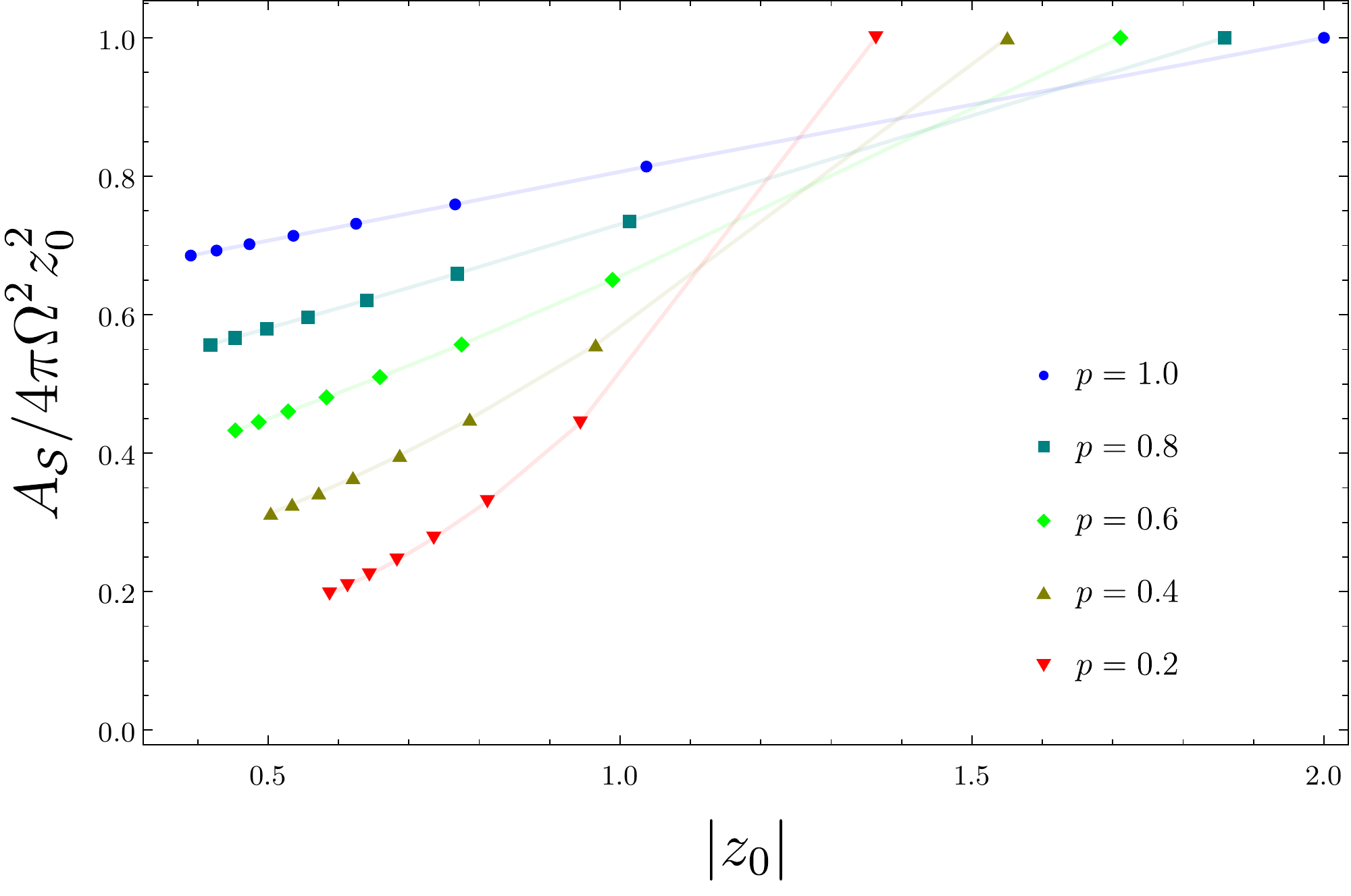}
\caption{Plot showing the area of MOTSs in the Schwarzschild black hole spacetime normalized by the area of a sphere of radius $z_0$. The symbols on the plots indicate the position of the MOTS, while we have connected these with lines of low opacity to aid in reading the plot.}
\label{Schw-areas}
\end{figure}

While the qualitative features are identical, some associated quantities of the MOTSs differ. The initial $r_0$ value is closer to the origin to generate the same looping MOTS for lower values of $p$. The MOTS essentially shrink as $p$ decreases, which is evident when we calculate the area of these surfaces (normalized by a sphere of their size) in Fig.~\ref{Schw-areas}. The MOTSs shown in Fig.~\ref{GenPG-loops} seem to show the opposite behaviour of what we just described. However, this is an illusion due to the fact that the different plots live in different coordinate systems. What is shown here is the MOTSs plotted in the conformally flat coordinates~\eqref{conform}. As such, it is sensible to compare angles between the two different curves visually, but the relative sizes cannot be reliably determined. To properly compare the sizes of the different curves requires a direct computation of the area, as shown in Fig.~\ref{Schw-areas}.

\begin{figure*}[htp]
\centering
\includegraphics[width=0.3\textwidth]{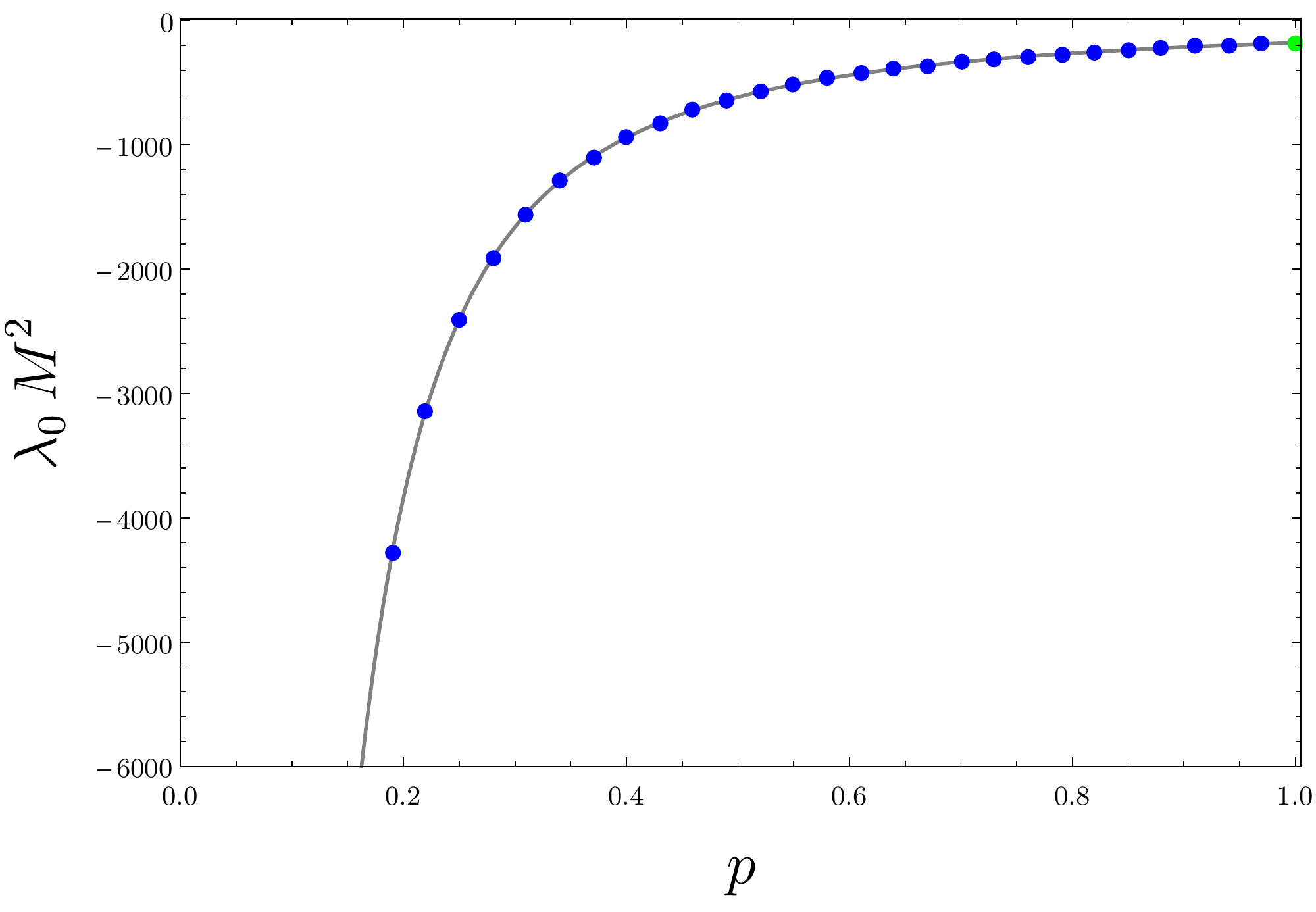}
\quad
\includegraphics[width=0.3\textwidth]{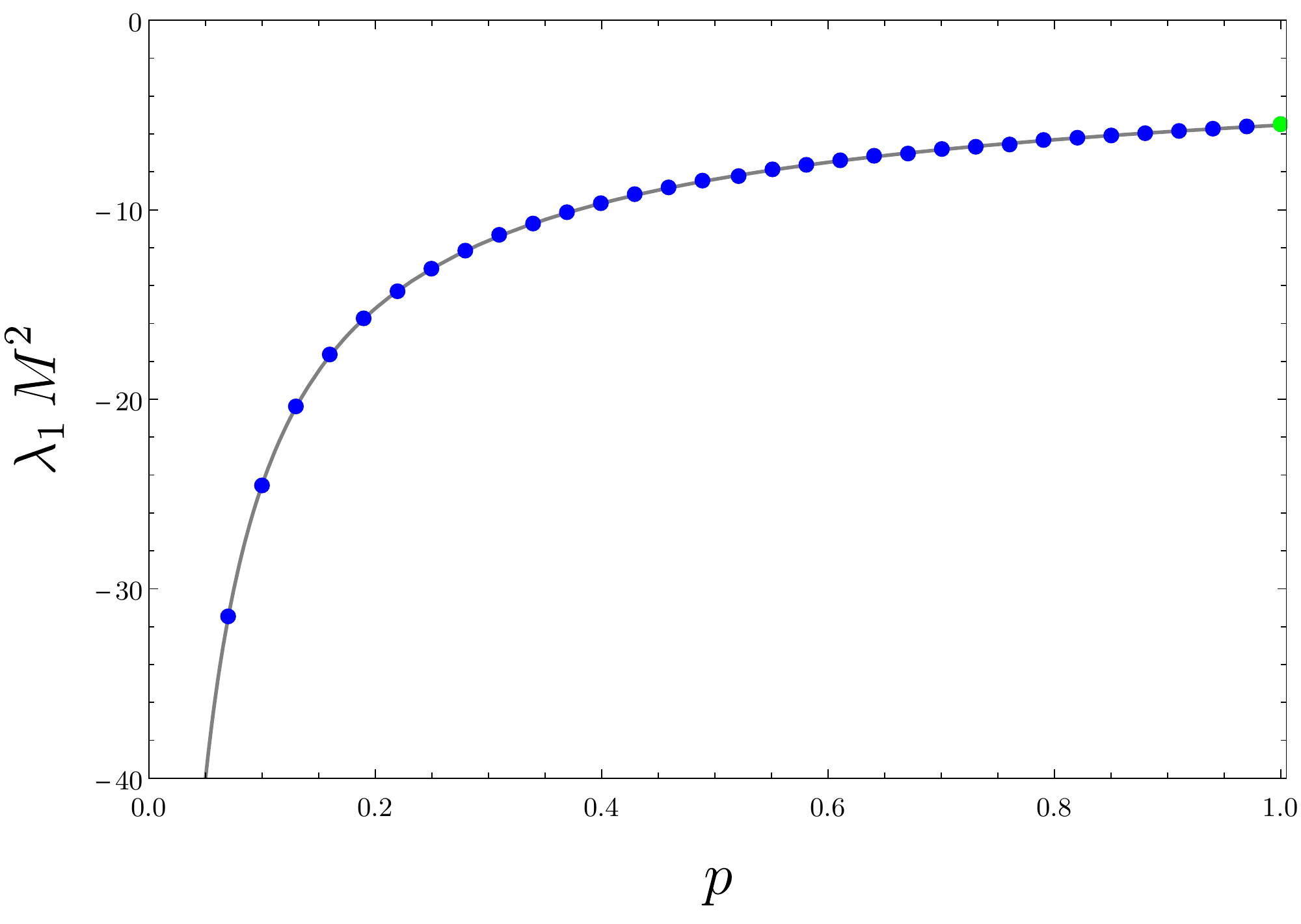}
\quad
\includegraphics[width=0.3\textwidth]{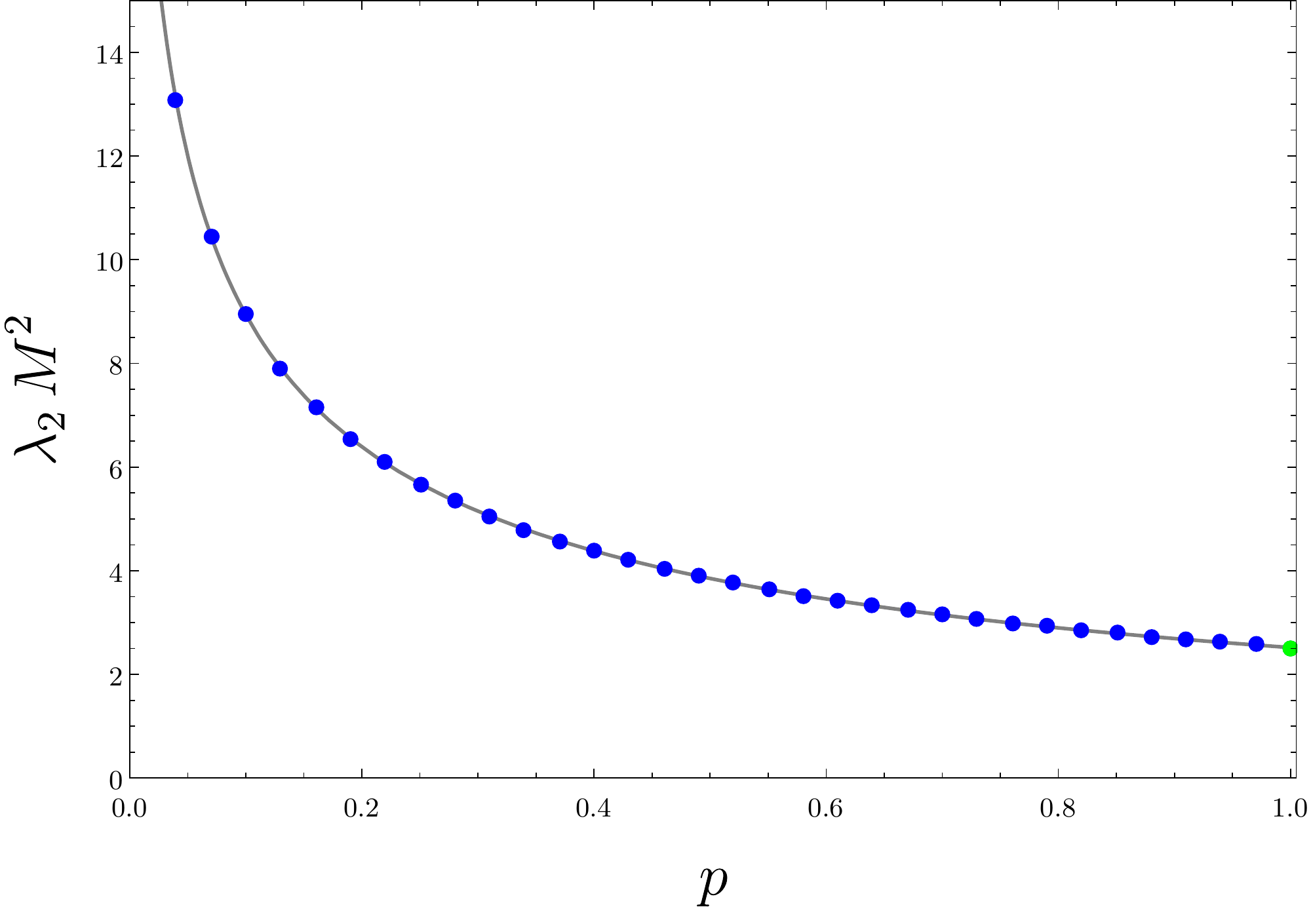}
\caption{A plot of the first three $m=0$ eigenvalues of the stability operator for the one-looped surface plotted here as a function of the parameter $p$. The shown curve is an interpolation of raw data, with a few selected points shown (in blue) for illustrative purposes.}
\label{genPG-eVals}
\end{figure*}

We also discuss the stability of the looped MOTSs as a function of the parameter $p$. As discussed earlier in the manuscript, the spectrum of the stability operator on the event horizon sections is invariant. However, the same is not true of the interior, looped MOTSs, where the numerical value of the eigenvalues depends on the parameter $p$.  As an example, we show in Figure~\ref{genPG-eVals} the spectrum of the one-looped surface as a function of $p$. Here we see that varying $p$ leads to some quantitative differences, but does not lead to any qualitatively new behaviour. For example, the one-looped surface continues to have exactly two negative eigenvalues, and does not accrue any further as a function of $p$. The same is true of the higher looped surfaces, which each have $2 \times (\# \text{ loops})$ negative $m=0$ eigenvalues. The only significant difference is that, as $p \to 0$, which corresponds to a limit to Eddington-Finkelstein coordinates, all the eigenvalues diverge. This is consistent with the fact that we do not find any looping MOTSs in that coordinate system, only the event horizon. These observations are all consistent with the corresponding MOTSodesic deviation approach.

\section{Black Holes with Inner Horizons}

We turn our attention now to the case when the black hole possesses an inner horizon. 
Here we will see that the structure and variety of MOTSs present in the interior is highly sensitive to the internal structure of the black hole. We will illustrate this through two examples. First, we will consider the spherically symmetric black hole in four-dimensional Gauss-Bonnet gravity, followed by the Reissner-Nordstr\"om black hole. We have chosen to present the results in this order due to the relative complexity of the two cases. As we will show, the Reissner-Nordstr\"om black hole presents a very intricate structure of internal MOTSs.

\subsection{Four-Dimensional Gauss-Bonnet Black Hole}

We begin by considering a black hole solution of four-dimensional Gauss-Bonnet gravity~\cite{Hennigar:2020lsl}. We have chosen to present this example because the metric is well-suited to this analysis, and the structure of internal MOTSs are about as simple as one could imagine while still being more interesting than the Schwarzschild black hole.


\begin{figure}[htp]
\centering
\includegraphics[width=0.45\textwidth]{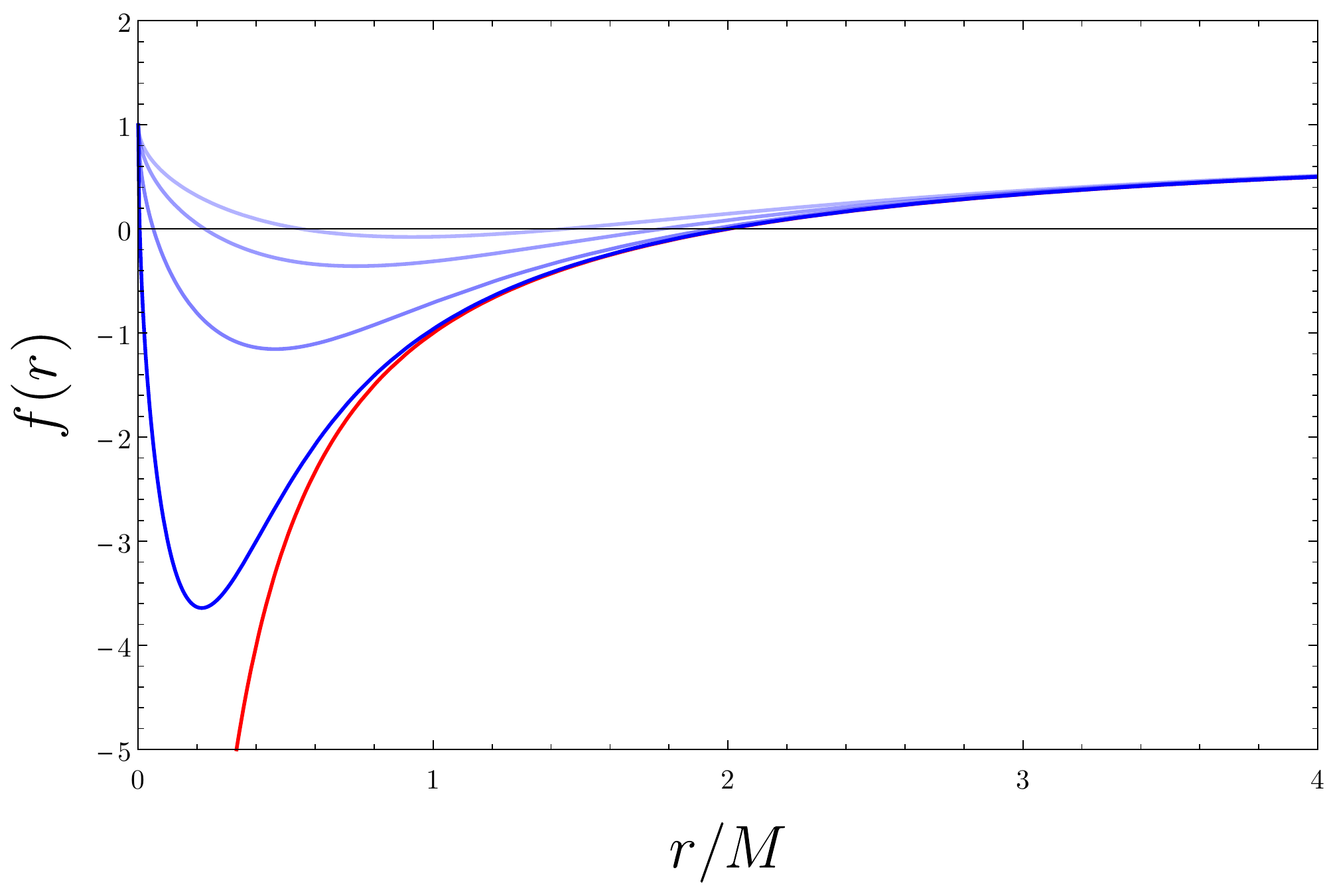}
\caption{Plots of the four-dimensional Gauss-Bonnet metric function for different values of the coupling constant. The blue curves correspond to $\alpha/M^2 = 1/100, 1/10, 4/10$ and $8/10$ in order of bottom to top (or decreasing opacity). The red curve is the Schwarzschild metric function.}
\label{GBmetric}
\end{figure}

The action for this theory reads
\begin{align} 
S =& \int d^4 x \sqrt{-g} \left[R + \alpha \big(\phi \mathcal{G} + 4 G^{ab} \partial_a \phi \partial_b \phi - 4 (\partial \phi)^2 \Box \phi \right.  \nonumber\\
&\left. 
 + 2 \left( (\partial \phi)^2 \right)^2 \big) \right]
\end{align}
where 
\be 
\mathcal{G} = R_{abcd}R^{abcd} - 4 R_{ab}R^{ab} + R^2
\ee
is the Gauss-Bonnet term. The theory admits static, spherically symmetric black holes characterized by the metric function
\be 
f = 1 + \frac{r^2}{2 \alpha} \left[1 - \sqrt{1 + \frac{8 \alpha M}{r^3}} \right] \, ,
\ee
where $\alpha$ is the Gauss-Bonnet coupling constant and has dimensions $[\text{Length}]^2$. The metric function describes an asymptotically flat black hole of mass $M$ and reduces to the Schwarzschild solution in the limit $\alpha \to 0$. Provided the coupling is positive, there is an inner and outer horizon located at
\be 
r_\pm = M\left[1 \pm \sqrt{1 - \frac{\alpha}{M^2}} \right]\, .
\ee 
For $\alpha =  M^2$ the horizons degenerate and the black hole is extremal. In this work, we will concern ourselves with $\alpha \in [0, M^2)$. There is a curvature singularity at the origin, but the metric function itself limits to a finite value there
\be 
f = 1 - \sqrt{\frac{2 M r}{\alpha}} + \mathcal{O}(r^{3/2}) \, .
\ee
As a result, the `ordinary' PG coordinate system~\eqref{ordinaryPG} extends all the way to $r=0$ in this geometry, making it an ideal toy model for exploring the effect of multiple horizons on the properties of self-intersecting MOTSs. Finally, we note that the metric function attains a minimum value at the point 
\be 
r_{\rm min} = (\alpha M)^{1/3} 
\ee
where the metric function takes its most negative value,
\be \label{GBmin}
1 \ge f(r) \ge f(r_{\rm min}) = 1 - \left(\frac{M}{\sqrt{\alpha}} \right)^{2/3} \, .
\ee
We show in Figure~\ref{GBmetric} plots of the metric function for several different values of the coupling. 


\begin{figure*}[htp]
\centering
\includegraphics[width=\textwidth]{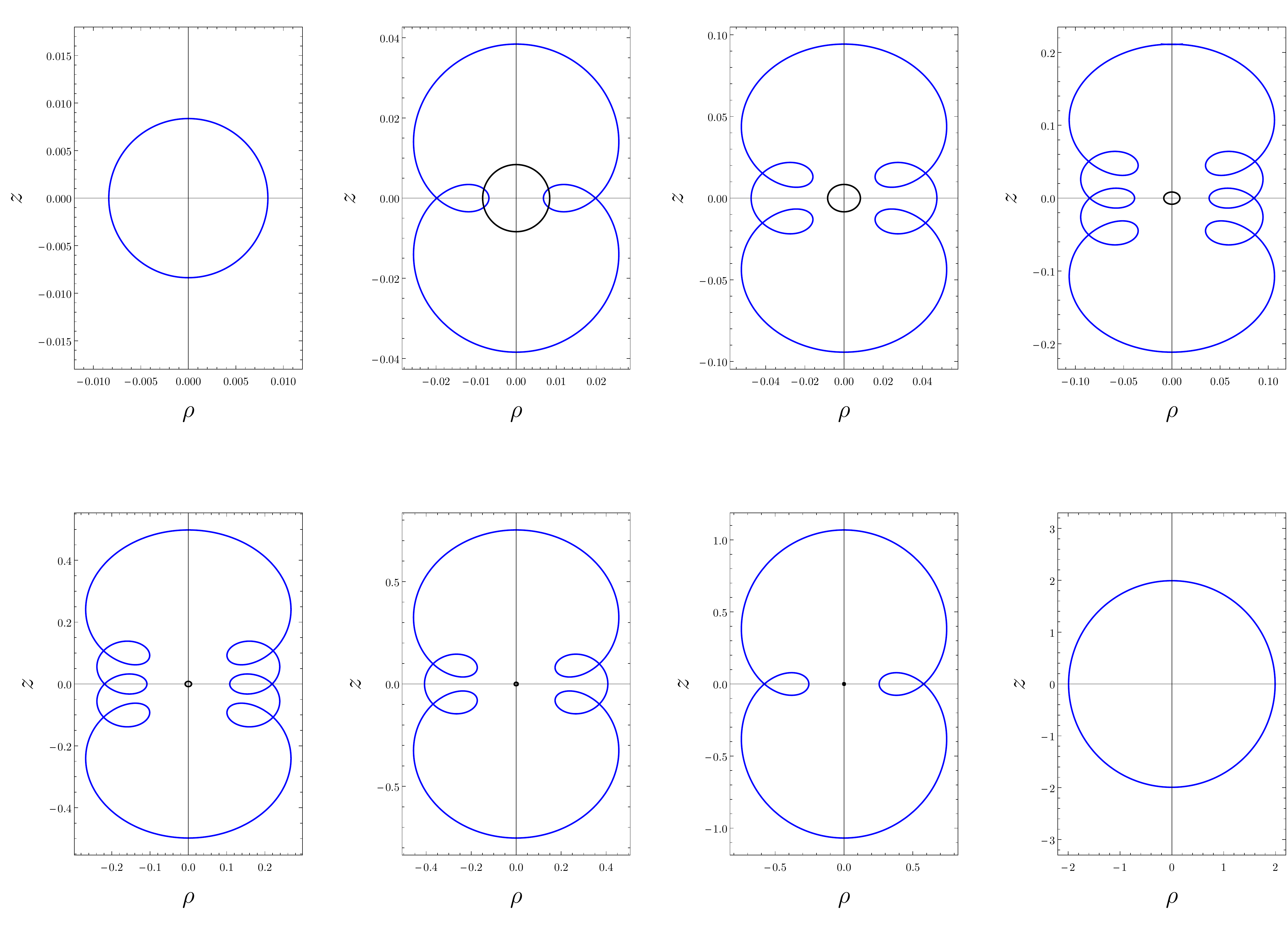}
\caption{Here we show all MOTSs present in the Gauss-Bonnet black hole spacetime in the case $\alpha/M^2 = 1/60$. Note that the scale changes in the plots for the purposes of visibility. In all cases, the inner horizon has been included as a black circle to provide scale.}
\label{GB-loops}
\end{figure*}

Let us begin our investigation of this metric by examining the implications of multiple horizons for the self-intersecting MOTSs. For this purpose, we will consider standard PG coordinates, setting $p = 1$. Of course, both the inner and outer horizon are MOTSs, but concerning the self-intersecting surfaces, the first observation is that there are no longer an infinite number of them. All the self-intersecting MOTSs are contained within the event horizon, and while all surfaces intersect the $z$-axis between the event and inner horizon, some portions of the MOTSs may extend inside of the inner horizon. They are symmetrically distributed in the sense that the surface closest to the event horizon has the same number of loops as the surface closest to the inner horizon, and similarly for the second-closest surfaces to these horizons, and so on. The number of loops the MOTSs have increases moving inward from the event horizon before reaching a maximum and then decreasing back to zero loops when the inner horizon is reached.  There are always two surfaces of a given number of loops.\footnote{The only exception to this is when a new surface forms, as discussed below.} We show an explicit example of this in Figure~\ref{GB-loops} for the case $\alpha/M^2 = 1/60$. In this case there are a total of 8 MOTSs, including 6 which exhibit self-intersections.

\begin{figure}[htp]
\centering
\includegraphics[width=0.45\textwidth]{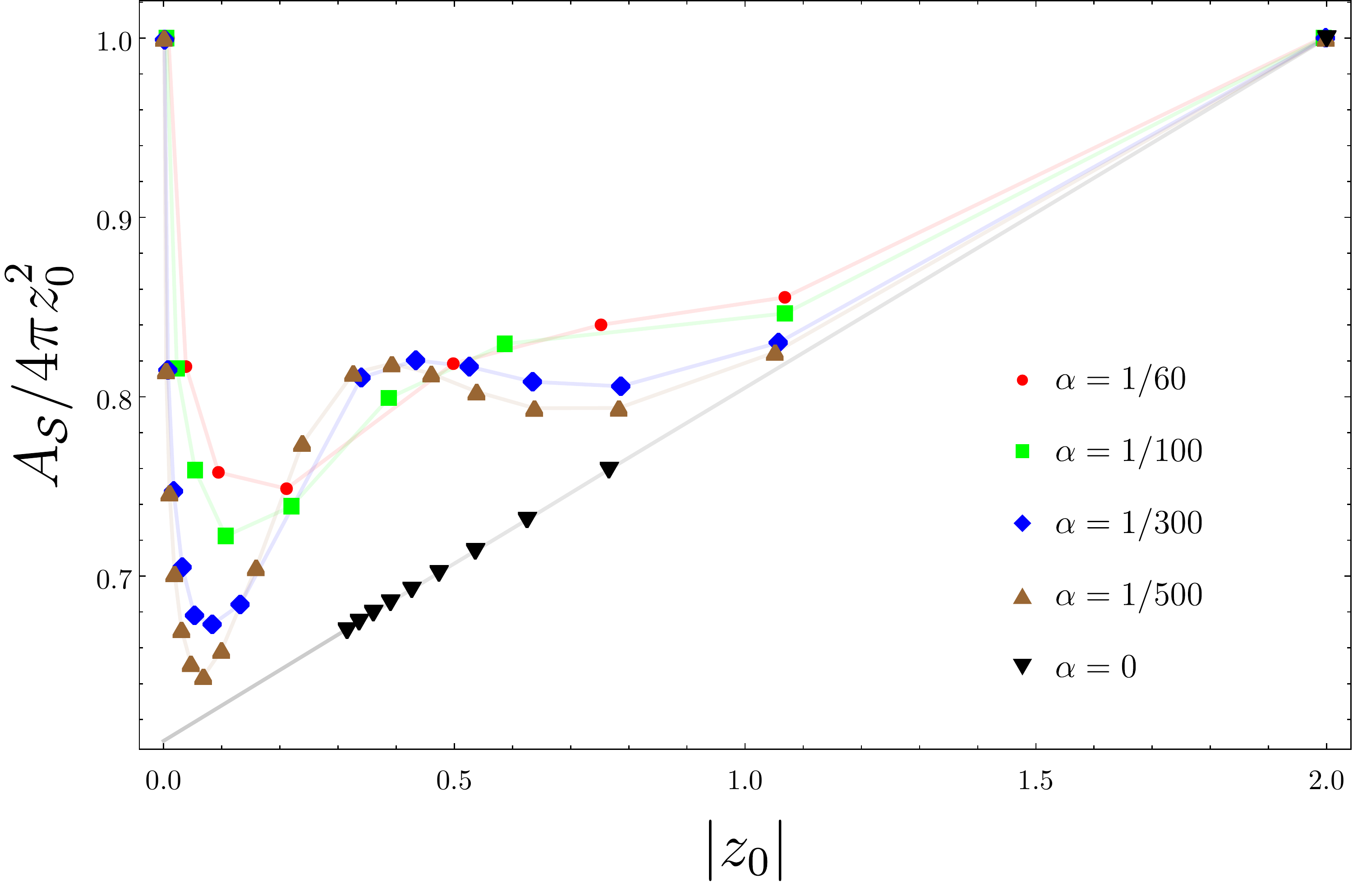}
\caption{Plot showing the area of MOTSs in the Gauss-Bonnet black hole spacetime normalized by the area of a sphere of radius $z_0$. The symbols on the plots indicate the position of the MOTS, while we have connected these with lines of low opacity to aid in reading the plot. The black upside-down triangles correspond to the Schwarzschild solution ($\alpha = 0$) and while in this case there are an infinite number of MOTSs, here we have illustrated only a finite number of them, along with the line of best fit.}
\label{GB-areas}
\end{figure}

Rather than displaying the direct plots for further values of $\alpha$, we believe it is more insightful to consider the comparison of the areas of the MOTSs for different values of $\alpha$. We show this for five values of $\alpha$ in Figure~\ref{GB-areas}, including also the result for the Schwarzschild solution for comparison purposes. In this plot, the plot markers indicate particular surfaces, while we have connected these with a line of low opacity to serve as a visual add in reading the plot. The first observation is that the area of the MOTSs in the Gauss-Bonnet spacetime is always larger than the area of the corresponding MOTSs in the Schwarzchild spacetime --- the $\alpha > 0$ curves lie entirely above the curve for Schwarzschild solution --- though still less than the area of a sphere of equivalent radius. The second observation is the the structure of areas for the Gauss-Bonnet case is much richer. Unlike the areas in the Schwarzschild case, the $\alpha > 0$ curves are not monotonic. In fact, as the number of MOTSs inside the horizon increases, it is observed that a `peak' forms around $z_0 \approx 0.4$. This all seems to suggest that the inner structure of the black hole influences the structure of the MOTSs in a non-perturbative manner. That is, even though the metric limits to the Schwarzschild solution as $\alpha \to 0$, there is a rich and nontrivial structure of different MOTSs within the horizon of the Gauss-Bonnet black hole that persists even at very small $\alpha$.

A more thorough investigation of the parameter space suggests that the formation of self-intersecting MOTSs is correlated with certain threshold `depths' of the metric function. We find that the smaller the value of $\alpha$, the larger the number of MOTSs found between the two horizons. This correlates with the fact that,  as seen in Eq.~\eqref{GBmin}, the metric function reaches more negative values the closer $\alpha$ is to zero. It is therefore interesting to consider a `dynamical' problem where $\alpha$ begins at some fixed large value and is then decreased, examining the behaviour of the MOTSs during this process.

We find that the threshold for the formation of the first self-intersecting MOTS is $\alpha/M^2 = 0.1359915193$, which is accurate to 10 decimal places (but a true specification would seem to require infinite precision) and intersects the $z$-axis at $z_0 \approx 0.59315$. Our observation is that when a self-intersecting MOTS forms it is `degenerate'. That is, immediately after the formation of a looped MOTS, it bifurcates into two distinct MOTSs. As $\alpha$ is further decreased, the two MOTSs move apart, one moving toward the event horizon and the other toward the inner horizon. Said in another way, if one considers the two $n$-looped surfaces and tracks them ``backward'' by increasing $\alpha$, it will be observed that they `merge' at some value of $\alpha$ before disappearing. It is interesting to note that there is not any obvious symmetric positioning of the MOTS when it first forms --- its location is neither at the minimum of $f$ nor is it symmetrically positioned between the inner and outer horizons. However surfaces with a larger number of loops always form between the two surfaces with one less loop.

\begin{figure*}[htp]
\centering
\includegraphics[width=0.3\textwidth]{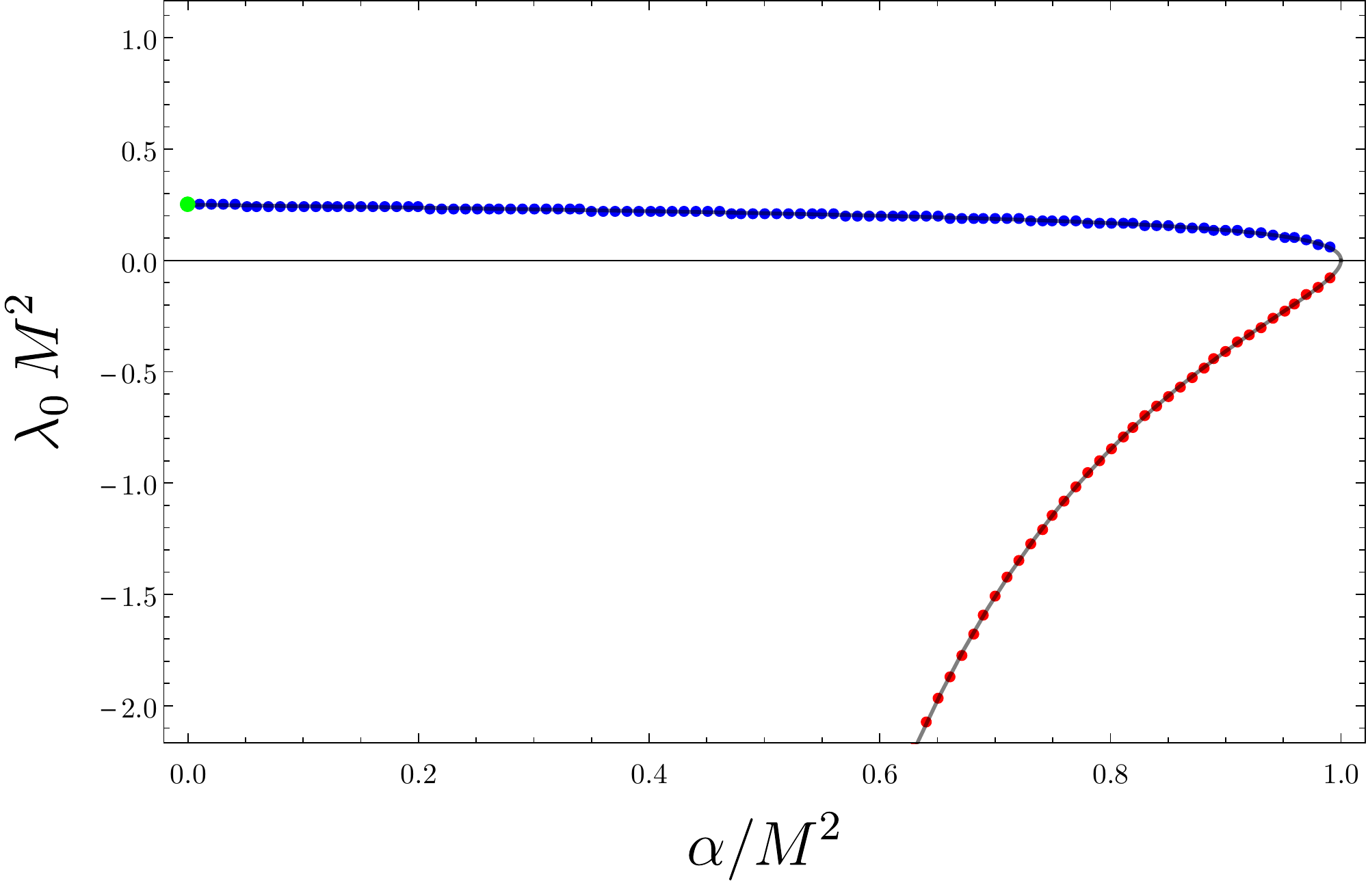}
\quad
\includegraphics[width=0.3\textwidth]{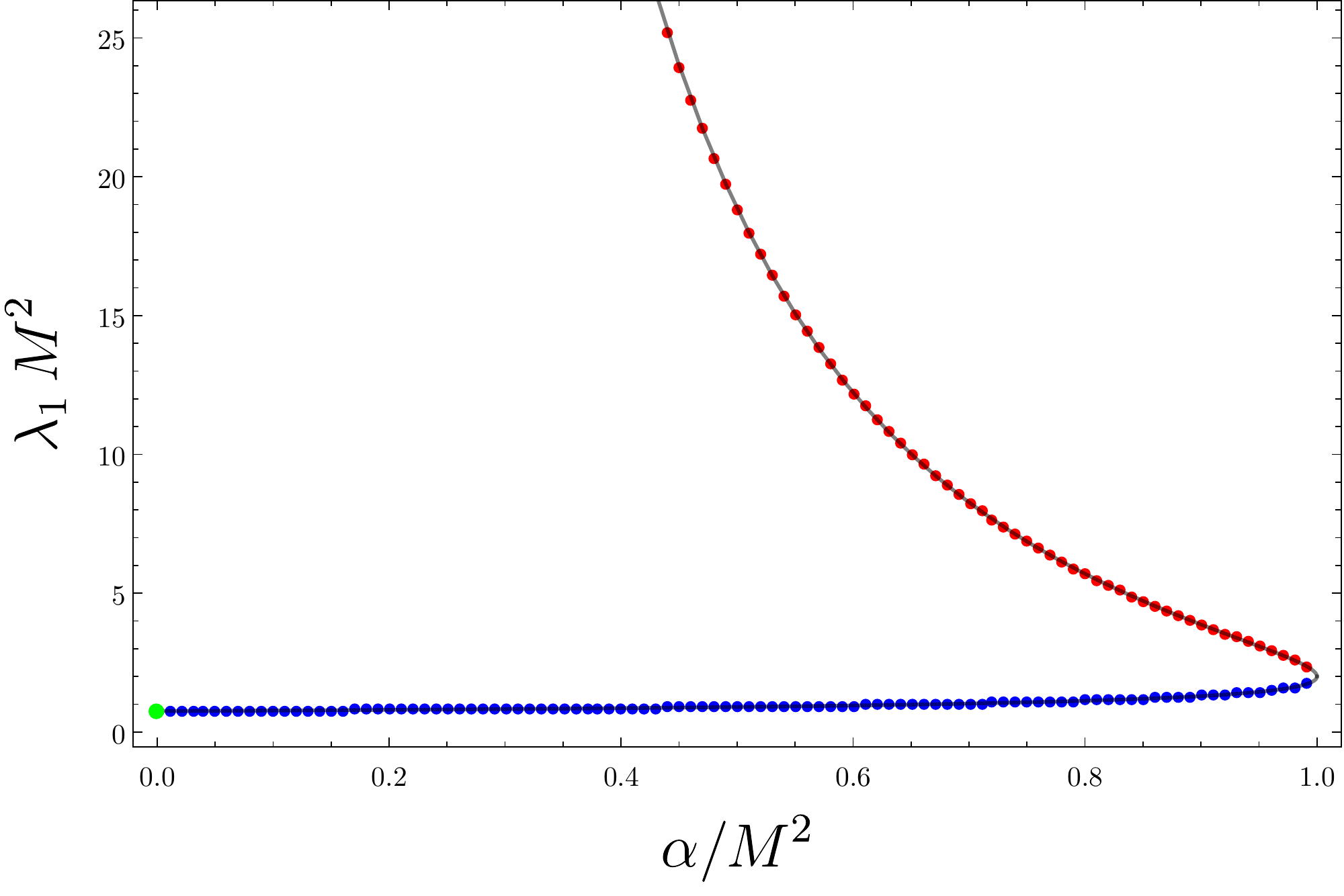}
\quad
\includegraphics[width=0.3\textwidth]{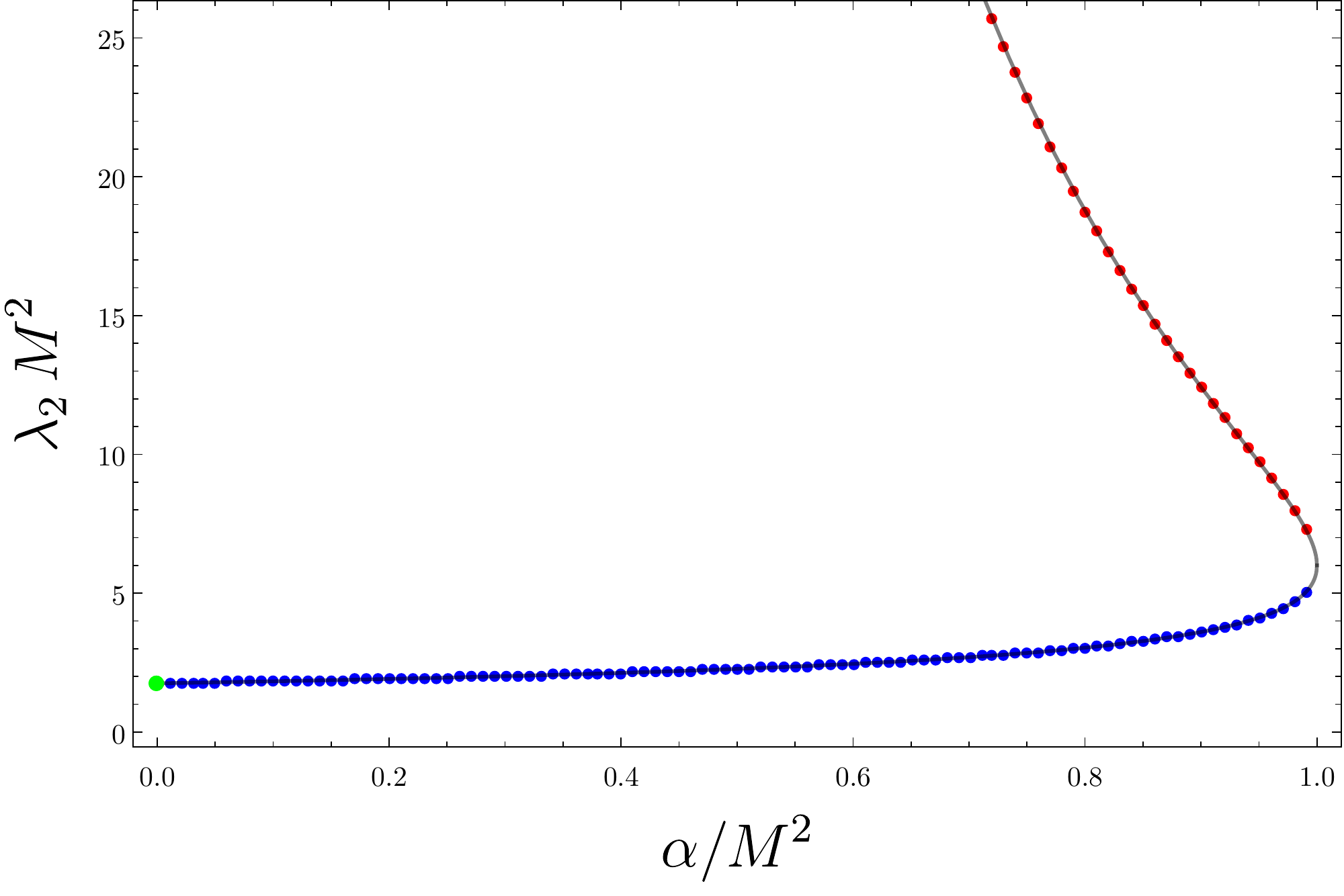}
\caption{First three $m=0$ eigenvalues of the stability operator as a function of coupling for the event horizon (blue) and inner horizon (red) for the Gauss-Bonnet spacetime. Since the surfaces are spherical, there is no splitting of the $m \neq 0$ eigenvalues. The dots correspond the the result of the numerical code, while the black curve is a plot of the exact result. The green dots show the corresponding, independently computed value of the eigenvalue for the Schwarzschild solution. }
\label{horizonSpectrum}
\end{figure*}

Let us now turn to an analysis of the stability properties of the MOTSs. We begin with the event and inner horizons, for which an exact solution is possible. This allows also for a validation of our numerical methods for determining the spectra. We show in Figure~\ref{horizonSpectrum} the first three $m=0$ eigenvalues for the event and inner horizons. We see that the numerics agree precisely with the analytical result. In the limit of extremality, the principal eigenvalue tends to zero. The eigenvalues of the event horizon slices are all positive, and approach the corresponding values from the Schwarzschild solution as $\alpha/M^2 \to 0$. The principal eigenvalue of the inner horizon is negative, while all higher-order eigenvalues are positive. Here we see a key difference from the Reissner-Nordstr\"om solution. In that case, one can arrange for arbitrarily many negative eigenvalues on the inner horizon by demanding that the quantity $r_- f'(r_-)$ be sufficiently negative~\cite{Booth:2017fob}.\footnote{This quantity behaves as $r_- f'(r_-) \sim M^2/Q^2$ for small charge in the Reissner-Nordstr\"om case.} Here, in the Gauss-Bonnet case, only the principal eigenvalue is negative. The reason for this is that here the quantity $r_- f'(r_-)$ is bounded below by $-1/2$, which is attained in the limit $\alpha/M^2 \to 0$.

\begin{figure*}[htp]
\centering
\includegraphics[width=0.3\textwidth]{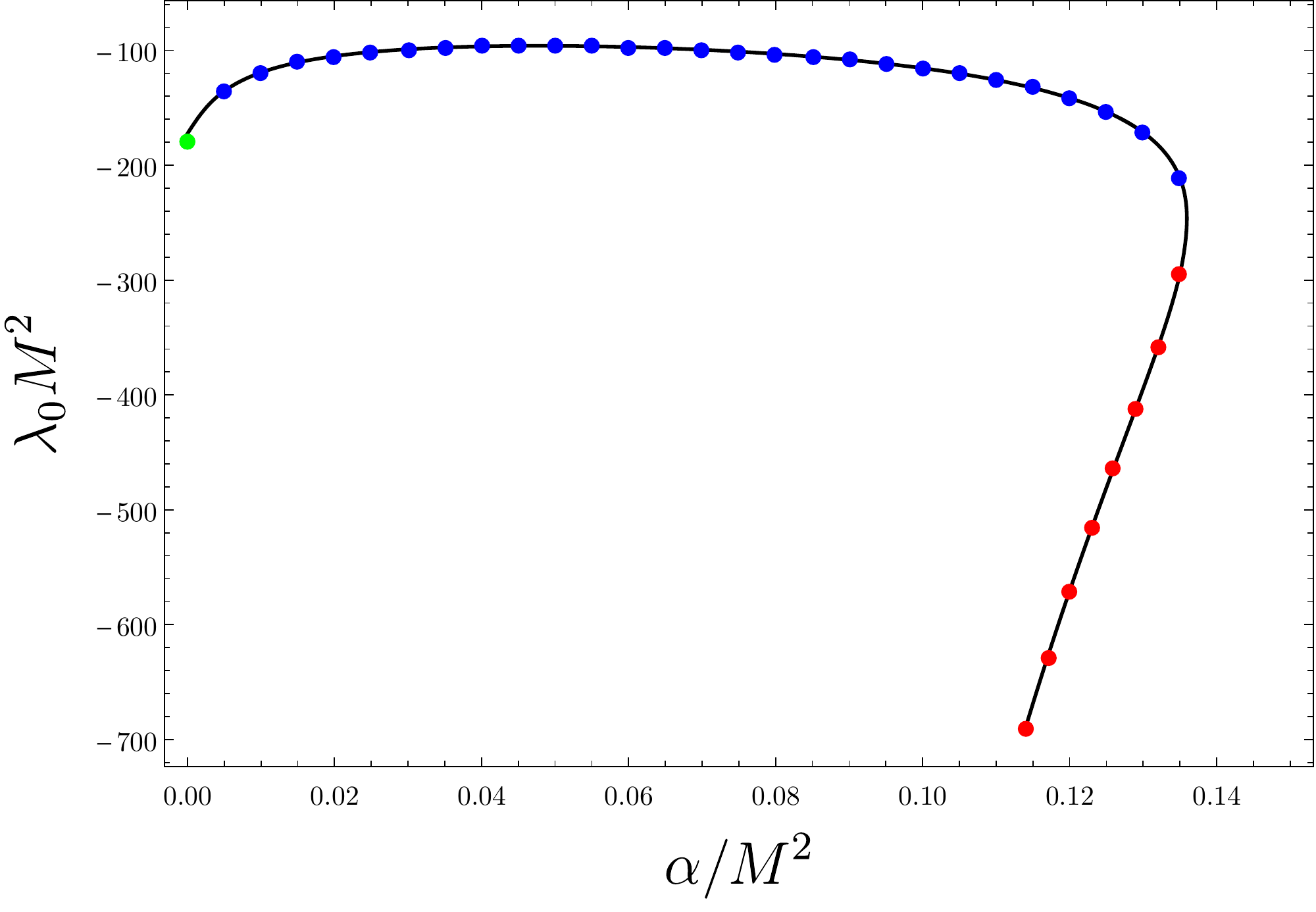}
\quad
\includegraphics[width=0.3\textwidth]{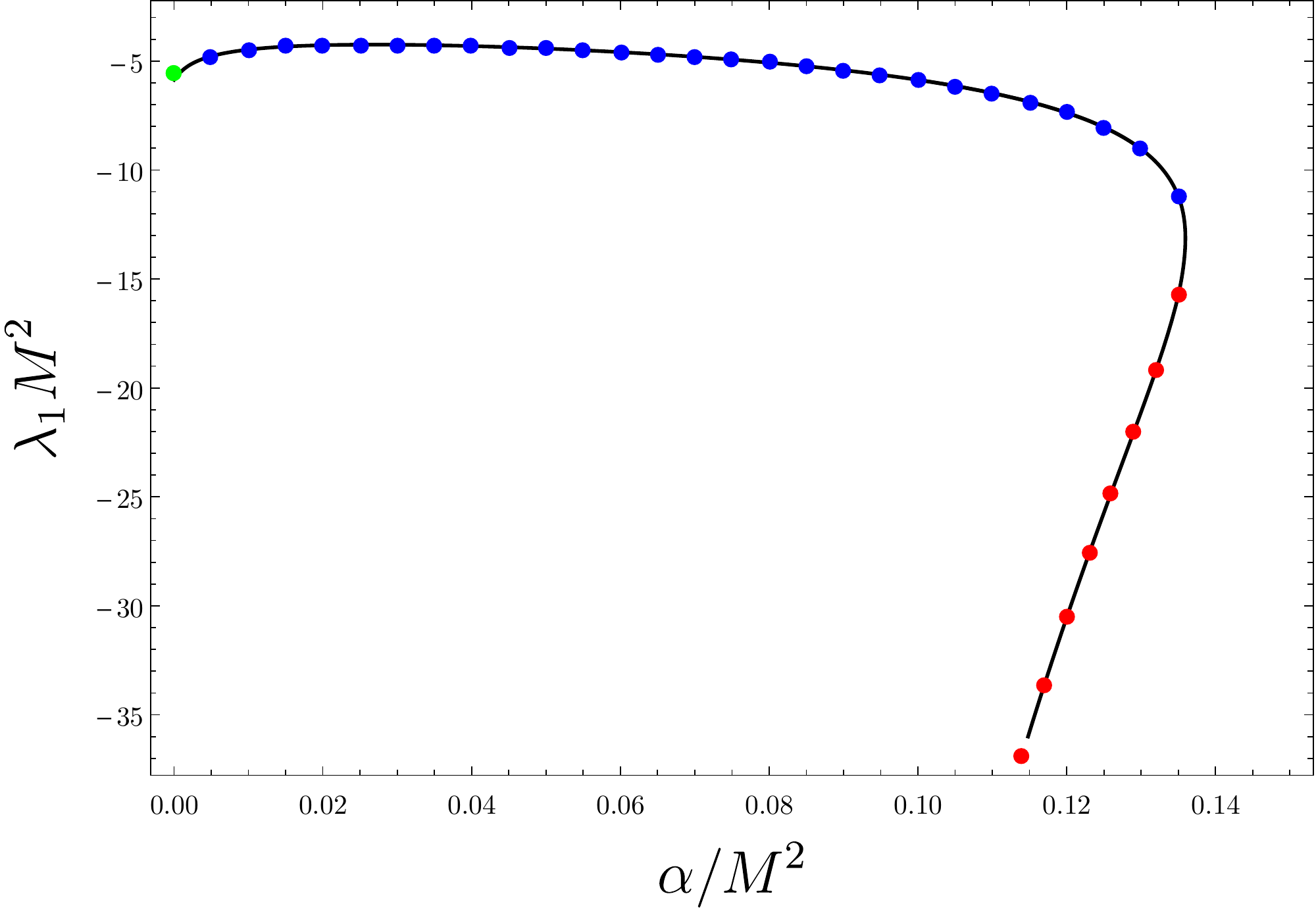}
\quad
\includegraphics[width=0.3\textwidth]{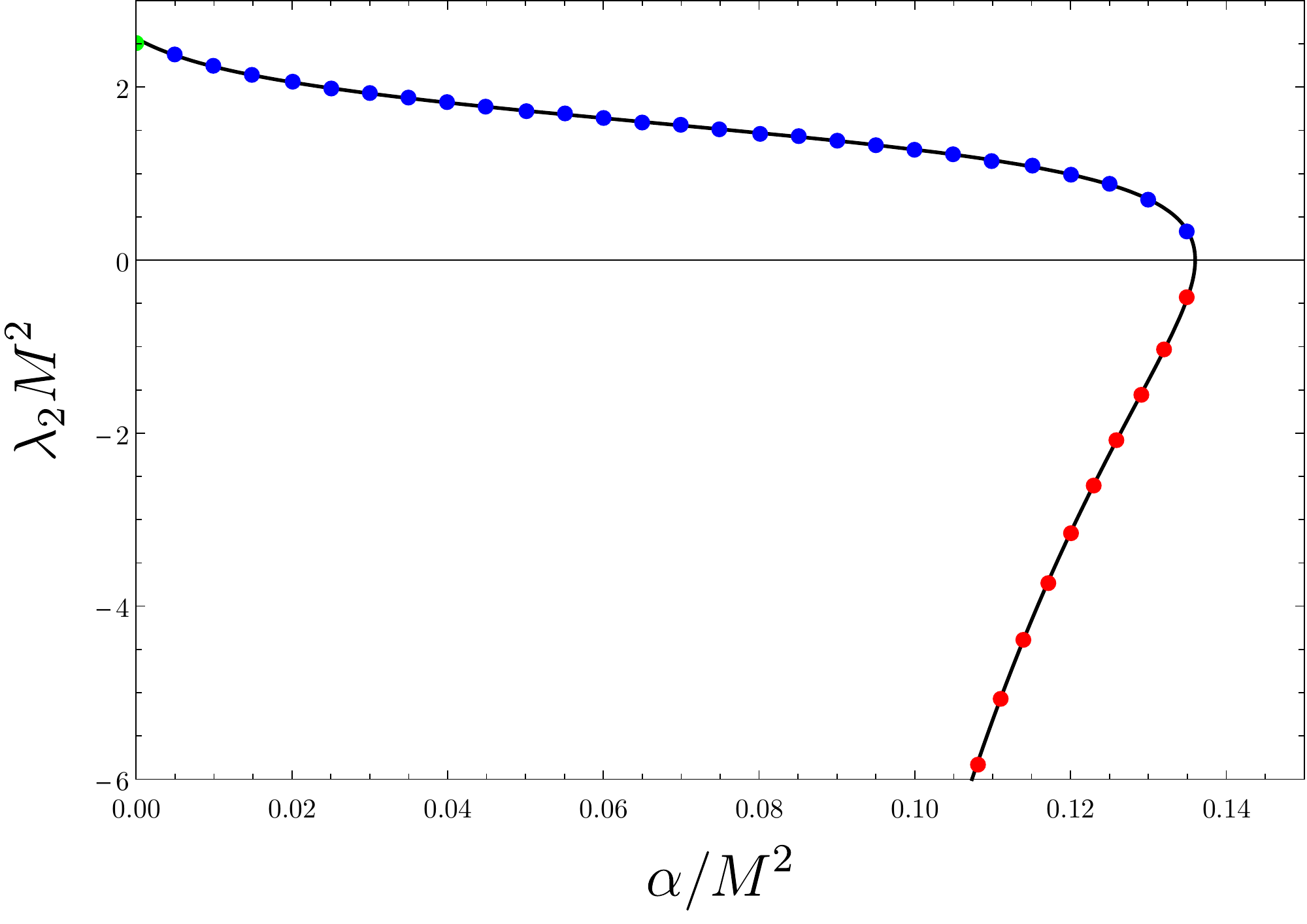}
\caption{First three $m=0$ eigenvalues of the stability operator as a function of coupling for the outer (blue) and inner (red) one-loop surface for the Gauss-Bonnet spacetime. The black curve is an interpolation of the numerically obtained data, while the dots are a few of the numerically determined values. The green dot corresponds to the respective eigenvalue in the Schwarzschild spacetime, as obtained via independent means.}
\label{oneLoopSpectrum}
\end{figure*}

\begin{figure}[htp]
\centering
\includegraphics[scale=0.25]{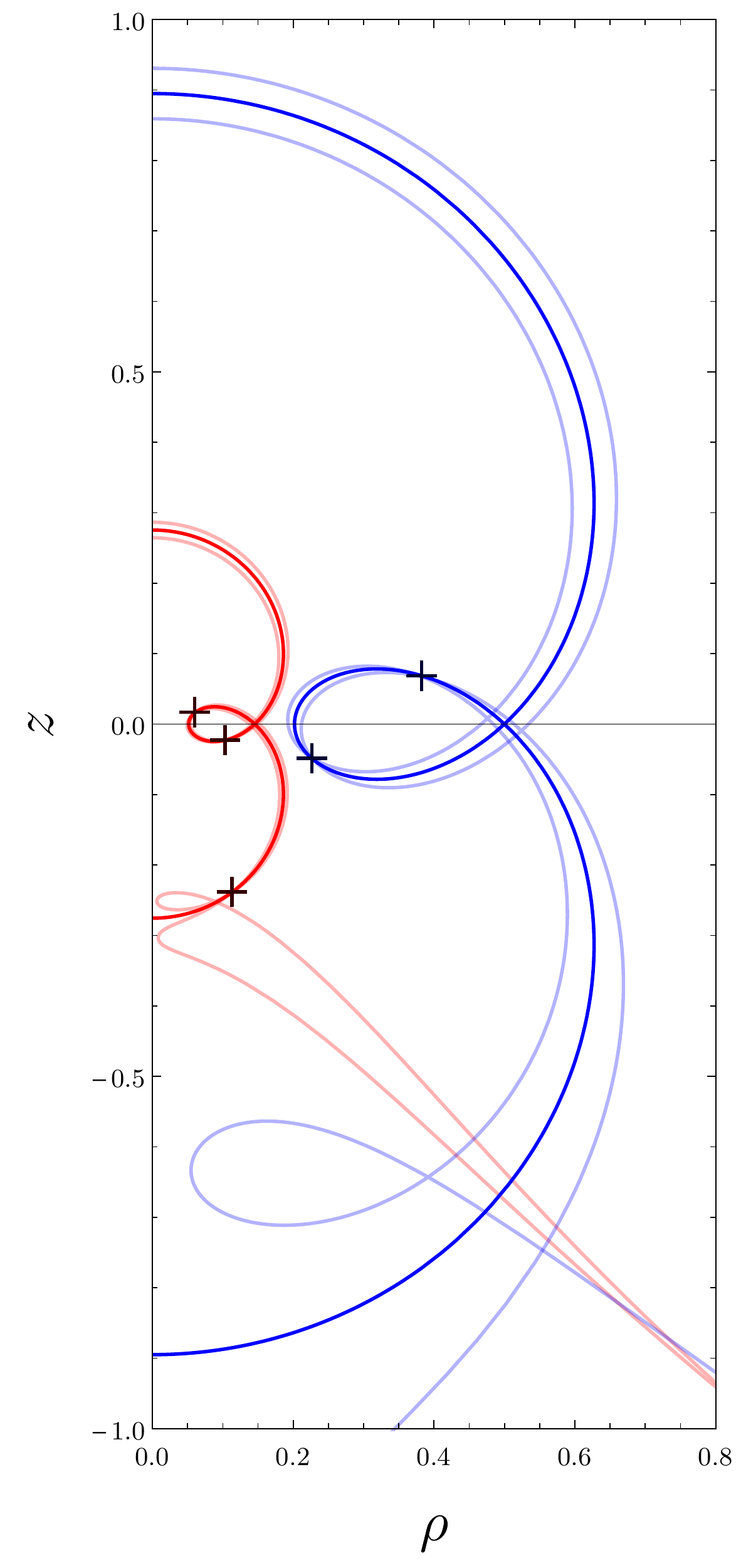}
\caption{MOTSodesic deviation for the one-loop MOTSs in the Gauss-Bonnet spacetime. The inner surface (red) has three negative $m=0$ eigenvalues, which corresponds to three intersections of nearby MOTSodesics, while the outer surface (blue) has two.}
\label{deviation-GB-1loop}
\end{figure}

The more interesting behaviour occurs for the eigenvalues of the surfaces with loops, which we show for the one-loop surface in Figure~\ref{oneLoopSpectrum}. In the Schwarzschild case, each loop gives rise to two negative $m=0$ eigenvalues of the stability operator. Here, we see the situation is a bit more complicated. Focusing on the one-loop surfaces we see that the outer MOTS has two negative $m=0$ eigenvalues, while the inner MOTS has three. When we study the dependence of these eigenvalues on the coupling, we find that the spectrum evolves smoothly as a function of $\alpha$ through the annihilation of the MOTSs. Interestingly, we find that precisely at the value of coupling for which the one-loop MOTSs annihilate, the eigenvalue $\lambda_{2,0}$ vanishes. This result is consistent with the general considerations of~\cite{Andersson:2007fh} (where it is shown that MOTSs can only appear/disappear when the stability operator is not invertible), and in line with the numerical observations of~\cite{Pook-Kolb:2021hvz,Pook-Kolb:2021jpd}. In those works, it was found that for unstable, axisymmetric MOTSs, it is one of the higher $m=0$ eigenvalues that vanishes at a point of bifurcation/annihilation. What we have found provides another example where this phenomenon occurs, here in a simple exact solution.   

A similar pattern occurs for the two-loop surface (though we do not show the plots for space reasons). In that case, we find that the outer surface behaves like the Schwarzschild counterpart, having four negative $m=0$ eigenvalues, while the inner surface has five negative eigenvalues. At the point of annihilation, we once again see that this coincides with the vanishing of a single eigenvalue, this time $\lambda_{4,0}$. We expect that the trend continues to surfaces of higher loops.

In~\cite{Booth:2021sow}, it was shown that there is a direct geometric relationship between the spectrum of the stability operator and the properties of MOTSodesics: The number of negative eigenvalues of the stability operator counts how many times a nearby MOTSodesic will intersect with a given MOTS. Here we see this behaviour borne out. Taking, for illustration, the one-loop surfaces, we see that nearby MOTSodesics intersect the outer surface twice, while they intersect the inner surface three times --- see Figure~\ref{deviation-GB-1loop}.\footnote{Note that this is a perturbative calculation, valid so long as the MOTSodesic is ``nearby'' the MOTS. Therefore, intersections that occur \textit{after} the MOTSodesic has deviated significantly from the original MOTS should not be counted. This is discussed at length in~\cite{Booth:2021sow}.} This provides additional support for the robustness of our numerical findings regarding the spectrum.

\subsection{Reissner-Nordstr\"om Black Hole}

As our second example of a black hole with inner horizon we take the quintessential case of the Riessner-Nordstrom spacetime. The metric function for this geometry is given by
\be 
f(r) = 1 - \frac{2 M}{r} + \frac{Q^2}{r^2} \, ,
\ee
where $M$ is the mass and $Q$ is the charge of the black hole. Without loss of generality, we consider only positive charges. The spacetime has two horizons, located at the roots $f(r_\pm) = 0$:
\be 
r_\pm = M \pm \sqrt{M^2 - Q^2} \, .
\label{RNrpm}
\ee
In the case $M = Q$ the black hole is extremal, while for $Q/M > 1$ there is a naked singularity. We will work only in the case $Q/M \le 1$. 

\begin{figure*}[htp]
\centering
\includegraphics[width=\textwidth]{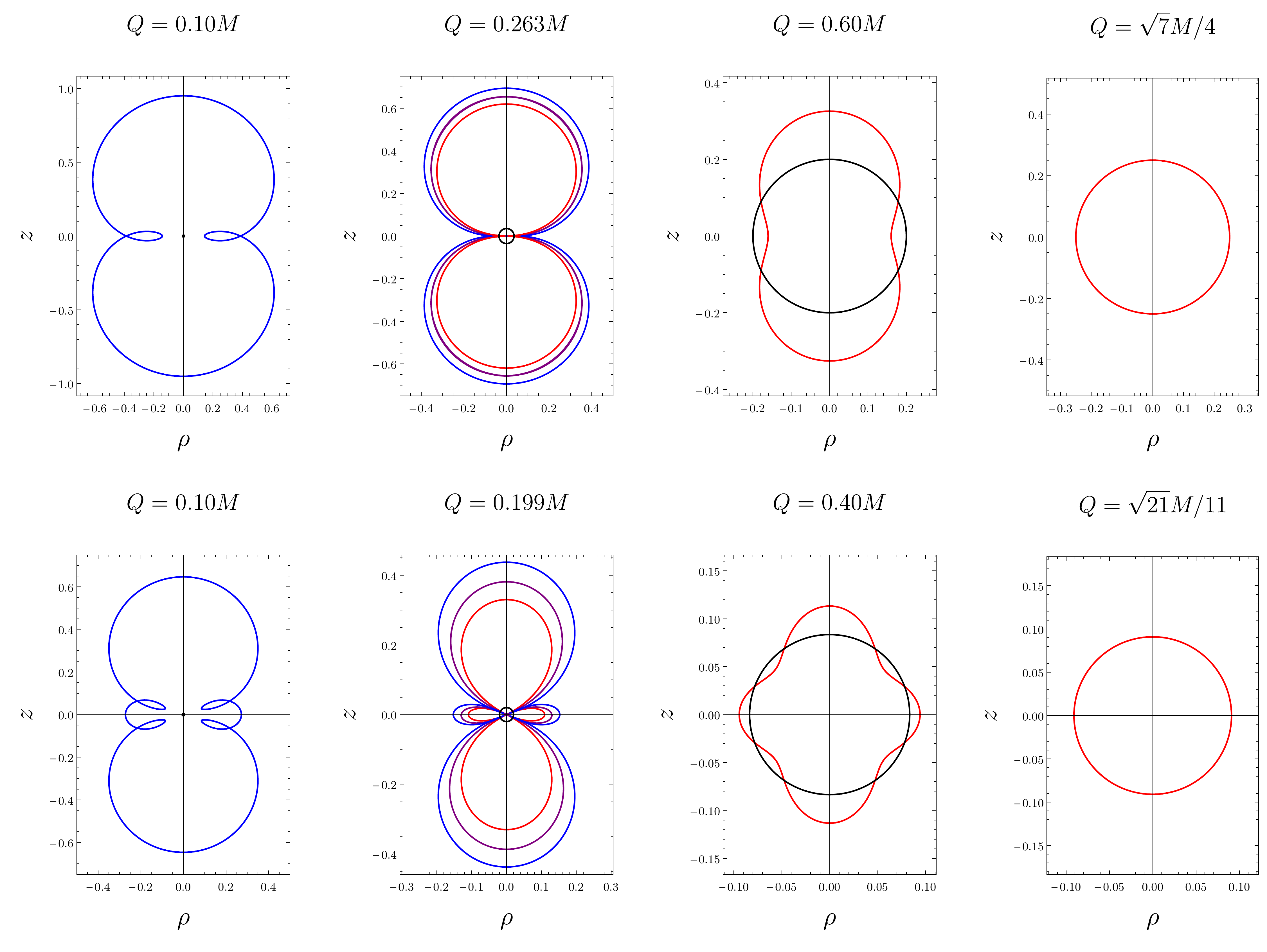}
\caption{Evolution of the first two self-intersecting Reissner-Nordstr\"om MOTSs 
for 
increasing  values of charge. In all cases, the black circle corresponds to the inner horizon. 
The figures show 
how the geometry of these surfaces changes, starting from small charge and continuing until the surface merges with the inner horizon. The MOTSs are plotted in coordinates $(\rho, z)  = (r \cos \theta, r \sin \theta)$. For comparison sake, note that $\sqrt{7}/4 \approx 0.661438$ and $\sqrt{21}/11 \approx 0.416598$.}
\label{RN_loops}
\end{figure*}


\begin{figure}
\centering
\includegraphics[width = 0.45\textwidth]{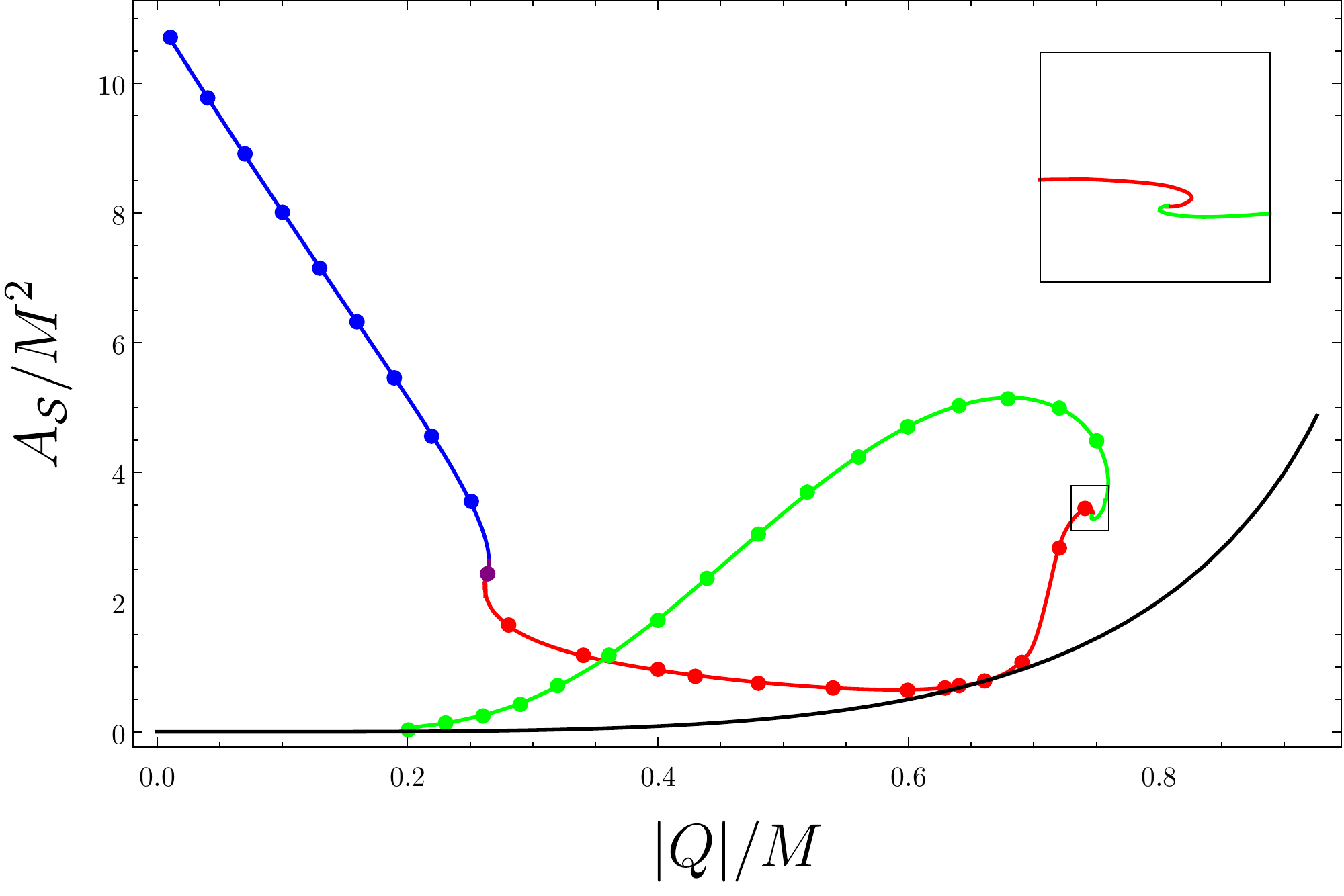}
\includegraphics[width = 0.45\textwidth]{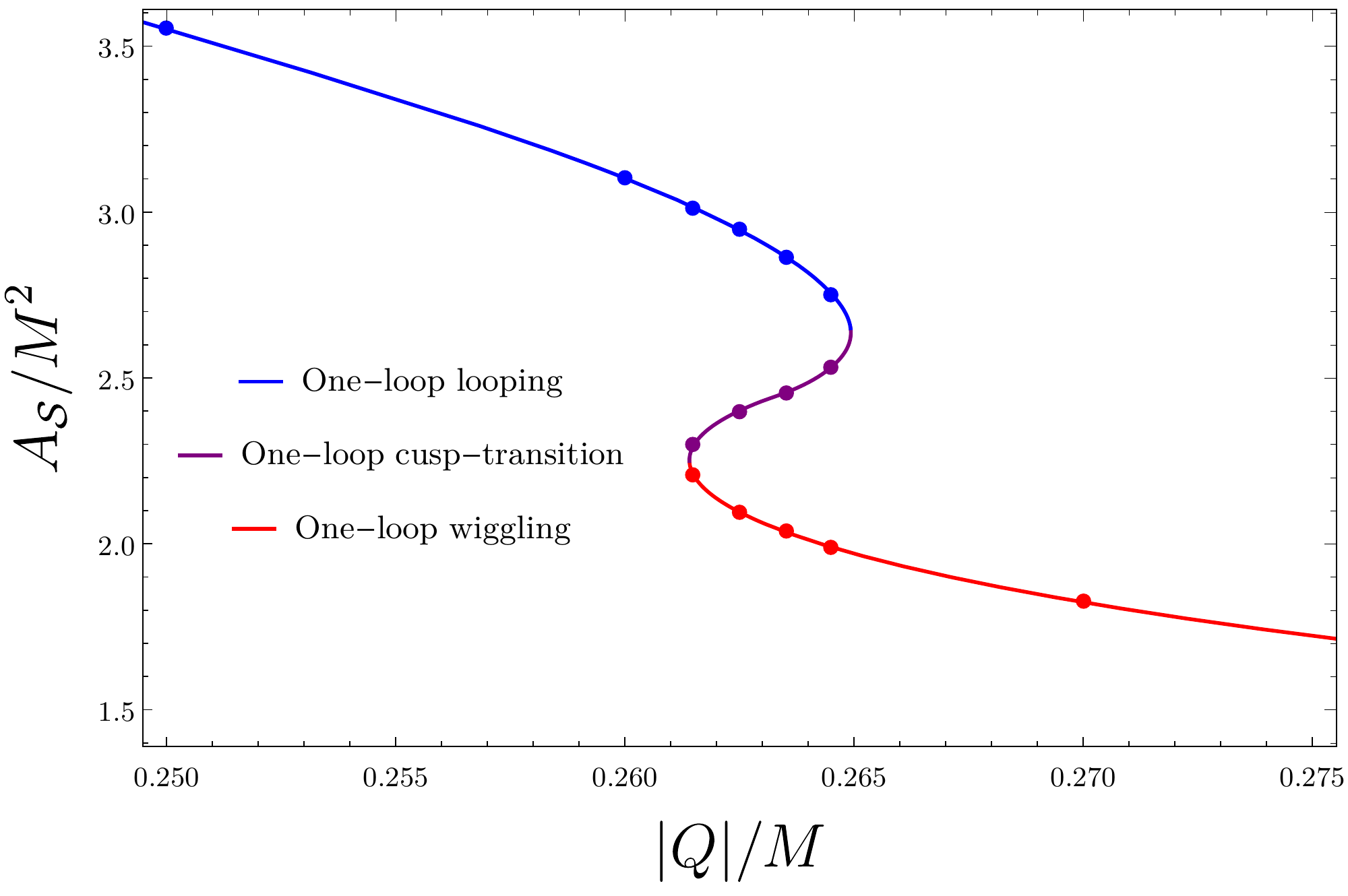}
\caption{Top: The area of the one loop MOTS, now considering its evolution past the merger with the inner horizon. The black curve gives the area of the inner horizon. The inset shows a zoomed-in version of the region highlighted with the small black hole. Bottom: A zoomed-in version of the top plot, focussing on the region where the one loop surface loses its loop.}
\label{fig:RNareafull}
\end{figure}

\begin{figure}
\centering
\includegraphics[width = 0.45\textwidth]{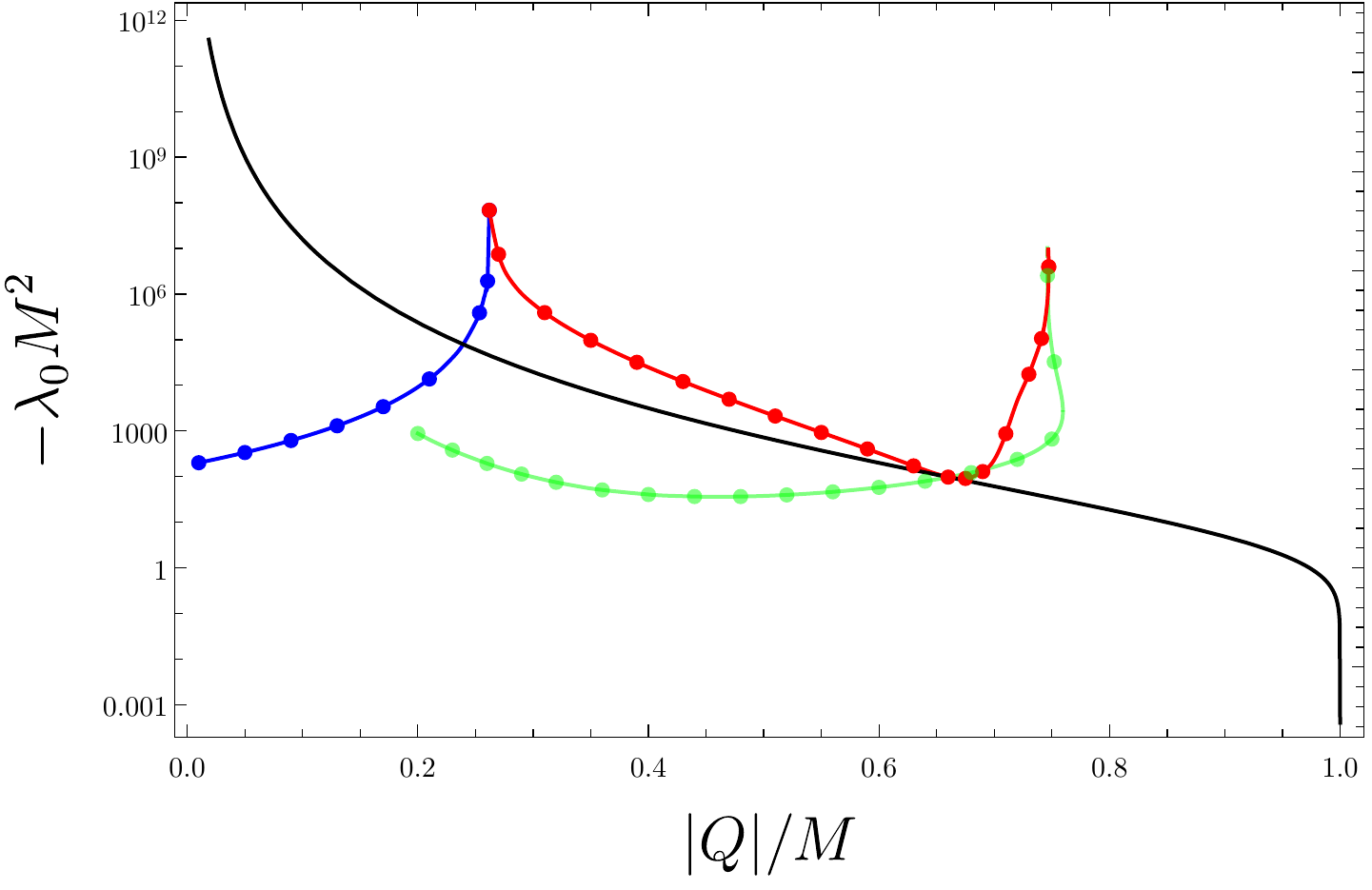}
\caption{A plot of (the negative of) the principal eigenvalue of the stability operator for the one-loop surface in the Reissner-Nordstr\"om spacetime. The principal eigenvalue exhibits strong peaks at the location of cusps. The black curve corresponds to the principal eigenvalue of the inner horizon of the black hole. }
\label{RN-principal}
\end{figure}

\begin{figure*}
\centering
\includegraphics[width = 0.45\textwidth]{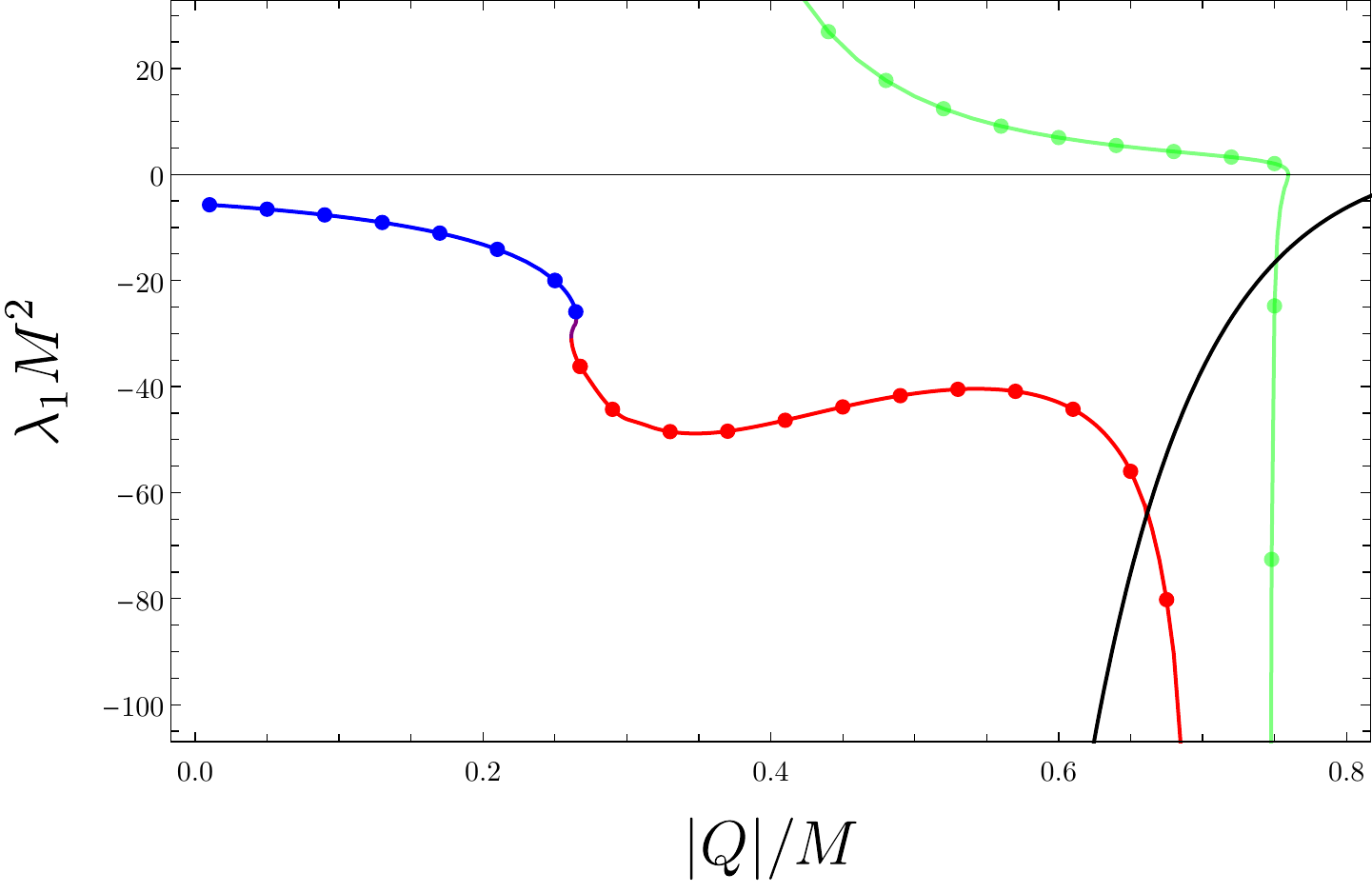}
\quad 
\includegraphics[width = 0.45\textwidth]{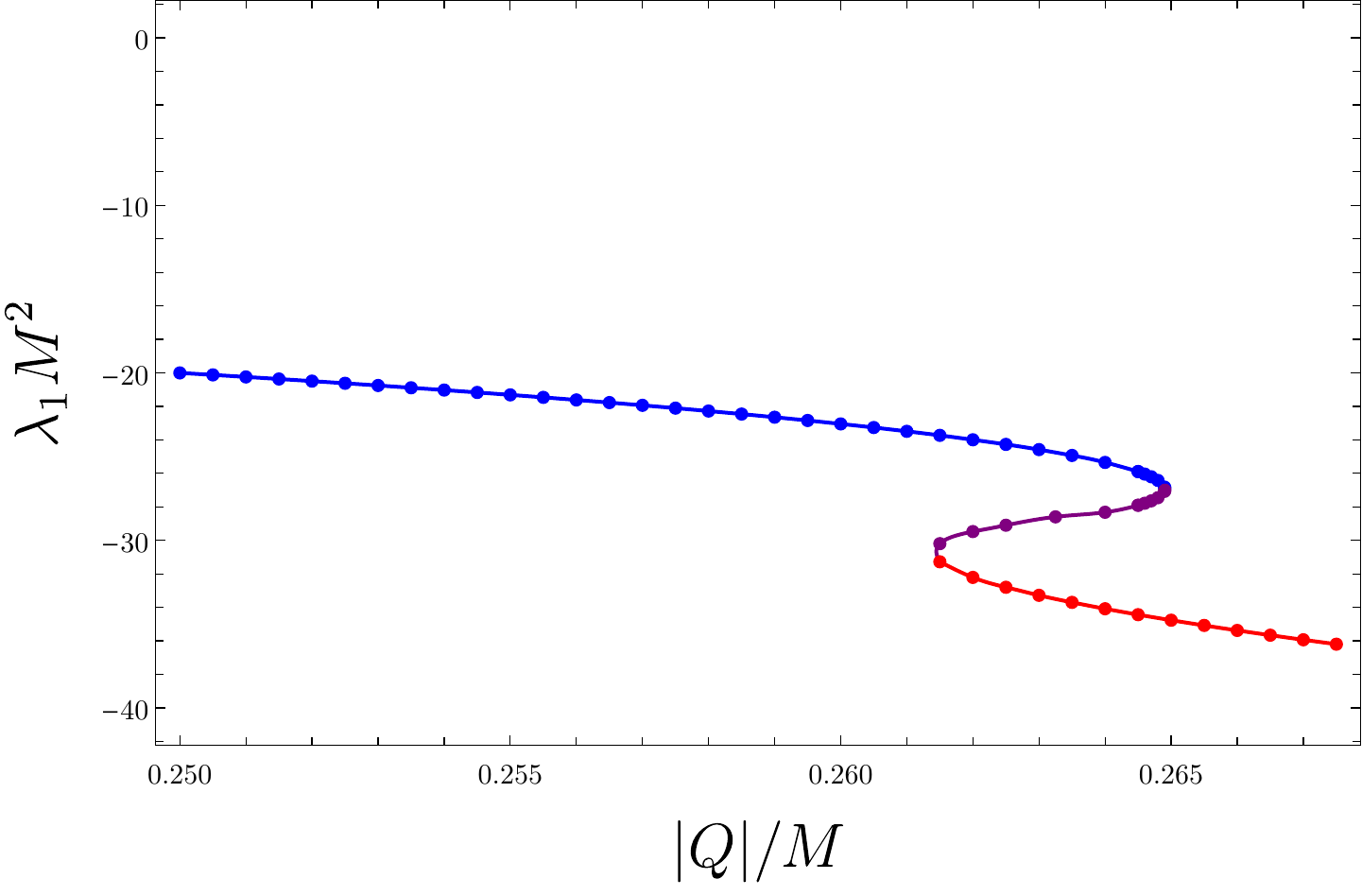}
\includegraphics[width = 0.45\textwidth]{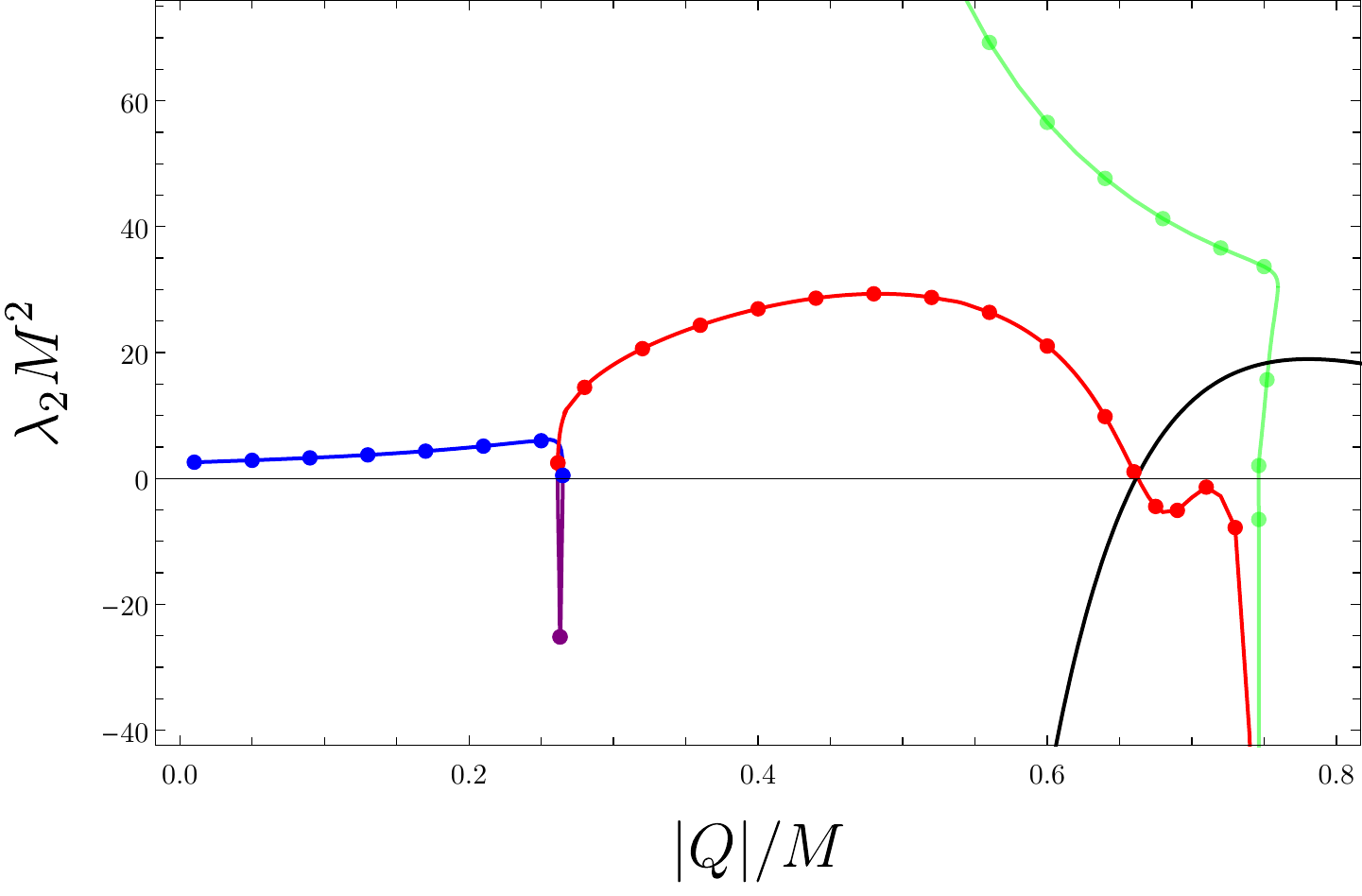}
\quad
\includegraphics[width = 0.45\textwidth]{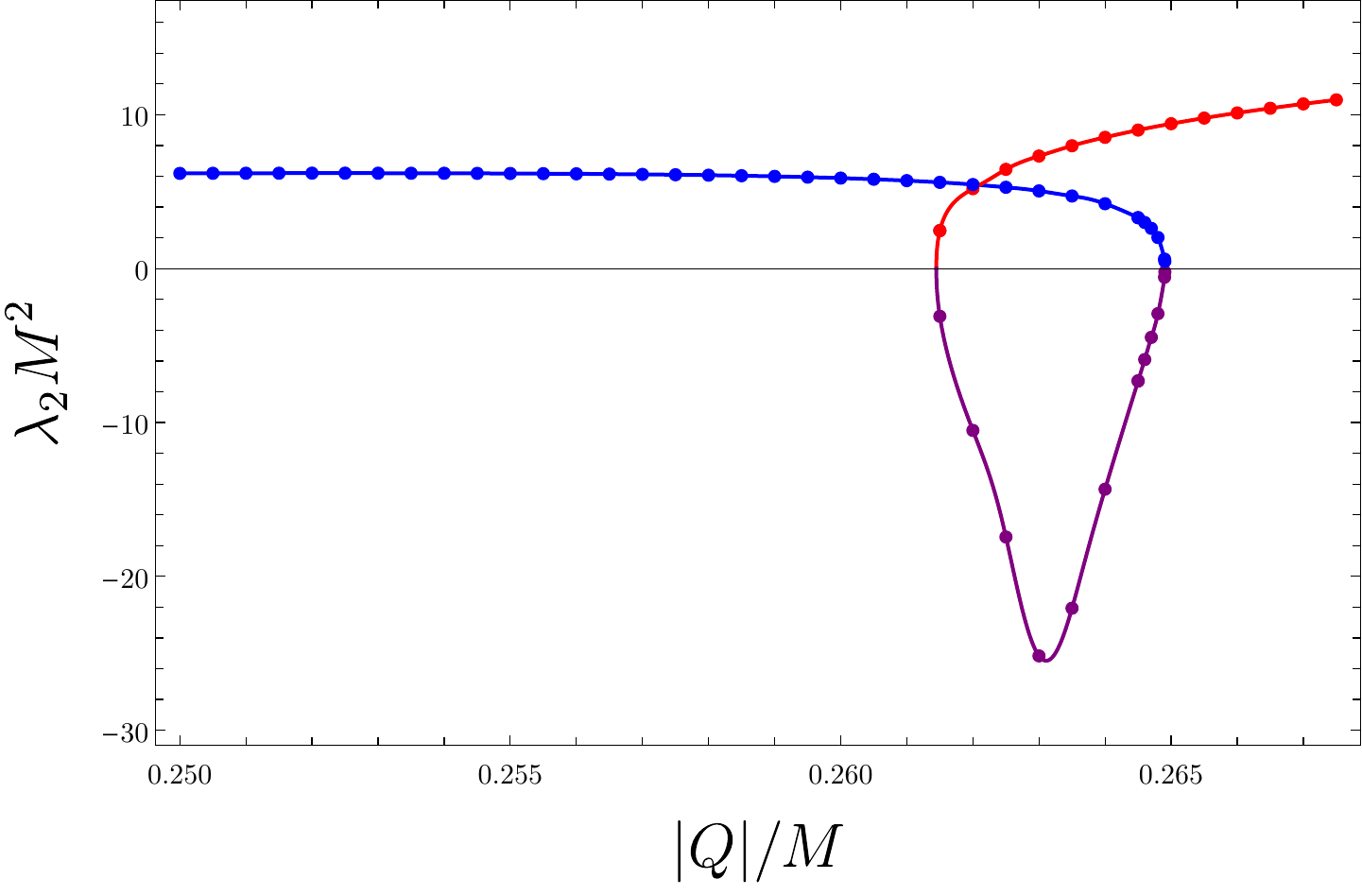}
\caption{Plots of the $\ell = 1$ (top) and $\ell = 2$ (bottom) eigenvalues with $m=0$. The right plots provide a zoomed-in view of the corresponding left plot over a region of interest. The black curve corresponds to the corresponding eigenvalue of the inner horizon of the black hole. In the left panels, the red and green curves eventually meet, though this occurs at such large values of the eigenvalues that displaying it would obscure the more interesting structures shown here.}
\label{RN-eigs}
\end{figure*}

As discussed earlier, ordinary PG coordinates fail for this black hole because the metric function $f(r) \to + \infty$ as $r \to 0$. Thus it is not possible to probe all the way to the singularity with the ordinary PG coordinates. Furthermore, while the Martel-Poisson family of generalized PG coordinates allow for the coordinate system to extend `closer' to the singularity, ultimately for any value of $p > 0$ they will fail as well, with only the $p=0$  (Eddington-Finkelstein) coordinates extending all the way to the singularity. Therefore, if we wish to have a space-like three-metric, we must make a non-trivial choice for the function $p(r)$ in the PG-like metric~\eqref{GenPG_metric}. The results in this subsection have been produced using a the choice
\be 
p(r) = \frac{r^2}{(r+Q)^2} \, ,
\ee
which is a special case of~\eqref{LorentzP}, corresponding to an infalling charged particle with charge to mass ratio equal to unity. However, we have also verified that the qualitative picture is the same for other choices of $p(r)$, having reproduced the qualitative picture with $p(r) = r^2/(r^2+Q^2)$ as well. 

The situation for Reissner-Nordstr\"om is far more complicated and interesting than for the Gauss-Bonnet black hole. Once again, the inner horizon has important effects. For any finite value of the electric charge, there are a finite number of axisymmetric MOTSs found in the black hole interior. For small values of charge there are more MOTSs. If we fix a small value of electric charge, we can recover various MOTSs characteristic of those present in the Schwarzschild black hole --- axisymmetric MOTSs with multiple loops. It is interesting to understand the evolution of these surfaces as the electric charge is increased. We show in Fig.~\ref{RN_loops} a partial picture of this evolution for the MOTSs with one and two loops.

For sufficiently small charge, the MOTSs we find resemble those in the Schwarzschild spacetime. However, as the charge is increased they undergo interesting transformations. Consider the case of the one-loop MOTS that is shown in the top row of Fig.~\ref{RN_loops}. The area of this MOTS as a function of charge is shown in Fig.~\ref{fig:RNareafull} in blue, which is helpful to understand the evolution. As the charge is increased, the loop portion of the MOTS gradually shrinks in size. When $Q \approx 0.2615 M$, a bifurcation event occurs, and two new MOTSs appear --- these correspond to the purple and red curves in the lower panel of Fig.~\ref{fig:RNareafull}. Neither of these surfaces contain loops and instead exhibit `dimples' that bulge inward along the $\rho$-axis. These three MOTSs are shown coexisting in the second panel of the top row in Fig.~\ref{RN_loops}.  As the charge is further increased, one of these surfaces (shown in red) continues to decrease in area, approaching the inner horizon. Meanwhile, the other MOTS (shown in purple) grows in area and tends toward the initially present looped MOTSs (shown in blue). During this process, the purple MOTS transitions from having a dimple, through a cusp, to possessing a loop at $Q \approx  0.2631 M$. For charges larger than this, the purple MOTS is looped. Ultimately, the purple MOTS annhilates with the initially present blue MOTS at $Q \approx 0.2645 M$. As the charge is further increased, the red MOTS becomes more symmetric (see the third panel in the top row of Fig.~\ref{RN_loops}) and approaches the inner horizon. At precisely $Q = \sqrt{7}M/4$ the red MOTS coincides with the inner horizon. However, as we will explain below, this is not truly an annihilation event.


A similar fate besets the two-looped MOTSs. The surfaces begin resembling those present in the Schwarzschild interior then, as the charge is increased, transition through a cusp-formation event mediated by two additional MOTSs that form in a bifurcation event, lose their loops and become dimpled surfaces, and then ultimately coincide with the inner horizon at some critical charge $Q_\star$. Snapshots of this process are shown for the two-looped MOTS in the bottom row of Fig.~\ref{RN_loops}.

\begin{figure*}[htp]
\centering
\includegraphics[width=\linewidth]{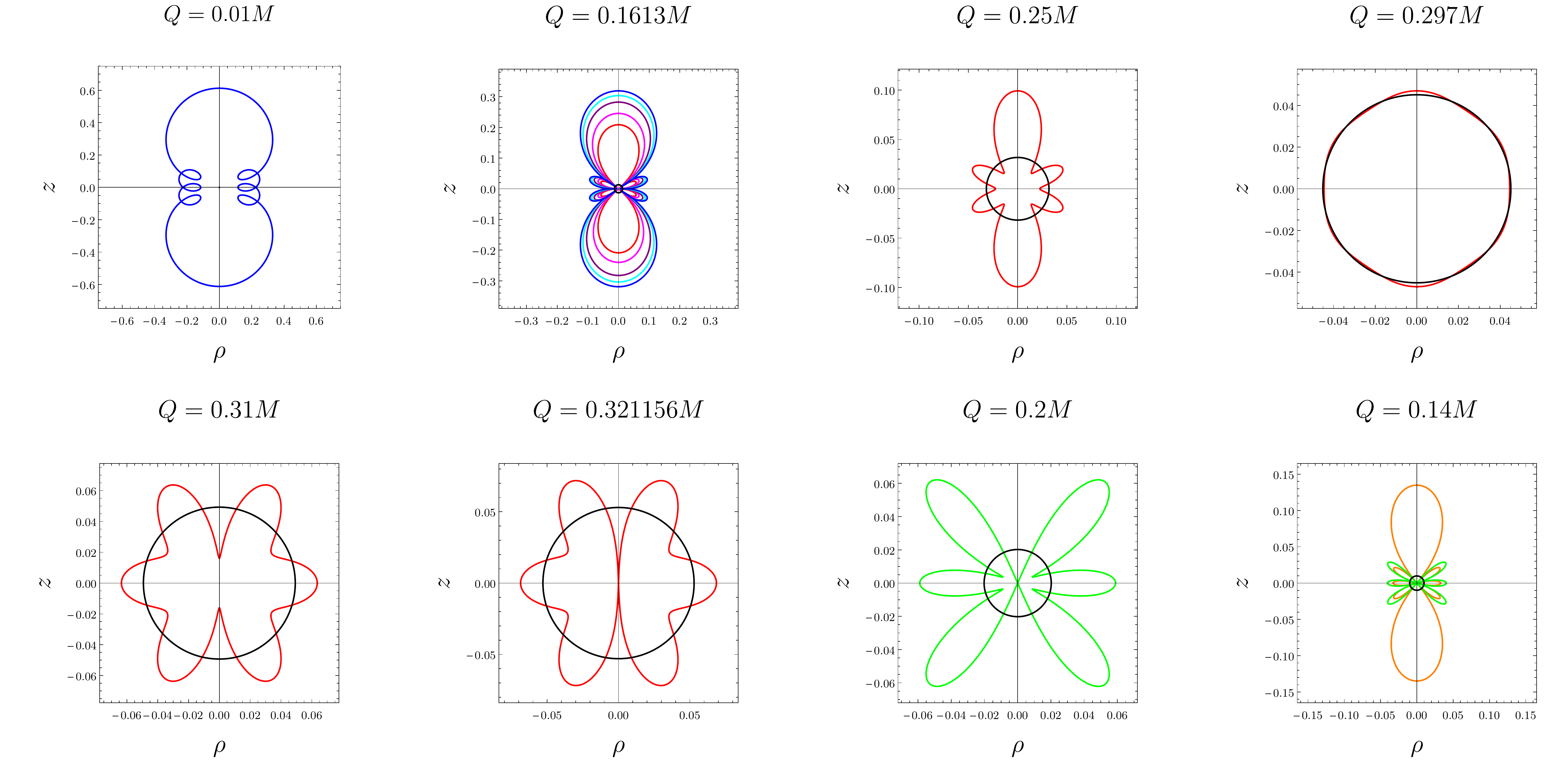}
\caption{Several snapshots of the MOTSs appearing along the `world-tube' traced out by the three loop surface. In all cases, the black circle corresponds to the inner horizon. The MOTSs are plotted in coordinates $(\rho, z)  = (r \cos \theta, r \sin \theta)$.}
\label{RN-3worldtube}
\end{figure*}

For higher-looped MOTSs, the number of MOTSs involved in the transition increases. In the three-looping case, the process involves five MOTS. This is seen in Fig.~\ref{RN-3worldtube} for the $Q=0.1613M$ slice. In this case, the loops transition to dimples in pairs. The loop that lies on the $\rho$-axis first transitions to a dimple, and then the two remaining loops  transition together, prompting two separate bifurcation-annihilation events. If this trend continues, we would expect that the four-loop MOTSs will show five MOTSs during the transition, the five-loop MOTSs will have seven MOTSs during the transition, and so on.

The stability operator 
provides extra
insight into the process just outlined, especially the coincidence of the various MOTSs with the inner horizon. The inner horizon, being a geometric sphere, allows for an explicit computation of the stability operator spectrum:
\be 
\lambda_\ell = \frac{2(\ell^2+\ell+1)M r_- - (\ell^2 + \ell + 2) Q^2}{r_-^4}
\ee
where $r_-$ is given in terms of $M$ and $Q$ in Eq.~\eqref{RNrpm}. A key point is that, because $r_- \sim Q^2/(2M)$ as $Q/M \to 0$, the inner horizon can have any number of negative eigenvalues. The principal, $\ell = 0$, eigenvalue is always negative, except for in the extremal limit where it vanishes. The higher-$\ell$ eigenvalues can be either positive or negative, depending on the black hole charge. The $N$-looped surface\footnote{Here, and throughout, we will use the term `N-looped surface' to refer to the surface that begins with $N$-loops and then is continuously connected to a particular dimpled surface, as explained in detail for the $1$-looped surface.} coincides with the inner horizon exactly when the inner horizon's $(2N+1$)'th eigenvalue vanishes. This leads to the explicit formula for the critical charge at the point of coincidence:
\be 
\frac{Q_\star}{M} = \frac{\sqrt{4 N^2 + 2 N + 1}}{2 N^2 + N + 1} \, .
\ee 
Note that this is independent of the choice of function $p(r)$,\footnote{Subject only to the condition that $p(r)$ allows for the generalized PG coordinates to extend to the inner horizon.} as the eigenvalue spectrum of the inner horizon is independent of the choice of $p(r)$.

\begin{figure*}[htp]
\centering
\includegraphics[width=\linewidth]{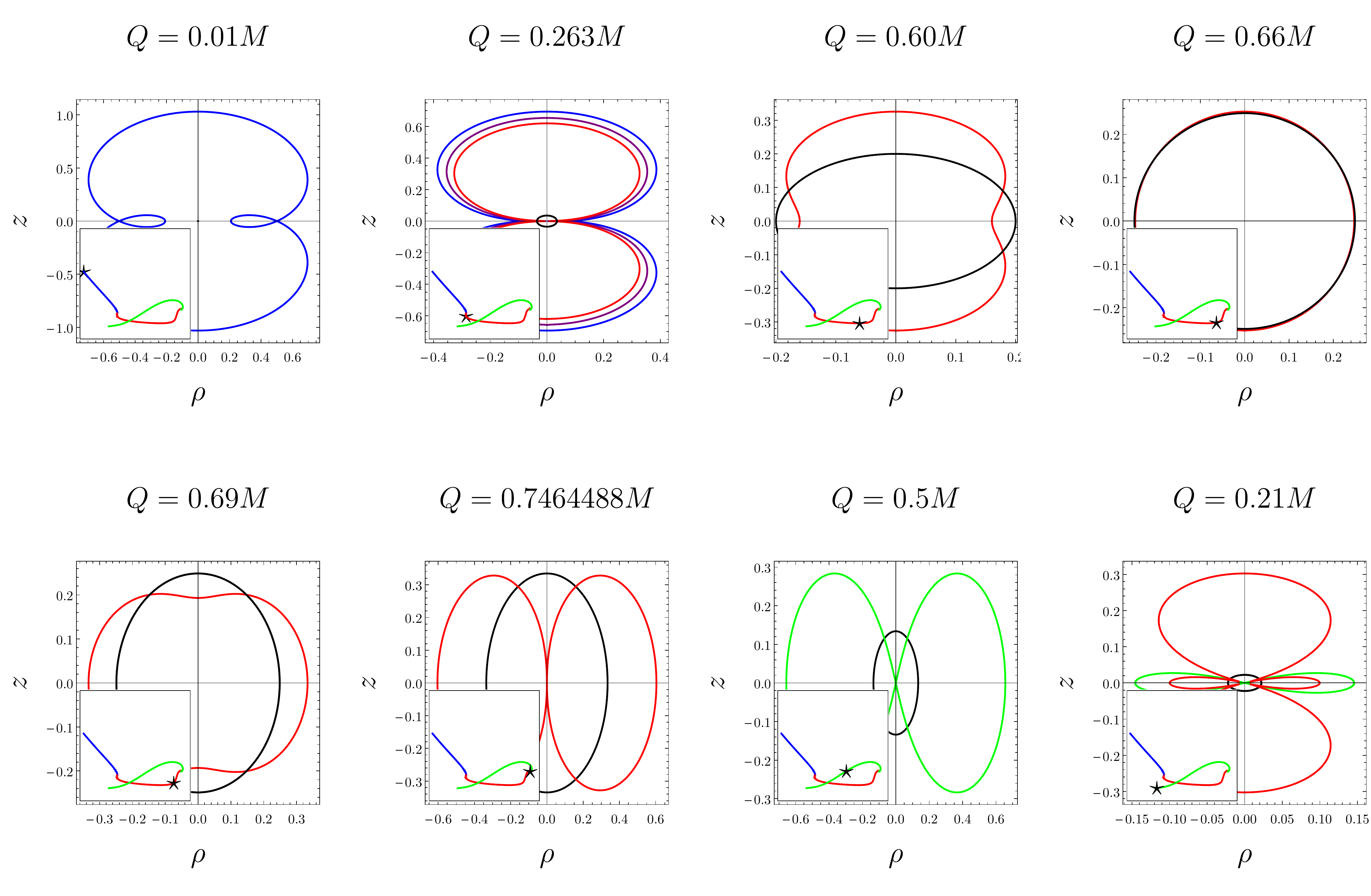}
\caption{Several snapshots of the MOTSs appearing along the `world-tube' traced out by the one loop surface. The star on the inset shows the placement along the area curve --- see Fig.~\ref{fig:RNareafull}. In all cases, the black circle  corresponds to the inner horizon. The MOTSs are plotted in coordinates $(\rho, z)  = (r \cos \theta, r \sin \theta)$. Note that the scales on the $\rho$ and $z$ axis are usually not the same and so the inner horizon often appears as an ellipse.}
\label{RN-worldtube}
\end{figure*}

As was alluded to before, the situation does not end when the surfaces coincide with the inner horizon. As can be seen from the area curves in Fig.~\ref{fig:RNareafull}, and also in the analysis of the stability operator's spectrum in Figs.~\ref{RN-principal} and \ref{RN-eigs}, the merger with the inner horizon is not smooth. In fact, the surfaces can be traced past this point. We will focus on the particular case of the one-loop surface to illustrate the idea, and show a more complete evolution of this surface's `world-tube' in Fig.~\ref{RN-worldtube}. After the coincidence with the inner horizon, the dimpled surface (shown in red) continues to evolve inward toward the singularity. It remains dimpled, but now along the opposite axis: before coinciding with the inner horizon, the surface bulged inward along the $\rho$-axis, while after the merger it bulges inward along the $z$-axis. As the charge is further increased, the 
surface bulges out more and more along the $\rho$-axis, and grows in area. As this happens, the point where it intersects with the $z$-axis gradually decreases toward zero. This process continues until $Q \approx 0.7464488 M$, at which point the point of intersection with the $z$-axis goes to zero. At this point, the MOTS is a two-lobed structure, where each lobe passes smoothly through the point $(\rho, z) = (0,0)$ --- this is shown in the second panel of the bottom row in Fig.~\ref{RN-worldtube}. The process by which this limit to zero happens involves another bifurcation/annihilation event, as seen in the inset of the top panel of Fig.~\ref{fig:RNareafull}. The red curve continues to a maximum charge, before it turns around to meet up with the green curve. The point where the red and green curves meet corresponds to the two-lobed MOTS just described. 

If we persist in following the surface, we find that it continues to exist, now intersecting the origin, yielding MOTSs that look like a `figure eight' --- see the third panel in the bottom row of Fig.~\ref{RN-worldtube}. All surfaces that appear in Fig.~\ref{fig:RNareafull} colour-coded as green have this structure.  As seen in the inset of Fig.~\ref{fig:RNareafull}, this MOTS exists for a small range of charges before it is annihilated by another MOTS of identical shape. We can follow this surface backward, to smaller charges and find that it ultimately coincides with the two-looped MOTS at the precise moment when its loops `pinch off' to form cusps --- see the final panel in the bottom row of Fig.~\ref{RN-worldtube} for the situation just before this coincidence occurs. In this sense, it is interesting to note that the one-loop and two-loop MOTSs in the Reissner-Nordstr\"om black hole are connected through a continuous sequence of intermediate MOTSs. We expect that this intricate behaviour continues also to higher-looped MOTSs, and it may very well be the case that all MOTSs present in the interior of the Reissner-Nordstr\"om  black hole can be connected to one another in this type of way.

The situation that we have described here is corroborated by an analysis of the spectrum of the stability operator. The first few $m=0$ eigenvalues are shown as a function of charge for the one-looped surface in Figs.~\ref{RN-principal} and~\ref{RN-eigs}. Specifically, Fig.~\ref{RN-principal} shows the principal eigenvalue. Note that this is a plot of the negative of the principal eigenvalue on a logarithmic scale. Note that the eigenvalue grows very large and displays a cusp at points where the surface itself does something dramatic. The two points in the plot where this happens corresponds to the cusp formation event of the one-looped MOTS --- when the purple MOTS displayed in Fig.~\ref{fig:RNareafull} transitions from being dimpled to being looped --- and when the dimpled MOTSs `pinches off' to yield the two-lobed MOTS shown in the second panel of the bottom row in Fig.~\ref{RN-worldtube}. This behaviour of the principal eigenvalue in the vicinity of cusp formation is consistent with previous observations of this phenomenon in the interior of a black hole merger~\cite{PhysRevLett.123.171102}. In Fig.~\ref{RN-eigs} the $\ell = 1$ and $\ell = 2$ eigenvalues are shown. In particular, here we can see how the purple surface smoothly interpolates between the looped and dimpled MOTS, with zeros of the $\ell = 2$ eigenvalue coinciding with the bifurcation/annihilation events. More broadly, we note that the eigenvalues of the surfaces never smoothly connect with the inner horizon. Of course, at the point where the surfaces coincide with the inner horizon the eigenvalues all match, but it is coincidence at a point, rather than a smooth connection. This adds further support to our argument that the one-looped MOTS is smoothly connected to the dimpled MOTS that intersects the $z$-axis inside the inner horizon.



\section{Conclusions}

In this work, we have studied the behaviour of MOTSs in the interior of spherically symmetric black holes. We have performed this study in a family of generalized Painlev{\'e}-Gullstrand that we have introduced. These coordinates may be of independent interest, beyond the application used for here. The coordinates are adapted to accelerated observers, rather the geodesic observers as is typical for  Painlev{\'e}-Gullstrand  coordinates. Nonetheless, the coordinates still have clear physical motivation in many cases. For example, in the case of the Reissner-Nordstr\"om black hole, the coordinates can be adapted to an infalling charged test particle. The main advantage of the generalized coordinates we have introduced is that they can continue to be valid in regimes where ordinary Painlev{\'e}-Gullstrand break down, such as inside the inner horizon of the Reissner-Nordstr\"om black hole. It may be interesting to construct analogous coordinate systems for asymptotically (A)dS black holes, where a similar failure of standard Painlev{\'e}-Gullstrand occurs.

We then considered the interior MOTSs of the Schwarzschild spacetime in a variety of different slicings within the family of generalized Painlev{\'e}-Gullstrand  coordinates. We found that the features observed in~\cite{Booth:2020qhb} --- namely, the presence of an infinite number of closed, self-intersecting MOTSs --- are robust, at least within this family of coordinate systems. 

We then considered how an inner horizon affects the structure of MOTSs present inside the interior, finding the results to be highly sensitive to the interior structure. We considered two examples: The four-dimensional Gauss-Bonnet black hole, and the Reissner-Nordstr\"om solution. In the first case, the internal structure was rather simple, involving pairs of looped MOTSs that lie between the event and inner horizons. Viewed as a function of the Gauss-Bonnet coupling, the pairs of looped MOTSs can be understood to bifurcate/annihilate. The Reissner-Nordstr\"om black hole exhibited a much more intricate structure. In that case, the looped MOTSs form and lose their loops through cusp formation events mediated through bifurcation/annihilations of additional MOTS. Rather than annihilating with a pair member, the MOTSs weave their way forward and back in `charge space', with the one-loop MOTS linking up in a non-trivial way with the two-looped MOTSs. The differences present in these two simple examples are surprising, and suggests that the structure of MOTSs in black hole interiors is rich. 

Our results provide exact solution examples where a number of the results discussed in~\cite{Pook-Kolb:2021hvz,Booth:2021sow, Pook-Kolb:2021jpd} can be observed in a much simpler context. For example, the connection between the bifurcation/annihilation of unstable MOTSs and the vanishing of a higher $m=0$ eigenvalue of the stability operator observed in~\cite{Pook-Kolb:2021jpd} was borne out in all cases studied here. More broadly, there is a remarkable similarity between the structure of MOTSs present in the interior of these exact solutions and the structures present in an axisymmetric merger of non-spinning black holes~\cite{PhysRevLett.123.171102, Booth:2020qhb, Pook-Kolb:2021hvz,Booth:2021sow, Pook-Kolb:2021jpd}. Carrying this idea forward, our observations here may offer guidance on what to expect in the interior of two merging electrically charged black holes. 

In future studies it would be interesting to extend our analysis to other exact solutions, for example rotating black holes or black holes with different asymptotic structure. It would also be desirable to develop a clear physical understanding of the internal MOTSs. Clearly, these surfaces are highly sensitive to the internal structure of the black hole, and it would be worthwhile to understand their physical implications. In this sense, understanding the development of these MOTSs in a collapse scenario may shed light on the physical meaning, as their development and evolution could be directly associated with physical properties such as density. We hope to return to some of these problems in the future.

\begin{acknowledgments}
The authors would like to thank Hari Kunduri, Sarah Muth, and Daniel Pook-Kolb for helpful discussions and comments. IB was supported by the Natural Science and Engineering Research Council of Canada Discovery Grant
2018-0473. The work of RAH was supported by the Natural Science and Engineering Research Council of Canada and by AOARD Grant FA2386-19-1-4077.
\end{acknowledgments}

\bibliography{thebib}{}

\begin{thebibliography}{40}%
\makeatletter
\providecommand \@ifxundefined [1]{%
 \@ifx{#1\undefined}
}%
\providecommand \@ifnum [1]{%
 \ifnum #1\expandafter \@firstoftwo
 \else \expandafter \@secondoftwo
 \fi
}%
\providecommand \@ifx [1]{%
 \ifx #1\expandafter \@firstoftwo
 \else \expandafter \@secondoftwo
 \fi
}%
\providecommand \natexlab [1]{#1}%
\providecommand \enquote  [1]{``#1''}%
\providecommand \bibnamefont  [1]{#1}%
\providecommand \bibfnamefont [1]{#1}%
\providecommand \citenamefont [1]{#1}%
\providecommand \href@noop [0]{\@secondoftwo}%
\providecommand \href [0]{\begingroup \@sanitize@url \@href}%
\providecommand \@href[1]{\@@startlink{#1}\@@href}%
\providecommand \@@href[1]{\endgroup#1\@@endlink}%
\providecommand \@sanitize@url [0]{\catcode `\\12\catcode `\$12\catcode
  `\&12\catcode `\#12\catcode `\^12\catcode `\_12\catcode `\%12\relax}%
\providecommand \@@startlink[1]{}%
\providecommand \@@endlink[0]{}%
\providecommand \url  [0]{\begingroup\@sanitize@url \@url }%
\providecommand \@url [1]{\endgroup\@href {#1}{\urlprefix }}%
\providecommand \urlprefix  [0]{URL }%
\providecommand \Eprint [0]{\href }%
\providecommand \doibase [0]{http://dx.doi.org/}%
\providecommand \selectlanguage [0]{\@gobble}%
\providecommand \bibinfo  [0]{\@secondoftwo}%
\providecommand \bibfield  [0]{\@secondoftwo}%
\providecommand \translation [1]{[#1]}%
\providecommand \BibitemOpen [0]{}%
\providecommand \bibitemStop [0]{}%
\providecommand \bibitemNoStop [0]{.\EOS\space}%
\providecommand \EOS [0]{\spacefactor3000\relax}%
\providecommand \BibitemShut  [1]{\csname bibitem#1\endcsname}%
\let\auto@bib@innerbib\@empty
\bibitem [{\citenamefont {Hawking}\ and\ \citenamefont
  {Ellis}(1973)}]{hawking_ellis_1973}%
  \BibitemOpen
  \bibfield  {author} {\bibinfo {author} {\bibfnamefont {S.~W.}\ \bibnamefont
  {Hawking}}\ and\ \bibinfo {author} {\bibfnamefont {G.~F.~R.}\ \bibnamefont
  {Ellis}},\ }\href {\doibase 10.1017/CBO9780511524646} {\emph {\bibinfo
  {title} {The Large Scale Structure of Space-Time}}},\ Cambridge Monographs on
  Mathematical Physics\ (\bibinfo  {publisher} {Cambridge University Press},\
  \bibinfo {year} {1973})\BibitemShut {NoStop}%
\bibitem [{\citenamefont {Bardeen}\ \emph {et~al.}(1973)\citenamefont
  {Bardeen}, \citenamefont {Carter},\ and\ \citenamefont
  {Hawking}}]{Bardeen:1973gs}%
  \BibitemOpen
  \bibfield  {author} {\bibinfo {author} {\bibfnamefont {James~M.}\
  \bibnamefont {Bardeen}}, \bibinfo {author} {\bibfnamefont {B.}~\bibnamefont
  {Carter}}, \ and\ \bibinfo {author} {\bibfnamefont {S.~W.}\ \bibnamefont
  {Hawking}},\ }\bibfield  {title} {\enquote {\bibinfo {title} {{The Four laws
  of black hole mechanics}},}\ }\href {\doibase 10.1007/BF01645742} {\bibfield
  {journal} {\bibinfo  {journal} {Commun. Math. Phys.}\ }\textbf {\bibinfo
  {volume} {31}},\ \bibinfo {pages} {161--170} (\bibinfo {year}
  {1973})}\BibitemShut {NoStop}%
\bibitem [{\citenamefont {Visser}(2014)}]{Visser:2014zqa}%
  \BibitemOpen
  \bibfield  {author} {\bibinfo {author} {\bibfnamefont {Matt}\ \bibnamefont
  {Visser}},\ }\bibfield  {title} {\enquote {\bibinfo {title} {{Physical
  observability of horizons}},}\ }\href {\doibase 10.1103/PhysRevD.90.127502}
  {\bibfield  {journal} {\bibinfo  {journal} {Phys. Rev. D}\ }\textbf {\bibinfo
  {volume} {90}},\ \bibinfo {pages} {127502} (\bibinfo {year} {2014})},\
  \Eprint {http://arxiv.org/abs/1407.7295} {arXiv:1407.7295 [gr-qc]}
  \BibitemShut {NoStop}%
\bibitem [{\citenamefont {Booth}(2005)}]{Booth:2005qc}%
  \BibitemOpen
  \bibfield  {author} {\bibinfo {author} {\bibfnamefont {Ivan}\ \bibnamefont
  {Booth}},\ }\bibfield  {title} {\enquote {\bibinfo {title} {{Black hole
  boundaries}},}\ }\href {\doibase 10.1139/p05-063} {\bibfield  {journal}
  {\bibinfo  {journal} {Can. J. Phys.}\ }\textbf {\bibinfo {volume} {83}},\
  \bibinfo {pages} {1073--1099} (\bibinfo {year} {2005})},\ \Eprint
  {http://arxiv.org/abs/gr-qc/0508107} {arXiv:gr-qc/0508107} \BibitemShut
  {NoStop}%
\bibitem [{\citenamefont {Penrose}(1969)}]{Penrose:1969pc}%
  \BibitemOpen
  \bibfield  {author} {\bibinfo {author} {\bibfnamefont {R.}~\bibnamefont
  {Penrose}},\ }\bibfield  {title} {\enquote {\bibinfo {title} {{Gravitational
  collapse: The role of general relativity}},}\ }\href {\doibase
  10.1023/A:1016578408204} {\bibfield  {journal} {\bibinfo  {journal} {Riv.
  Nuovo Cim.}\ }\textbf {\bibinfo {volume} {1}},\ \bibinfo {pages} {252--276}
  (\bibinfo {year} {1969})}\BibitemShut {NoStop}%
\bibitem [{\citenamefont {Dreyer}\ \emph {et~al.}(2003)\citenamefont {Dreyer},
  \citenamefont {Krishnan}, \citenamefont {Shoemaker},\ and\ \citenamefont
  {Schnetter}}]{Dreyer:2002mx}%
  \BibitemOpen
  \bibfield  {author} {\bibinfo {author} {\bibfnamefont {Olaf}\ \bibnamefont
  {Dreyer}}, \bibinfo {author} {\bibfnamefont {Badri}\ \bibnamefont
  {Krishnan}}, \bibinfo {author} {\bibfnamefont {Deirdre}\ \bibnamefont
  {Shoemaker}}, \ and\ \bibinfo {author} {\bibfnamefont {Erik}\ \bibnamefont
  {Schnetter}},\ }\bibfield  {title} {\enquote {\bibinfo {title} {{Introduction
  to isolated horizons in numerical relativity}},}\ }\href {\doibase
  10.1103/PhysRevD.67.024018} {\bibfield  {journal} {\bibinfo  {journal} {Phys.
  Rev. D}\ }\textbf {\bibinfo {volume} {67}},\ \bibinfo {pages} {024018}
  (\bibinfo {year} {2003})},\ \Eprint {http://arxiv.org/abs/gr-qc/0206008}
  {arXiv:gr-qc/0206008} \BibitemShut {NoStop}%
\bibitem [{\citenamefont {Krishnan}(2008)}]{Krishnan:2007va}%
  \BibitemOpen
  \bibfield  {author} {\bibinfo {author} {\bibfnamefont {Badri}\ \bibnamefont
  {Krishnan}},\ }\bibfield  {title} {\enquote {\bibinfo {title} {{Fundamental
  properties and applications of quasi-local black hole horizons}},}\ }\href
  {\doibase 10.1088/0264-9381/25/11/114005} {\bibfield  {journal} {\bibinfo
  {journal} {Class. Quant. Grav.}\ }\textbf {\bibinfo {volume} {25}},\ \bibinfo
  {pages} {114005} (\bibinfo {year} {2008})},\ \Eprint
  {http://arxiv.org/abs/0712.1575} {arXiv:0712.1575 [gr-qc]} \BibitemShut
  {NoStop}%
\bibitem [{\citenamefont {Gupta}\ \emph {et~al.}(2018)\citenamefont {Gupta},
  \citenamefont {Krishnan}, \citenamefont {Nielsen},\ and\ \citenamefont
  {Schnetter}}]{Gupta:2018znn}%
  \BibitemOpen
  \bibfield  {author} {\bibinfo {author} {\bibfnamefont {Anshu}\ \bibnamefont
  {Gupta}}, \bibinfo {author} {\bibfnamefont {Badri}\ \bibnamefont {Krishnan}},
  \bibinfo {author} {\bibfnamefont {Alex}\ \bibnamefont {Nielsen}}, \ and\
  \bibinfo {author} {\bibfnamefont {Erik}\ \bibnamefont {Schnetter}},\
  }\bibfield  {title} {\enquote {\bibinfo {title} {{Dynamics of marginally
  trapped surfaces in a binary black hole merger: Growth and approach to
  equilibrium}},}\ }\href {\doibase 10.1103/PhysRevD.97.084028} {\bibfield
  {journal} {\bibinfo  {journal} {Phys. Rev.}\ }\textbf {\bibinfo {volume}
  {D97}},\ \bibinfo {pages} {084028} (\bibinfo {year} {2018})},\ \Eprint
  {http://arxiv.org/abs/1801.07048} {arXiv:1801.07048 [gr-qc]} \BibitemShut
  {NoStop}%
\bibitem [{\citenamefont {Hayward}(1993)}]{Hayward:1993mw}%
  \BibitemOpen
  \bibfield  {author} {\bibinfo {author} {\bibfnamefont {Sean~A.}\ \bibnamefont
  {Hayward}},\ }\bibfield  {title} {\enquote {\bibinfo {title} {{Marginal
  surfaces and apparent horizons}},}\ }\href@noop {} {\  (\bibinfo {year}
  {1993})},\ \Eprint {http://arxiv.org/abs/gr-qc/9303006} {arXiv:gr-qc/9303006}
  \BibitemShut {NoStop}%
\bibitem [{\citenamefont {Ashtekar}\ and\ \citenamefont
  {Krishnan}(2002)}]{Ashtekar:2002ag}%
  \BibitemOpen
  \bibfield  {author} {\bibinfo {author} {\bibfnamefont {Abhay}\ \bibnamefont
  {Ashtekar}}\ and\ \bibinfo {author} {\bibfnamefont {Badri}\ \bibnamefont
  {Krishnan}},\ }\bibfield  {title} {\enquote {\bibinfo {title} {{Dynamical
  horizons: Energy, angular momentum, fluxes and balance laws}},}\ }\href
  {\doibase 10.1103/PhysRevLett.89.261101} {\bibfield  {journal} {\bibinfo
  {journal} {Phys. Rev. Lett.}\ }\textbf {\bibinfo {volume} {89}},\ \bibinfo
  {pages} {261101} (\bibinfo {year} {2002})},\ \Eprint
  {http://arxiv.org/abs/gr-qc/0207080} {arXiv:gr-qc/0207080} \BibitemShut
  {NoStop}%
\bibitem [{\citenamefont {Ashtekar}\ and\ \citenamefont
  {Krishnan}(2003)}]{Ashtekar:2003hk}%
  \BibitemOpen
  \bibfield  {author} {\bibinfo {author} {\bibfnamefont {Abhay}\ \bibnamefont
  {Ashtekar}}\ and\ \bibinfo {author} {\bibfnamefont {Badri}\ \bibnamefont
  {Krishnan}},\ }\bibfield  {title} {\enquote {\bibinfo {title} {{Dynamical
  horizons and their properties}},}\ }\href {\doibase
  10.1103/PhysRevD.68.104030} {\bibfield  {journal} {\bibinfo  {journal} {Phys.
  Rev.}\ }\textbf {\bibinfo {volume} {D68}},\ \bibinfo {pages} {104030}
  (\bibinfo {year} {2003})},\ \Eprint {http://arxiv.org/abs/gr-qc/0308033}
  {arXiv:gr-qc/0308033} \BibitemShut {NoStop}%
\bibitem [{\citenamefont {Ashtekar}\ and\ \citenamefont
  {Krishnan}(2004)}]{Ashtekar:2004cn}%
  \BibitemOpen
  \bibfield  {author} {\bibinfo {author} {\bibfnamefont {Abhay}\ \bibnamefont
  {Ashtekar}}\ and\ \bibinfo {author} {\bibfnamefont {Badri}\ \bibnamefont
  {Krishnan}},\ }\bibfield  {title} {\enquote {\bibinfo {title} {{Isolated and
  dynamical horizons and their applications}},}\ }\href@noop {} {\bibfield
  {journal} {\bibinfo  {journal} {Living Rev. Rel.}\ }\textbf {\bibinfo
  {volume} {7}},\ \bibinfo {pages} {10} (\bibinfo {year} {2004})},\ \Eprint
  {http://arxiv.org/abs/gr-qc/0407042} {arXiv:gr-qc/0407042} \BibitemShut
  {NoStop}%
\bibitem [{\citenamefont {Engelhardt}\ and\ \citenamefont
  {Wall}(2018)}]{Engelhardt:2017aux}%
  \BibitemOpen
  \bibfield  {author} {\bibinfo {author} {\bibfnamefont {Netta}\ \bibnamefont
  {Engelhardt}}\ and\ \bibinfo {author} {\bibfnamefont {Aron~C.}\ \bibnamefont
  {Wall}},\ }\bibfield  {title} {\enquote {\bibinfo {title} {{Decoding the
  Apparent Horizon: Coarse-Grained Holographic Entropy}},}\ }\href {\doibase
  10.1103/PhysRevLett.121.211301} {\bibfield  {journal} {\bibinfo  {journal}
  {Phys. Rev. Lett.}\ }\textbf {\bibinfo {volume} {121}},\ \bibinfo {pages}
  {211301} (\bibinfo {year} {2018})},\ \Eprint
  {http://arxiv.org/abs/1706.02038} {arXiv:1706.02038 [hep-th]} \BibitemShut
  {NoStop}%
\bibitem [{\citenamefont {Andersson}\ \emph {et~al.}(2005)\citenamefont
  {Andersson}, \citenamefont {Mars},\ and\ \citenamefont
  {Simon}}]{Andersson:2005gq}%
  \BibitemOpen
  \bibfield  {author} {\bibinfo {author} {\bibfnamefont {Lars}\ \bibnamefont
  {Andersson}}, \bibinfo {author} {\bibfnamefont {Marc}\ \bibnamefont {Mars}},
  \ and\ \bibinfo {author} {\bibfnamefont {Walter}\ \bibnamefont {Simon}},\
  }\bibfield  {title} {\enquote {\bibinfo {title} {{Local existence of
  dynamical and trapping horizons}},}\ }\href {\doibase
  10.1103/PhysRevLett.95.111102} {\bibfield  {journal} {\bibinfo  {journal}
  {Phys.Rev.Lett.}\ }\textbf {\bibinfo {volume} {95}},\ \bibinfo {pages}
  {111102} (\bibinfo {year} {2005})},\ \Eprint
  {http://arxiv.org/abs/gr-qc/0506013} {arXiv:gr-qc/0506013 [gr-qc]}
  \BibitemShut {NoStop}%
\bibitem [{\citenamefont {Andersson}\ \emph {et~al.}(2008)\citenamefont
  {Andersson}, \citenamefont {Mars},\ and\ \citenamefont
  {Simon}}]{Andersson:2007fh}%
  \BibitemOpen
  \bibfield  {author} {\bibinfo {author} {\bibfnamefont {Lars}\ \bibnamefont
  {Andersson}}, \bibinfo {author} {\bibfnamefont {Marc}\ \bibnamefont {Mars}},
  \ and\ \bibinfo {author} {\bibfnamefont {Walter}\ \bibnamefont {Simon}},\
  }\bibfield  {title} {\enquote {\bibinfo {title} {{Stability of marginally
  outer trapped surfaces and existence of marginally outer trapped tubes}},}\
  }\href@noop {} {\bibfield  {journal} {\bibinfo  {journal}
  {Adv.Theor.Math.Phys.}\ }\textbf {\bibinfo {volume} {12}} (\bibinfo {year}
  {2008})},\ \Eprint {http://arxiv.org/abs/0704.2889} {arXiv:0704.2889 [gr-qc]}
  \BibitemShut {NoStop}%
\bibitem [{\citenamefont {Booth}\ \emph {et~al.}(2021)\citenamefont {Booth},
  \citenamefont {Hennigar},\ and\ \citenamefont {Pook-Kolb}}]{Booth:2021sow}%
  \BibitemOpen
  \bibfield  {author} {\bibinfo {author} {\bibfnamefont {Ivan}\ \bibnamefont
  {Booth}}, \bibinfo {author} {\bibfnamefont {Robie~A.}\ \bibnamefont
  {Hennigar}}, \ and\ \bibinfo {author} {\bibfnamefont {Daniel}\ \bibnamefont
  {Pook-Kolb}},\ }\bibfield  {title} {\enquote {\bibinfo {title} {{Ultimate
  fate of apparent horizons during a binary black hole merger. I. Locating and
  understanding axisymmetric marginally outer trapped surfaces}},}\ }\href
  {\doibase 10.1103/PhysRevD.104.084083} {\bibfield  {journal} {\bibinfo
  {journal} {Phys. Rev. D}\ }\textbf {\bibinfo {volume} {104}},\ \bibinfo
  {pages} {084083} (\bibinfo {year} {2021})},\ \Eprint
  {http://arxiv.org/abs/2104.11343} {arXiv:2104.11343 [gr-qc]} \BibitemShut
  {NoStop}%
\bibitem [{\citenamefont {Pook-Kolb}\ \emph
  {et~al.}(2021{\natexlab{a}})\citenamefont {Pook-Kolb}, \citenamefont
  {Hennigar},\ and\ \citenamefont {Booth}}]{Pook-Kolb:2021hvz}%
  \BibitemOpen
  \bibfield  {author} {\bibinfo {author} {\bibfnamefont {Daniel}\ \bibnamefont
  {Pook-Kolb}}, \bibinfo {author} {\bibfnamefont {Robie~A.}\ \bibnamefont
  {Hennigar}}, \ and\ \bibinfo {author} {\bibfnamefont {Ivan}\ \bibnamefont
  {Booth}},\ }\bibfield  {title} {\enquote {\bibinfo {title} {{What Happens to
  Apparent Horizons in a Binary Black Hole Merger?}}}\ }\href {\doibase
  10.1103/physrevlett.127.181101} {\bibfield  {journal} {\bibinfo  {journal}
  {Phys. Rev. Lett.}\ }\textbf {\bibinfo {volume} {127}},\ \bibinfo {pages}
  {181101} (\bibinfo {year} {2021}{\natexlab{a}})},\ \Eprint
  {http://arxiv.org/abs/2104.10265} {arXiv:2104.10265 [gr-qc]} \BibitemShut
  {NoStop}%
\bibitem [{\citenamefont {Pook-Kolb}\ \emph
  {et~al.}(2019{\natexlab{a}})\citenamefont {Pook-Kolb}, \citenamefont
  {Birnholtz}, \citenamefont {Krishnan},\ and\ \citenamefont
  {Schnetter}}]{pook-kolb:2018igu}%
  \BibitemOpen
  \bibfield  {author} {\bibinfo {author} {\bibfnamefont {Daniel}\ \bibnamefont
  {Pook-Kolb}}, \bibinfo {author} {\bibfnamefont {Ofek}\ \bibnamefont
  {Birnholtz}}, \bibinfo {author} {\bibfnamefont {Badri}\ \bibnamefont
  {Krishnan}}, \ and\ \bibinfo {author} {\bibfnamefont {Erik}\ \bibnamefont
  {Schnetter}},\ }\bibfield  {title} {\enquote {\bibinfo {title} {Existence and
  stability of marginally trapped surfaces in black-hole spacetimes},}\ }\href
  {\doibase 10.1103/PhysRevD.99.064005} {\bibfield  {journal} {\bibinfo
  {journal} {Phys. Rev. D}\ }\textbf {\bibinfo {volume} {99}},\ \bibinfo
  {pages} {064005} (\bibinfo {year} {2019}{\natexlab{a}})}\BibitemShut
  {NoStop}%
\bibitem [{\citenamefont {Pook-Kolb}\ \emph
  {et~al.}(2019{\natexlab{b}})\citenamefont {Pook-Kolb}, \citenamefont
  {Birnholtz}, \citenamefont {Krishnan},\ and\ \citenamefont
  {Schnetter}}]{PhysRevLett.123.171102}%
  \BibitemOpen
  \bibfield  {author} {\bibinfo {author} {\bibfnamefont {Daniel}\ \bibnamefont
  {Pook-Kolb}}, \bibinfo {author} {\bibfnamefont {Ofek}\ \bibnamefont
  {Birnholtz}}, \bibinfo {author} {\bibfnamefont {Badri}\ \bibnamefont
  {Krishnan}}, \ and\ \bibinfo {author} {\bibfnamefont {Erik}\ \bibnamefont
  {Schnetter}},\ }\bibfield  {title} {\enquote {\bibinfo {title} {Interior of a
  binary black hole merger},}\ }\href {\doibase 10.1103/PhysRevLett.123.171102}
  {\bibfield  {journal} {\bibinfo  {journal} {Phys. Rev. Lett.}\ }\textbf
  {\bibinfo {volume} {123}},\ \bibinfo {pages} {171102} (\bibinfo {year}
  {2019}{\natexlab{b}})}\BibitemShut {NoStop}%
\bibitem [{\citenamefont {Pook-Kolb}\ \emph
  {et~al.}(2019{\natexlab{c}})\citenamefont {Pook-Kolb}, \citenamefont
  {Birnholtz}, \citenamefont {Krishnan},\ and\ \citenamefont
  {Schnetter}}]{PhysRevD.100.084044}%
  \BibitemOpen
  \bibfield  {author} {\bibinfo {author} {\bibfnamefont {Daniel}\ \bibnamefont
  {Pook-Kolb}}, \bibinfo {author} {\bibfnamefont {Ofek}\ \bibnamefont
  {Birnholtz}}, \bibinfo {author} {\bibfnamefont {Badri}\ \bibnamefont
  {Krishnan}}, \ and\ \bibinfo {author} {\bibfnamefont {Erik}\ \bibnamefont
  {Schnetter}},\ }\bibfield  {title} {\enquote {\bibinfo {title}
  {Self-intersecting marginally outer trapped surfaces},}\ }\href {\doibase
  10.1103/PhysRevD.100.084044} {\bibfield  {journal} {\bibinfo  {journal}
  {Phys. Rev. D}\ }\textbf {\bibinfo {volume} {100}},\ \bibinfo {pages}
  {084044} (\bibinfo {year} {2019}{\natexlab{c}})}\BibitemShut {NoStop}%
\bibitem [{\citenamefont {Pook-Kolb}\ \emph
  {et~al.}(2020{\natexlab{a}})\citenamefont {Pook-Kolb}, \citenamefont
  {Birnholtz}, \citenamefont {Jaramillo}, \citenamefont {Krishnan},\ and\
  \citenamefont {Schnetter}}]{pook-kolb2020I}%
  \BibitemOpen
  \bibfield  {author} {\bibinfo {author} {\bibfnamefont {Daniel}\ \bibnamefont
  {Pook-Kolb}}, \bibinfo {author} {\bibfnamefont {Ofek}\ \bibnamefont
  {Birnholtz}}, \bibinfo {author} {\bibfnamefont {Jos{\'e}~Luis}\ \bibnamefont
  {Jaramillo}}, \bibinfo {author} {\bibfnamefont {Badri}\ \bibnamefont
  {Krishnan}}, \ and\ \bibinfo {author} {\bibfnamefont {Erik}\ \bibnamefont
  {Schnetter}},\ }\bibfield  {title} {\enquote {\bibinfo {title} {{Horizons in
  a binary black hole merger I: Geometry and area increase}},}\ }\href@noop {}
  {\  (\bibinfo {year} {2020}{\natexlab{a}})},\ \Eprint
  {http://arxiv.org/abs/2006.03939} {arXiv:2006.03939 [gr-qc]} \BibitemShut
  {NoStop}%
\bibitem [{\citenamefont {Pook-Kolb}\ \emph
  {et~al.}(2020{\natexlab{b}})\citenamefont {Pook-Kolb}, \citenamefont
  {Birnholtz}, \citenamefont {Jaramillo}, \citenamefont {Krishnan},\ and\
  \citenamefont {Schnetter}}]{pook-kolb2020II}%
  \BibitemOpen
  \bibfield  {author} {\bibinfo {author} {\bibfnamefont {Daniel}\ \bibnamefont
  {Pook-Kolb}}, \bibinfo {author} {\bibfnamefont {Ofek}\ \bibnamefont
  {Birnholtz}}, \bibinfo {author} {\bibfnamefont {Jos{\'e}~Luis}\ \bibnamefont
  {Jaramillo}}, \bibinfo {author} {\bibfnamefont {Badri}\ \bibnamefont
  {Krishnan}}, \ and\ \bibinfo {author} {\bibfnamefont {Erik}\ \bibnamefont
  {Schnetter}},\ }\bibfield  {title} {\enquote {\bibinfo {title} {{Horizons in
  a binary black hole merger II: Fluxes, multipole moments and stability}},}\
  }\href@noop {} {\  (\bibinfo {year} {2020}{\natexlab{b}})},\ \Eprint
  {http://arxiv.org/abs/2006.03940} {arXiv:2006.03940 [gr-qc]} \BibitemShut
  {NoStop}%
\bibitem [{\citenamefont {Pook-Kolb}\ \emph
  {et~al.}(2021{\natexlab{b}})\citenamefont {Pook-Kolb}, \citenamefont
  {Booth},\ and\ \citenamefont {Hennigar}}]{Pook-Kolb:2021jpd}%
  \BibitemOpen
  \bibfield  {author} {\bibinfo {author} {\bibfnamefont {Daniel}\ \bibnamefont
  {Pook-Kolb}}, \bibinfo {author} {\bibfnamefont {Ivan}\ \bibnamefont {Booth}},
  \ and\ \bibinfo {author} {\bibfnamefont {Robie~A.}\ \bibnamefont
  {Hennigar}},\ }\bibfield  {title} {\enquote {\bibinfo {title} {{Ultimate fate
  of apparent horizons during a binary black hole merger. II. The vanishing of
  apparent horizons}},}\ }\href {\doibase 10.1103/PhysRevD.104.084084}
  {\bibfield  {journal} {\bibinfo  {journal} {Phys. Rev. D}\ }\textbf {\bibinfo
  {volume} {104}},\ \bibinfo {pages} {084084} (\bibinfo {year}
  {2021}{\natexlab{b}})},\ \Eprint {http://arxiv.org/abs/2104.11344}
  {arXiv:2104.11344 [gr-qc]} \BibitemShut {NoStop}%
\bibitem [{\citenamefont {Ben-Dov}(2004)}]{BenDov:2004gh}%
  \BibitemOpen
  \bibfield  {author} {\bibinfo {author} {\bibfnamefont {Ishai}\ \bibnamefont
  {Ben-Dov}},\ }\bibfield  {title} {\enquote {\bibinfo {title} {{The Penrose
  inequality and apparent horizons}},}\ }\href {\doibase
  10.1103/PhysRevD.70.124031} {\bibfield  {journal} {\bibinfo  {journal} {Phys.
  Rev. D}\ }\textbf {\bibinfo {volume} {70}},\ \bibinfo {pages} {124031}
  (\bibinfo {year} {2004})},\ \Eprint {http://arxiv.org/abs/gr-qc/0408066}
  {arXiv:gr-qc/0408066} \BibitemShut {NoStop}%
\bibitem [{\citenamefont {Booth}\ \emph {et~al.}(2006)\citenamefont {Booth},
  \citenamefont {Brits}, \citenamefont {Gonzalez},\ and\ \citenamefont {Van
  Den~Broeck}}]{Booth:2005ng}%
  \BibitemOpen
  \bibfield  {author} {\bibinfo {author} {\bibfnamefont {Ivan}\ \bibnamefont
  {Booth}}, \bibinfo {author} {\bibfnamefont {Lionel}\ \bibnamefont {Brits}},
  \bibinfo {author} {\bibfnamefont {Jose~A.}\ \bibnamefont {Gonzalez}}, \ and\
  \bibinfo {author} {\bibfnamefont {Chris}\ \bibnamefont {Van Den~Broeck}},\
  }\bibfield  {title} {\enquote {\bibinfo {title} {{Marginally trapped tubes
  and dynamical horizons}},}\ }\href {\doibase 10.1088/0264-9381/23/2/009}
  {\bibfield  {journal} {\bibinfo  {journal} {Class. Quant. Grav.}\ }\textbf
  {\bibinfo {volume} {23}},\ \bibinfo {pages} {413--440} (\bibinfo {year}
  {2006})},\ \Eprint {http://arxiv.org/abs/gr-qc/0506119} {arXiv:gr-qc/0506119}
  \BibitemShut {NoStop}%
\bibitem [{\citenamefont {Booth}\ \emph {et~al.}(2020)\citenamefont {Booth},
  \citenamefont {Hennigar},\ and\ \citenamefont {Mondal}}]{Booth:2020qhb}%
  \BibitemOpen
  \bibfield  {author} {\bibinfo {author} {\bibfnamefont {Ivan}\ \bibnamefont
  {Booth}}, \bibinfo {author} {\bibfnamefont {Robie~A.}\ \bibnamefont
  {Hennigar}}, \ and\ \bibinfo {author} {\bibfnamefont {Saikat}\ \bibnamefont
  {Mondal}},\ }\bibfield  {title} {\enquote {\bibinfo {title} {{Marginally
  outer trapped surfaces in the Schwarzschild spacetime: Multiple
  self-intersections and extreme mass ratio mergers}},}\ }\href {\doibase
  10.1103/PhysRevD.102.044031} {\bibfield  {journal} {\bibinfo  {journal}
  {Phys. Rev. D}\ }\textbf {\bibinfo {volume} {102}},\ \bibinfo {pages}
  {044031} (\bibinfo {year} {2020})},\ \Eprint
  {http://arxiv.org/abs/2005.05350} {arXiv:2005.05350 [gr-qc]} \BibitemShut
  {NoStop}%
\bibitem [{\citenamefont {Emparan}\ and\ \citenamefont
  {Martinez}(2016)}]{Emparan:2016ylg}%
  \BibitemOpen
  \bibfield  {author} {\bibinfo {author} {\bibfnamefont {Roberto}\ \bibnamefont
  {Emparan}}\ and\ \bibinfo {author} {\bibfnamefont {Marina}\ \bibnamefont
  {Martinez}},\ }\bibfield  {title} {\enquote {\bibinfo {title} {{Exact Event
  Horizon of a Black Hole Merger}},}\ }\href {\doibase
  10.1088/0264-9381/33/15/155003} {\bibfield  {journal} {\bibinfo  {journal}
  {Class. Quant. Grav.}\ }\textbf {\bibinfo {volume} {33}},\ \bibinfo {pages}
  {155003} (\bibinfo {year} {2016})},\ \Eprint
  {http://arxiv.org/abs/1603.00712} {arXiv:1603.00712 [gr-qc]} \BibitemShut
  {NoStop}%
\bibitem [{\citenamefont {Glavan}\ and\ \citenamefont
  {Lin}(2020)}]{Glavan:2019inb}%
  \BibitemOpen
  \bibfield  {author} {\bibinfo {author} {\bibfnamefont {Dra\v{z}en}\
  \bibnamefont {Glavan}}\ and\ \bibinfo {author} {\bibfnamefont {Chunshan}\
  \bibnamefont {Lin}},\ }\bibfield  {title} {\enquote {\bibinfo {title}
  {{Einstein-Gauss-Bonnet Gravity in Four-Dimensional Spacetime}},}\ }\href
  {\doibase 10.1103/PhysRevLett.124.081301} {\bibfield  {journal} {\bibinfo
  {journal} {Phys. Rev. Lett.}\ }\textbf {\bibinfo {volume} {124}},\ \bibinfo
  {pages} {081301} (\bibinfo {year} {2020})},\ \Eprint
  {http://arxiv.org/abs/1905.03601} {arXiv:1905.03601 [gr-qc]} \BibitemShut
  {NoStop}%
\bibitem [{\citenamefont {Lu}\ and\ \citenamefont {Pang}(2020)}]{Lu:2020iav}%
  \BibitemOpen
  \bibfield  {author} {\bibinfo {author} {\bibfnamefont {H.}~\bibnamefont
  {Lu}}\ and\ \bibinfo {author} {\bibfnamefont {Yi}~\bibnamefont {Pang}},\
  }\bibfield  {title} {\enquote {\bibinfo {title} {{Horndeski gravity as $D
  \rightarrow 4$ limit of Gauss-Bonnet}},}\ }\href {\doibase
  10.1016/j.physletb.2020.135717} {\bibfield  {journal} {\bibinfo  {journal}
  {Phys. Lett. B}\ }\textbf {\bibinfo {volume} {809}},\ \bibinfo {pages}
  {135717} (\bibinfo {year} {2020})},\ \Eprint
  {http://arxiv.org/abs/2003.11552} {arXiv:2003.11552 [gr-qc]} \BibitemShut
  {NoStop}%
\bibitem [{\citenamefont {Fernandes}\ \emph {et~al.}(2020)\citenamefont
  {Fernandes}, \citenamefont {Carrilho}, \citenamefont {Clifton},\ and\
  \citenamefont {Mulryne}}]{Fernandes:2020nbq}%
  \BibitemOpen
  \bibfield  {author} {\bibinfo {author} {\bibfnamefont {Pedro G.~S.}\
  \bibnamefont {Fernandes}}, \bibinfo {author} {\bibfnamefont {Pedro}\
  \bibnamefont {Carrilho}}, \bibinfo {author} {\bibfnamefont {Timothy}\
  \bibnamefont {Clifton}}, \ and\ \bibinfo {author} {\bibfnamefont {David~J.}\
  \bibnamefont {Mulryne}},\ }\bibfield  {title} {\enquote {\bibinfo {title}
  {{Derivation of Regularized Field Equations for the Einstein-Gauss-Bonnet
  Theory in Four Dimensions}},}\ }\href {\doibase 10.1103/PhysRevD.102.024025}
  {\bibfield  {journal} {\bibinfo  {journal} {Phys. Rev. D}\ }\textbf {\bibinfo
  {volume} {102}},\ \bibinfo {pages} {024025} (\bibinfo {year} {2020})},\
  \Eprint {http://arxiv.org/abs/2004.08362} {arXiv:2004.08362 [gr-qc]}
  \BibitemShut {NoStop}%
\bibitem [{\citenamefont {Hennigar}\ \emph {et~al.}(2020)\citenamefont
  {Hennigar}, \citenamefont {Kubiz\v{n}\'ak}, \citenamefont {Mann},\ and\
  \citenamefont {Pollack}}]{Hennigar:2020lsl}%
  \BibitemOpen
  \bibfield  {author} {\bibinfo {author} {\bibfnamefont {Robie~A.}\
  \bibnamefont {Hennigar}}, \bibinfo {author} {\bibfnamefont {David}\
  \bibnamefont {Kubiz\v{n}\'ak}}, \bibinfo {author} {\bibfnamefont {Robert~B.}\
  \bibnamefont {Mann}}, \ and\ \bibinfo {author} {\bibfnamefont {Christopher}\
  \bibnamefont {Pollack}},\ }\bibfield  {title} {\enquote {\bibinfo {title}
  {{On taking the D \textrightarrow{} 4 limit of Gauss-Bonnet gravity: theory
  and solutions}},}\ }\href {\doibase 10.1007/JHEP07(2020)027} {\bibfield
  {journal} {\bibinfo  {journal} {JHEP}\ }\textbf {\bibinfo {volume} {07}},\
  \bibinfo {pages} {027} (\bibinfo {year} {2020})},\ \Eprint
  {http://arxiv.org/abs/2004.09472} {arXiv:2004.09472 [gr-qc]} \BibitemShut
  {NoStop}%
\bibitem [{\citenamefont {Faraoni}\ and\ \citenamefont
  {Vachon}(2020)}]{Faraoni:2020ehi}%
  \BibitemOpen
  \bibfield  {author} {\bibinfo {author} {\bibfnamefont {Valerio}\ \bibnamefont
  {Faraoni}}\ and\ \bibinfo {author} {\bibfnamefont {Genevi\`eve}\ \bibnamefont
  {Vachon}},\ }\bibfield  {title} {\enquote {\bibinfo {title} {{When
  Painlev\'e\textendash{}Gullstrand coordinates fail}},}\ }\href {\doibase
  10.1140/epjc/s10052-020-8345-4} {\bibfield  {journal} {\bibinfo  {journal}
  {Eur. Phys. J. C}\ }\textbf {\bibinfo {volume} {80}},\ \bibinfo {pages} {771}
  (\bibinfo {year} {2020})},\ \Eprint {http://arxiv.org/abs/2006.10827}
  {arXiv:2006.10827 [gr-qc]} \BibitemShut {NoStop}%
\bibitem [{\citenamefont {Volovik}(2003)}]{Volovik:2003ga}%
  \BibitemOpen
  \bibfield  {author} {\bibinfo {author} {\bibfnamefont {G.~E.}\ \bibnamefont
  {Volovik}},\ }\bibfield  {title} {\enquote {\bibinfo {title} {{What can the
  quantum liquid say on the brane black hole, the entropy of extremal black
  hole and the vacuum energy?}}}\ }\href {\doibase 10.1023/A:1023762013553}
  {\bibfield  {journal} {\bibinfo  {journal} {Found. Phys.}\ }\textbf {\bibinfo
  {volume} {33}},\ \bibinfo {pages} {349--368} (\bibinfo {year} {2003})},\
  \Eprint {http://arxiv.org/abs/gr-qc/0301043} {arXiv:gr-qc/0301043}
  \BibitemShut {NoStop}%
\bibitem [{\citenamefont {Nielsen}\ and\ \citenamefont
  {Visser}(2006)}]{Nielsen:2005af}%
  \BibitemOpen
  \bibfield  {author} {\bibinfo {author} {\bibfnamefont {Alex~B.}\ \bibnamefont
  {Nielsen}}\ and\ \bibinfo {author} {\bibfnamefont {Matt}\ \bibnamefont
  {Visser}},\ }\bibfield  {title} {\enquote {\bibinfo {title} {{Production and
  decay of evolving horizons}},}\ }\href {\doibase 10.1088/0264-9381/23/14/006}
  {\bibfield  {journal} {\bibinfo  {journal} {Class. Quant. Grav.}\ }\textbf
  {\bibinfo {volume} {23}},\ \bibinfo {pages} {4637--4658} (\bibinfo {year}
  {2006})},\ \Eprint {http://arxiv.org/abs/gr-qc/0510083} {arXiv:gr-qc/0510083}
  \BibitemShut {NoStop}%
\bibitem [{\citenamefont {Lin}\ and\ \citenamefont {Soo}(2009)}]{Lin:2008xg}%
  \BibitemOpen
  \bibfield  {author} {\bibinfo {author} {\bibfnamefont {Chun-Yu}\ \bibnamefont
  {Lin}}\ and\ \bibinfo {author} {\bibfnamefont {Chopin}\ \bibnamefont {Soo}},\
  }\bibfield  {title} {\enquote {\bibinfo {title} {{Generalized
  Painleve-Gullstrand metrics}},}\ }\href {\doibase
  10.1016/j.physletb.2008.12.051} {\bibfield  {journal} {\bibinfo  {journal}
  {Phys. Lett. B}\ }\textbf {\bibinfo {volume} {671}},\ \bibinfo {pages}
  {493--495} (\bibinfo {year} {2009})},\ \Eprint
  {http://arxiv.org/abs/0810.2161} {arXiv:0810.2161 [gr-qc]} \BibitemShut
  {NoStop}%
\bibitem [{\citenamefont {Martel}\ and\ \citenamefont
  {Poisson}(2001)}]{Martel:2000rn}%
  \BibitemOpen
  \bibfield  {author} {\bibinfo {author} {\bibfnamefont {Karl}\ \bibnamefont
  {Martel}}\ and\ \bibinfo {author} {\bibfnamefont {Eric}\ \bibnamefont
  {Poisson}},\ }\bibfield  {title} {\enquote {\bibinfo {title} {{Regular
  coordinate systems for Schwarzschild and other spherical space-times}},}\
  }\href {\doibase 10.1119/1.1336836} {\bibfield  {journal} {\bibinfo
  {journal} {Am. J. Phys.}\ }\textbf {\bibinfo {volume} {69}},\ \bibinfo
  {pages} {476--480} (\bibinfo {year} {2001})},\ \Eprint
  {http://arxiv.org/abs/gr-qc/0001069} {arXiv:gr-qc/0001069} \BibitemShut
  {NoStop}%
\bibitem [{\citenamefont {Boyd}(2001)}]{Boyd}%
  \BibitemOpen
  \bibfield  {author} {\bibinfo {author} {\bibfnamefont {J.~P.}\ \bibnamefont
  {Boyd}},\ }\href@noop {} {\emph {\bibinfo {title} {{Chebyshev and Fourier
  Spectral Methods}}}}\ (\bibinfo  {publisher} {{Dover Publications}},\
  \bibinfo {year} {2001})\BibitemShut {NoStop}%
\bibitem [{\citenamefont {Canuto}\ \emph {et~al.}(2007)\citenamefont {Canuto},
  \citenamefont {Hussaini}, \citenamefont {Quarteroni},\ and\ \citenamefont
  {Zang}}]{canuto2007spectral}%
  \BibitemOpen
  \bibfield  {author} {\bibinfo {author} {\bibfnamefont {C.}~\bibnamefont
  {Canuto}}, \bibinfo {author} {\bibfnamefont {M.Y.}\ \bibnamefont {Hussaini}},
  \bibinfo {author} {\bibfnamefont {A.}~\bibnamefont {Quarteroni}}, \ and\
  \bibinfo {author} {\bibfnamefont {T.A.}\ \bibnamefont {Zang}},\ }\href
  {https://books.google.ca/books?id=DFJB0kiq0CQC} {\emph {\bibinfo {title}
  {Spectral Methods: Fundamentals in Single Domains}}},\ Scientific
  Computation\ (\bibinfo  {publisher} {Springer Berlin Heidelberg},\ \bibinfo
  {year} {2007})\BibitemShut {NoStop}%
\bibitem [{\citenamefont {Pook-Kolb}\ \emph
  {et~al.}(2021{\natexlab{c}})\citenamefont {Pook-Kolb}, \citenamefont
  {Birnholtz}, \citenamefont {Booth}, \citenamefont {Hennigar}, \citenamefont
  {Jaramillo}, \citenamefont {Krishnan}, \citenamefont {Schnetter},\ and\
  \citenamefont {Zhang}}]{pook_kolb_daniel_2021_4687700}%
  \BibitemOpen
  \bibfield  {author} {\bibinfo {author} {\bibfnamefont {Daniel}\ \bibnamefont
  {Pook-Kolb}}, \bibinfo {author} {\bibfnamefont {Ofek}\ \bibnamefont
  {Birnholtz}}, \bibinfo {author} {\bibfnamefont {Ivan}\ \bibnamefont {Booth}},
  \bibinfo {author} {\bibfnamefont {Robie~A.}\ \bibnamefont {Hennigar}},
  \bibinfo {author} {\bibfnamefont {Jos{\'e}~Luis}\ \bibnamefont {Jaramillo}},
  \bibinfo {author} {\bibfnamefont {Badri}\ \bibnamefont {Krishnan}}, \bibinfo
  {author} {\bibfnamefont {Erik}\ \bibnamefont {Schnetter}}, \ and\ \bibinfo
  {author} {\bibfnamefont {Victor}\ \bibnamefont {Zhang}},\ }\href {\doibase
  10.5281/zenodo.4687700} {\enquote {\bibinfo {title} {{MOTS Finder} version
  1.5},}\ } (\bibinfo {year} {2021}{\natexlab{c}})\BibitemShut {NoStop}%
\bibitem [{\citenamefont {Booth}\ \emph {et~al.}(2017)\citenamefont {Booth},
  \citenamefont {Kunduri},\ and\ \citenamefont {O'Grady}}]{Booth:2017fob}%
  \BibitemOpen
  \bibfield  {author} {\bibinfo {author} {\bibfnamefont {Ivan}\ \bibnamefont
  {Booth}}, \bibinfo {author} {\bibfnamefont {Hari~K.}\ \bibnamefont
  {Kunduri}}, \ and\ \bibinfo {author} {\bibfnamefont {Anna}\ \bibnamefont
  {O'Grady}},\ }\bibfield  {title} {\enquote {\bibinfo {title} {{Unstable
  marginally outer trapped surfaces in static spherically symmetric
  spacetimes}},}\ }\href {\doibase 10.1103/PhysRevD.96.024059} {\bibfield
  {journal} {\bibinfo  {journal} {Phys. Rev. D}\ }\textbf {\bibinfo {volume}
  {96}},\ \bibinfo {pages} {024059} (\bibinfo {year} {2017})},\ \Eprint
  {http://arxiv.org/abs/1705.03063} {arXiv:1705.03063 [gr-qc]} \BibitemShut
  {NoStop}%
\end{thebibliography}%
\end{document}